\documentclass[useAMS,usegraphicx,usenatbib]{mn2e}

\newcommand{\alm}{a_{\ell m}}
\newcommand{\var}{\mathop{\rm Var}\nolimits}
\newcommand{\trace}{\mathop{\rm Tr}\nolimits}
\newcommand{\mat}[1]{\mathsf{#1}}
\newcommand{\unit}[1]{\;\mathrm{#1}}
\renewcommand{\vec}[1]{\bmath{#1}}

\let\oldhat\hat 
\def\hat#1{\oldhat{\vec{#1}}}

\newcommand{\iimag}{\mathrm{i}}
\newcommand{\dderiv}{\mathrm{d}}

\title[On the large-angle anomalies of the microwave sky]{
  On the large-angle anomalies of the microwave sky}
\author[C.J. Copi, D. Huterer, D.J. Schwarz and G.D. Starkman]
{Craig J. Copi$^{1}$\thanks{E-mail: cjc5@cwru.edu},
  Dragan Huterer$^{2}$\thanks{E-mail: dhuterer@kicp.uchicago.edu},
  Dominik J. Schwarz$^{3}$\thanks{E-mail: dschwarz@physik.uni-bielefeld.de}
  and 
  Glenn D. Starkman$^{1}$\thanks{E-mail: glenn.starkman@case.edu}\\
  $^{1}$Department of Physics, Case Western Reserve University, Cleveland, 
  OH 44106-7079, USA\\
  $^{2}$ Kavli Institute for Cosmological Physics and
  Department of Astronomy and Astrophysics,
  University of Chicago, Chicago, IL~~60637, USA\\
  $^{3}$Fakult\"at f\"ur Physik, Universit\"at Bielefeld, Postfach 100131, 
  33501 Bielefeld, Germany}

\begin{document}

\date{Accepted xxxx. Received xxxx; in original form xxxx}

\pagerange{\pageref{firstpage}--\pageref{lastpage}} \pubyear{2005}

\maketitle

\label{firstpage}

\begin{abstract}
We apply the multipole vector framework to full-sky maps derived from the first
year WMAP data.  We significantly extend our earlier work showing that the two
lowest cosmologically interesting multipoles, $\ell=2$ and $3$, are not
statistically isotropic.  These results are compared to the findings obtained
using related methods.  In particular, we show that the planes of the
quadrupole and the octopole are unexpectedly aligned. Moreover, the combined
quadrupole plus octopole is surprisingly aligned with the geometry and
direction of motion of the solar system: the plane they define is perpendicular
to the ecliptic plane and to the plane defined by the dipole direction, and the
ecliptic plane carefully separates stronger from weaker extrema, running within
a couple of degrees of the null-contour between a maximum and a minimum over
more than $120\degr$ of the sky.  Even given the alignment of the quadrupole
and octopole with each other, we find that their alignment with the ecliptic is
unlikely at $>98$\% C.L., and argue that it is in fact unlikely at
$>99.9$\% C.L\@.  Most of the $\ell=2$ and $3$ multipole vectors of the known
Galactic foregrounds are located far from those of the observed sky, strongly
suggesting that residual contamination by such foregrounds is unlikely to be
the cause of the observed correlations.  Multipole vectors, like individual
$a_{\ell m}$, are very sensitive to sky cuts, and we demonstrate that analyses
using cut skies induce relatively large errors, thus weakening the observed
correlations but preserving their consistency with the full-sky
results. Similarly, the analysis of COBE cut-sky maps shows increased errors
but is consistent with WMAP full-sky results.  We briefly extend these
explorations to higher multipoles, noting again anomalous deviations from
statistical isotropy and comparing with ecliptic asymmetry manifest in the WMAP
team's own analysis.  If the correlations we observe are indeed a signal of
non-cosmic origin, then the lack of low-$\ell$ power will very likely be
exacerbated, with important consequences for our understanding of cosmology on
large scales.
\end{abstract}


\section{Introduction}

Following a series of increasingly successful experiments measuring the
anisotropy of the cosmic microwave background (CMB), full-sky maps obtained by
the Wilkinson Microwave Anisotropy Probe (WMAP) in its first year of
observation have revolutionized the study of the CMB sky \citep{Bennett_foregr,
Bennett2003, Hinshaw2003}. In particular, a number of cosmological
parameters have been determined with high accuracy
\citep{Spergel2003}. Moreover, a WMAP-type survey opens a unique window to the
physics of the early universe \citep{Peiris2003} and enables tests of the
standard inflationary picture, which predicts a CMB temperature anisotropy
pattern that is nearly scale-free, statistically isotropic and, to the accuracy
of all current or planned CMB experiments, Gaussian random (terms that we shall
define more carefully below).  Consequently, WMAP data led to many studies of
Gaussianity
\citep{Komatsu2003,Park2003,Chiang2003,Magueijo_Medeiros,Vielva2003,
Mukherjee2004,McEwen2004,Cruz2004,Eriksen_Npt,Larson2004,Naselsky2005a,
Naselsky2005b,Tojeiro2005,Cayon2005}
and statistical isotropy \citep{Eriksen_asym,Hajian2003,Hajian2004,Hajian2005,
Hansen_isotropy,Hansen_asym,Prunet,Donoghue} of the CMB. While there was no
significant evidence for the violation of Gaussianity and statistical isotropy
in the pre-WMAP era, some of the aforementioned post-WMAP studies found
evidence for violation of either Gaussianity or statistical isotropy, or both.

Arguably, the biggest surprises were to be found in temperature anisotropies on
the largest angular scales. Most prominent among the ``low-$\ell$ anomalies''
is the near vanishing of the two-point angular correlation function $C(\theta)$
at angular separations greater than about 60 degrees \citep{Spergel2003},
confirming what was first measured using the Cosmic Background Explorer's
Differential Microwave Radiometer (COBE-DMR) a decade ago \citep{DMR4_Ctheta}.
Beyond this long-standing anomaly in the overall amplitude of the large-angle
fluctuations, it has also been noted that the octopole of the CMB 
is planar and oriented parallel to the quadrupole
\citep{TOH,deOliveira2004}. Furthermore, three of the four planes determined by
the quadrupole and octopole are orthogonal to the ecliptic at a level
inconsistent with Gaussian random, statistically isotropic skies at $99.8$\%
C.L., while the normals to these planes are aligned at $99.9$\% C.L. with the
direction of the cosmological dipole and with the equinoxes
\citep{Schwarz2004}.  These peculiar correlations presumably are connected to
north-south asymmetry in the angular power spectrum
\citep{Eriksen_asym} and in the statistics of the extrema \citep{Wandelt:GEM}.
Finally, there is a suggestion that the presence of preferred directions in the
microwave background multipoles extends beyond the octopole to higher
multipoles \citep{Land2005a} and that there is an associated mirror symmetry \citep{Land2005c}.

The correct explanation of these unexpected correlations of the low-$\ell$
features of the microwave background with each other and with the solar system
is currently not known. There are four possibilities: (1)~there is a systematic
error (an error in the data analysis or instrument characterization), (2)~the
source is astrophysical (i.e.\ an unexpected foreground), (3)~it is
cosmological in nature (e.g.\ an anisotropic universe, such as one with
nontrivial topology), or (4)~the observed correlations are a pure statistical
fluke. The evidence for many of these low-$\ell$ correlations is strong ($>
99$\% C.L.) and presented with a variety of different methods (see, for
example, \citet{TOH,Schwarz2004,Land2005a,Hansen_asym,Eriksen_asym}).  It is
therefore unlikely that all of them are mere accidents.  In this paper, we will
at least attempt to clarify which of these and other correlations is likely to
be a statistical fluke.  Whatever the source of these correlations, until they
are understood, cosmological inferences drawn from low-$\ell$ WMAP data
(including polarization data), such as the possibility of early reionization,
should be viewed with a healthy dose of skepticism.

This paper has two principal goals. First, we would like to further develop the
theory of multipole vectors, a new representation of microwave background
anisotropy introduced by \citet{Copi2004}. In particular, we derive several new
results, and comment on various ideas and results on the tests of Gaussianity
and statistical isotropy that were subsequently presented by others using the
multipole vectors or their variants.  Our second major goal is to present a
detailed analysis of the large-angle correlations discussed in
\citet{Schwarz2004} and extend them in several directions.

The paper is organized as follows.  In Sec.~\ref{sec:standard}, we introduce
the WMAP maps and their treatment used in our analysis.  In Sec.~\ref{sec:MV},
we review the theory behind the multipole vectors, derive several new results,
and comment on other recent work on this topic.  Section \ref{sec:quad_oct}
deals in more detail with the morphology of the quadrupole and octopole.  In
Sec.~\ref{sec:stat}, we introduce various statistics to quantify the low-$\ell$
correlations, extend our analysis to higher $\ell$, and present the results.
Section~\ref{sec:foregr} includes a detailed analysis of the issue of
foregrounds and Sec.~\ref{sec:cutsky} describes the cut-sky reconstruction
algorithm and results based on reconstructed maps.  In Sec.~\ref{sec:COBE}, we
present a comparison to COBE.  The correlations of the WMAP angular power
spectrum with the ecliptic plane is discussed in Sec.~\ref{sec:plane_poles}.
We conclude in Sec.~\ref{sec:conclusion}.

\section{The microwave background sky}
\label{sec:standard}

The temperature fluctuations of the microwave background $\Delta
T(\theta,\phi;\vec x,t)$ are, in principle, functions of both the direction of
observation $(\theta,\phi)$, and the location, $\vec x$, and time, $t$, of the
observation.  In practice, essentially all our observations occur within the
solar system and over a cosmologically short period of time, so we can ignore
any local spatial and temporal variations in the microwave background.  Our
observations in each wavelength band are therefore the intensity of the
microwave background radiation as a function of direction on the celestial
sphere.

\subsection{The standard representation}

The most common representation of a real scalar function, $f(\theta,\phi)$,
on the sphere is as an expansion in terms of multipole moments 
\begin{equation}
  f(\theta,\phi) = \sum_{\ell=0}^{\infty} f_{\ell}(\theta,\phi) 
  \label{eqn:multipole-decomp}
\end{equation}
where the
$\ell$-th multipole, $f_\ell(\theta,\phi)$, is typically given by
\begin{equation}
  f_{\ell}(\theta,\phi) = \sum_{m=-\ell}^{\ell} a_{\ell m} Y_{\ell
    m}(\theta,\phi)  .
  \label{eqn:Ylm-decomp}
\end{equation}
Here $Y_{\ell m}(\theta,\phi)$ are the standard spherical harmonic
functions.  For $f_\ell (\theta,\phi)$ real the complex coefficients
$a_{\ell m}$ need to satisfy the reality condition $a_{\ell m}^* = (-1)^m
a_{\ell, -m}$ and for fixed $\ell$ we have $2\ell + 1$ independent, real
degrees of freedom.

Standard cosmological theory predicts that the CMB fluctuations sample a
statistically isotropic, Gaussian random field of zero mean.  
``Gaussian random'' means that the real and imaginary parts of the
$\alm$ are each an independent random variable that is distributed according to
a Gaussian distribution of zero mean.  In principle, the variances of these
Gaussian distributions (which, by the nature of a Gaussian, fully characterize
the distribution) could be different for each $\ell$ and $m$ and for both the
real and imaginary parts.  ``Statistically isotropic'' means that, instead,
these variances depend only on $\ell$.  Thus, the expectation of any pair of
$a_{\ell m}$ is
\begin{equation}
\left<a_{\ell'm'}^*\alm \right> = {\cal C}_\ell \delta_{\ell'\ell}\delta_{m'm},
\end{equation}
where ${\cal C}_\ell$ is the expected power in the $\ell$th multipole and 
its (standard) observable estimator is
\begin{equation}
  C_\ell \equiv \frac1{2\ell+1} \sum_{m=-\ell}^\ell \left| \alm \right|^2.
  \label{eqn:Cl}
\end{equation}
The set of estimators  $\{C_\ell\mid \ell=0,\ldots,\infty\}$ is called the 
angular power spectrum.  The cosmic variance in the estimators is
\begin{equation}
{\rm Var}(C_\ell) = 2 {\cal C}_\ell^2 /(2\ell + 1). 
\end{equation}

Since the distributions of Gaussian variables are completely determined by
their means and variances (and since the $a_{\ell m}$ have zero means), if the
microwave background sky is indeed a realization of a Gaussian random,
statistically isotropic process, then all of the accessible information in a
microwave background temperature map about the underlying physics is contained
in the angular power spectrum.  (Non-linear growth of fluctuations will cause
departures from Gaussianity.  However, these departures are small at the large
angular scales we will be considering.)

As mentioned in the introduction, many studies have been done looking for
evidence of non-Gaussianity or deviations from statistical isotropy in the
microwave background.  A significant difficulty is that in the absence of
particular models the range of possible manifestations of non-Gaussianity and
statistical-anisotropy is enormous.  Statistics that measure one manifestation
well, can be entirely insensitive to another.  Limits on particular statistics
should be viewed in that light.
Another conceptual and practical difficulty is separating tests of Gaussianity
from tests of statistical isotropy. If statistical isotropy,
$\left<a_{\ell'm'}^* \alm \right> \propto \delta_{\ell'\ell}\delta_{m'm}$,
is violated, then it is unclear how one can test Gaussianity, since each
$a_{\ell m}$ could have its own independent distribution for which we are
provided just one sample.

\subsection{Full-sky maps}
\label{sec:full-sky}

In this work, we use three full-sky maps based on the original single-frequency
WMAP maps. The first two, the Internal Linear Combination (ILC) map and the
Lagrange Internal Linear Combination (LILC) map, are minimum-variance maps
obtained from WMAP's single-frequency maps by \citet{Bennett_foregr} (the WMAP
team) and by \citet{LILC}, respectively. The third map is the cleaned full-sky
map of \citet{TOH} (henceforth the TOH map).  The full-sky maps may have
residual foreground contamination that is mainly due to imperfect subtraction
of the Galactic signal.  Furthermore, the full-sky maps have complicated noise
properties \citep{Bennett_foregr} that make them less than ideal for
cosmological tests. While one can, in principle, straightforwardly compute the
true (full-sky) multipole vectors from the single-frequency maps with the sky
cut (as explicitly shown for an isolatitude sky cut in Sec.~\ref{sec:cutsky}),
a Galaxy cut larger than a few degrees will introduce a significant uncertainty
in the reconstructed multipole vectors and consequently any statistics.
 
Nevertheless, there are compelling reasons why we believe the results of the
analysis of the full-sky maps.  First, all of the results are robust with
respect to choice of the map, despite the fact that the ILC and LILC maps were
obtained using a different method than the TOH map. This suggests that the
full-sky maps indeed resemble the true Galaxy-subtracted microwave sky.
Furthermore, one can show that the dominant component of the bias inherent in
creating an ILC-type map has a quadrupolar pattern, and in particular looks
like the spherical harmonic $Y_{20}$ with an amplitude of $\la 20\unit{\mu K}$
(H.K. Eriksen, private communication). For the quadrupole and octopole the
dominant contributions to the Galactic foregrounds are a linear combination of
$Y_{20}$ and ${\rm Re}(Y_{31})$, which is easily seen from the symmetry of the
Galaxy (approximately north-south and east-west symmetric with a hot spot at
the Galactic center and a cold spot at the anti-center). Indeed, the sum of the
synchrotron, free-free and dust WMAP foreground maps \citep{Bennett_foregr} in
the V-band gives $a_{20} = -217 \unit{\mu K}$ and ${\rm Re}(a_{31}) = 88
\unit{\mu K}$. These two modes make up about $90$\% of the power in the
quadrupole and octopole foregrounds. The next important modes in the WMAP
foreground maps turn out to be $\mathrm{Re}(Y_{22})$ and
$\mathrm{Re}(Y_{33})$. The quoted $20 \unit{\mu K}$ uncertainty in the ILC-type
map for the quadrupole can thus be understood as a $10$\% uncertainty in the
understanding of our galaxy.  The uncertainty in the octopole is considerably
smaller ($\la 10 \unit{\mu K}$).  As shown in \citet{Schwarz2004} and further
demonstrated in Sec.~\ref{sec:foregr}, this Galactic contamination, even if
present, would lead to Galactic and not the observed ecliptic (Solar System)
correlations.

\subsection{Methodological differences between the minimal variance maps}

The guiding principle in the construction of the WMAP ILC, TOH and LILC 
full-sky maps is to search for a temperature map with minimal variance.  
For the convenience of the reader we summarize the essential steps that 
lead to the cleaned full-sky maps (see the original papers for full details).

In all three approaches, the input are the five WMAP frequency maps (K, Ka,
Q, V and W band). The bands differ in a number of ways including noise
properties and angular resolution. While the ILC and LILC map reduce all
five bands to the K band resolution, the TOH map makes use of the higher
resolution of the higher frequency bands.  Each map can be written as
\begin{equation}
T(\nu_i) = T_{\rm CMB} + T_{\rm residual}(\nu_i) . 
\end{equation}
A combined map is created by the linear combination
\begin{equation}
T =  T_{\rm CMB} + \sum_{i=1}^5 w_i  T_{\rm residual}(\nu_i) ,
\end{equation} 
with $\sum_i w_i = 1$. Now it is assumed that the residuals 
(noise and foregrounds) and the CMB are uncorrelated, so that
\begin{equation}
{\rm Var}(T) = {\rm Var}(T_{\rm CMB}) +  
{\rm Var}\left(\sum_{i=1}^5 w_i  T_{\rm residual}(\nu_i)\right).
\end{equation}
The idea is now to determine the 4 independent weights by minimizing the
variance with respect to all pixels of a region of the sky.  The ILC and LILC
maps use the same 12 regions of the sky, whereas the TOH map uses 9
regions. For the ILC and LILC maps the weights are constants within a region of
the sky, whereas for the TOH map the weights depend on multipole number $\ell$
(the minimization is done in spherical harmonic space instead of pixel
space). The regions are selected according to the level of foreground
contamination. To produce the final maps a Gaussian smoothing is applied to
soften the edges. We see that the TOH map on the one hand and the ILC and LILC
maps on the other hand use different procedures.  The ILC and LILC maps differ
mainly in the detailed implementation of the method.  The drawback of the
minimal variance method is that large CMB fluctuations tend to be canceled by
large artificial foregrounds in order to obtain a small variance.

We shall see that, despite the differences between the three maps,
the results of our analysis are very similar.
This robustness argues strongly in favor of the legitimacy of using these full-sky maps 
for multipole vector analysis.

\subsection{Kinetic quadrupole correction}\label{sec:DQ}

By far the largest signal in the microwave background anisotropy is the dipole,
recently measured by WMAP \citep{Bennett2003} to be $(3.346\pm 0.017)\unit{mK}$
in the direction $(l=263 \fdg 85\pm0 \fdg 1, b=48 \fdg 25\pm0 \fdg 04)$ in
Galactic coordinates.  This is nearly two orders of magnitude larger than the
root-mean-square (rms) anisotropy in the dipole-subtracted sky, and so thought
not to be of cosmological origin, but rather to be caused by the motion of the
solar system with respect to the rest frame defined by the CMB. As shown by
\citet{dipole}, the dipole induced by a velocity $v$ is $\bar{T} (v/c)
\cos\theta$, where $\theta$ is measured from the direction of motion. Given
$\bar{T}=(2.725\pm0.002)\unit{K}$ \citep{Mather}, one infers that $v\simeq 370
\unit{km\,s^{-1}}$.

The solar motion also implies the presence of a kinematically induced Doppler
quadrupole (DQ; \citealt{dipole,kam_knox}). To second order in $\beta \equiv v/c
\simeq 10^{-3}$, the specific intensity of the CMB for an observer moving with
respect to the CMB rest-frame includes the usual monopole term with a
black-body spectrum ($\propto x^3/[e^x-1]$, where $x=h\nu/[k_{\rm B} T]$); a
dipolar term $\propto \cos\theta$, linear in $\beta$, with a dipole spectrum
($\propto x^4e^x/[e^x-1]^2$, the same as for primordial anisotropies); and a
quadrupolar term $\propto 3\cos^2\theta - 1$, quadratic in $\beta$, with a
quadrupole spectrum ($\propto x^5e^{x}[e^x+1]/[e^x-1]^3$).  Higher multipoles
are induced only at higher order in $\beta$ and so can be neglected.

To first approximation the quadrupole spectrum differs very little from the
dipole spectrum across the frequency range probed by WMAP. The DQ is itself a
small contribution to the quadrupole. It has a total band-power of only
$3.6\unit{\mu K}^2$ compared to $123.4\unit{\mu K}^2$ from the cut-sky WMAP
analysis \citep{Hinshaw2003}, $195.1\unit{\mu K}^2$ extracted
\citep{deOliveira2004} from the ILC map \citep{Bennett2003}, $201.6\unit{\mu
K}^2$ from the TOH map (\citet{TOH}) or $350.6\unit{\mu K}^2$ from the LILC map
(\citet{LILC}). Therefore, it is a good approximation to treat the
Doppler-quadrupole as having a dipole spectrum plus a small spectral distortion
which we shall ignore.  We can then readily subtract the DQ from any microwave
background map.

The kinetic quadrupole is a very small contribution to the total theoretical
power in the quadrupole, however, it gives rise to non-negligible contributions
to some of the $a_{2 m}$.  This is due partially to the low power in the
observed quadrupole and partially to the ``orthogonality'' of the correction
(the correction is $m$ dependent and the largest corrections are not to the
largest $a_{2 m}$).  This is a well known, well understood physical correction
to the quadrupole that is often ignored.  This correction \textit{must} be
applied when studying the alignment of the quadrupole, leaving it out
introduces a correctable systematic error.  Though it has little effect on the
power in the quadrupole, it has a noticeable effect on the quadrupole
orientation as shown below.  We again stress that it is the orientation of the
correction that makes it important.  A quadrupole correction of this size
pointing in a random direction would typically not lead to noticeable alignment
changes.

Figures \ref{fig:map:tegmark:2}, \ref{fig:map:tegmark:3}, and 
\ref{fig:map:tegmark:2+3} show the Doppler corrected quadrupole, the octopole,
and their sum, for the TOH map in Galactic coordinates in Mollweide projection. 
The corresponding multipoles from the ILC and LILC maps are very similar.  

\section{Multipole vectors}\label{sec:MV}

Multipole vectors provide an alternative representation of the complete set of
information represented by the full-sky map.  They represent a radical
reorganization of the information that nevertheless retains the integrity of
the individual multipoles. Moreover, the information is arranged in such away
as to be independent of the choice of coordinate system. Therefore the
multipole vectors may be a superior representation for probing the null
hypothesis of statistical isotropy, and for looking for signatures of specific
effects that might pick out special directions on the sky due to nonstandard
inflationary physics, systematic artifacts in the map-making, unexpected
foregrounds, cosmic topology, deviations from General Relativity, or other
unknown effects.

\subsection{Definition}

In the multipole vector representation, $f_\ell(\theta,\phi)$ is written in
terms of a scalar, $A^{(\ell)}$
and $\ell$ unit vectors, $\left\{\hat v^{(\ell,j)} \mid
j=1,\ldots,\ell\right\}$:
\begin{equation}
  \label{eqn:MVE}
  f_\ell(\theta,\phi) = A^{(\ell)} \left[ \prod_{i=1}^\ell
    \left(\hat v^{(\ell,i)}\cdot {\hat e}\right) - {\cal T}_\ell \right]  .
\end{equation}
Here $\hat e$ is the (radial) unit vector in the $(\theta,\phi)$ direction.  
(Henceforth, we will use  $\hat e$ and $(\theta,\phi)$ interchangeably.)
In Cartesian coordinates, $\hat e=(\sin\theta\cos\phi, \sin\theta\sin\phi, \cos\theta)$.
$\mathcal{T}_\ell$ is the sum of all possible traces of the first term;
rendering the full expression traceless.  In this context, a trace
means replacing a product of dipoles
$\left(\hat v^{(\ell,i)}\cdot {\hat e}\right)\left(\hat v^{(\ell,j)}
\cdot {\hat e}\right)$,
by
$\left(\hat v^{(\ell,i)}\cdot \hat v^{(\ell,j)}\right)$.
Equation (\ref{eqn:MVE}) can also be written as
\begin{equation}
  f_\ell(\hat e) = A^{(\ell)}  \left[ \hat v^{(\ell,1)} \cdots 
    \hat v^{(\ell,\ell)}\right]_{TF}^{i_1\cdots i_\ell}
  \left[{\hat e} \cdots {\hat e}  \right]^{TF}_{i_1\cdots i_\ell}
\end{equation}
where $\left[\cdots\right]_{TF}$ denote the trace free tensor product, and
the sum over repeated indices is assumed. These unit vectors
$\hat v^{(\ell,j)}$ are only defined up to a sign (and are thus ``headless
vectors''), as a change in sign of the vector can always be absorbed by the
scalar $A^{(\ell)}$.  We have chosen the convention that multipole vectors
point in the northern Galactic hemisphere, although when plotting them we
often instead show the southern counterpart for clarity.  

Note that only $A^{(\ell)}$ depends on the total power $C_\ell$.  The
multipole vectors are independent of $C_\ell$ --- if all the $a_{\ell m}$ of a given $\ell$
are multiplied by a common factor, then $A^\ell$ too will be multiplied  by that factor
and the $\hat v^{(\ell,i)}$ will remain unchanged.  In particular, let
\begin{equation}
{\tilde a}_{\ell m} \equiv a_{\ell m}/\sqrt{C_\ell}  ;
\end{equation}
then, the multipole vectors, $\hat v^{(\ell ,i)}$, depend only on the ${\tilde
a}_{\ell m}$ and not on $C_\ell$.  This is true independent of any assumptions
about Gaussianity and statistical isotropy.  Note, however, that $C_\ell$ and
$A^{(\ell)}$ do not contain identical information; $C_\ell$ is the two-point
correlation function while $A^{(\ell)}$ contains the two-point correlation as
well as a particular combination of the higher order moments.  An equivalent
definition of the estimator (\ref{eqn:Cl}) over the full sky is
\begin{equation}
C_\ell \equiv \frac{1}{2\ell +1} \int \dderiv \Omega \left[
  f_\ell(\theta,\phi) \right]^2.
\end{equation}
In terms of the multipole vectors~(\ref{eqn:MVE}), we find
\begin{equation}
  (2\ell+1) C_\ell = \left[ A^{(\ell)}\right]^2 \int \left[ \prod_j \hat
    v^{(\ell,j)}\cdot\hat e - \mathcal{T}_\ell \right]^2 \dderiv\Omega.
\end{equation}
Evaluating this for the monopole, dipole and quadrupole we obtain
\begin{eqnarray}
    C_0 & = & 4\pi \left[ A^{(0)} \right]^2 \\
  3 C_1 & = & \frac{4\pi}{3} \left[ A^{(1)} \right]^2, \\
  5 C_2 & = & \frac{4\pi}{15} \left[ A^{(2)} \right]^2 \left[ 1 + \frac13
    \left( \hat v^{(2,1)}\cdot \hat v^{(2,2)} \right)^2 \right].
\end{eqnarray}
Notice the presence of the term $\hat v^{(2,1)}\cdot \hat
v^{(2,2)}$ in the quadrupole expression. This implies that 
$A^{(2)}$ cannot be extracted solely from the angular power spectrum,
but requires additional information, such as higher order correlation functions.

A simple algorithm for constructing these vectors has been provided by
\citet{Copi2004} which builds upon the standard spherical harmonic
decomposition.  The algorithm relies on the observation that a dipole
defines a direction in space, that is, a vector, and, in general, the
$\ell$-th multipole is a rank $\ell$, symmetric, traceless tensor.  
This tensor can be written as the symmetric trace-free product of 
a vector and a symmetric trace-free rank $\ell-1$ tensor.  
This procedure can be repeated recursively.  It leads to sets of coupled
quadratic equations for the components of the vectors and the remaining
tensor which can be solved numerically.  The details of this are given in
\citet{Copi2004}.  We have used the freely available implementation of this
algorithm\footnote{See http://www.phys.cwru.edu/projects/mpvectors/ for
access to the code and other information.} for our work here. 

Recently the multipole vector representation has been studied and used in
various ways.  The multipole vectors can be understood in the context of
harmonic polynomials \citep{Katz2004,Lachieze-Rey:2004zh}, which has led to an
alternative algorithm for determining the components of the vectors as roots of
a polynomial.  Expressions for $N$-point correlation functions of these vectors
(for Gaussian random $a_{\ell m}$) have been derived analytically by
\citet{Dennis2005}.  Finally the application of these multipole vectors to the
cosmic microwave background is being actively pursued (see for example
\citet{Copi2004, Schwarz2004, SS2004, Weeks, Land2004b, Bielewicz2005}).

The multipole decomposition includes \textit{all} the available information.  
Some representations related to the multipole vectors do not fully encode this 
information as discussed below.

\subsection{Maxwell multipole vectors}

James Clerk Maxwell, in his study of the properties of the spherical harmonics,
introduced his own vector representation \citep{Maxwell}.  In this
representation, the $\ell$-th multipole of a function is written as
\begin{equation}
  f_{\ell}(\theta,\phi) = \left[A_{\rm M}^{(\ell)} \left( \hat v^{(\ell,1)}\cdot\vec\nabla
  \right) \cdots \left( \hat v^{(\ell,\ell)}\cdot\vec\nabla \right) \frac{1}{r}\right]_{r=1}
\end{equation}
where the unit vectors $\hat v^{(\ell,j)}$ are known as the Maxwell
multipole vectors.  It is well known that this representation is unique
(see \citealt{Dennis2004} and references therein).  It can be shown that
these vectors are precisely the multipole vectors constructed by
\citet{Copi2004}.  To see this we can rewrite Maxwell's representation
using standard Fourier integration techniques (see appendix A of
\citealt{Dennis2004}) as
\begin{equation}
  f_{\ell}(\theta,\phi) = A^{(\ell)} \left(v^{(\ell,1)} \cdot
      \hat e \right) \cdots \left( \hat v^{(\ell,\ell)} \cdot \hat e \right)
    + B^{(\ell)}
    \label{eqn:MMPV-decomp}
\end{equation}
where $A^{(\ell)} = (-1)^\ell\left[(2\ell -1)!!\right] A^{(\ell)}_{\rm M}$ and 
$B^{(\ell)}$ is an object with maximum angular momentum $\ell-2$ that ensures the
traceless nature of $f_\ell$.  That is, $B^{(\ell)} = - A^{(\ell)}
\mathcal{T}_\ell$.  (Note that the Maxwell multipole vector representation
is manifestly symmetric in the $\hat v^{(\ell,j)}$.)  This should be
compared to the discussion in Sec.~III and Appendix A of
\citet{Copi2004}.  Thus, the construction outlined above and given in
detail in \citet{Copi2004} is actually an algorithm for quickly and easily
converting a standard spherical harmonic
decomposition~(\ref{eqn:Ylm-decomp}) into a Maxwell multipole vector
decomposition~(\ref{eqn:MMPV-decomp}).

\begin{figure*}
  \includegraphics[width=4in,angle=-90]{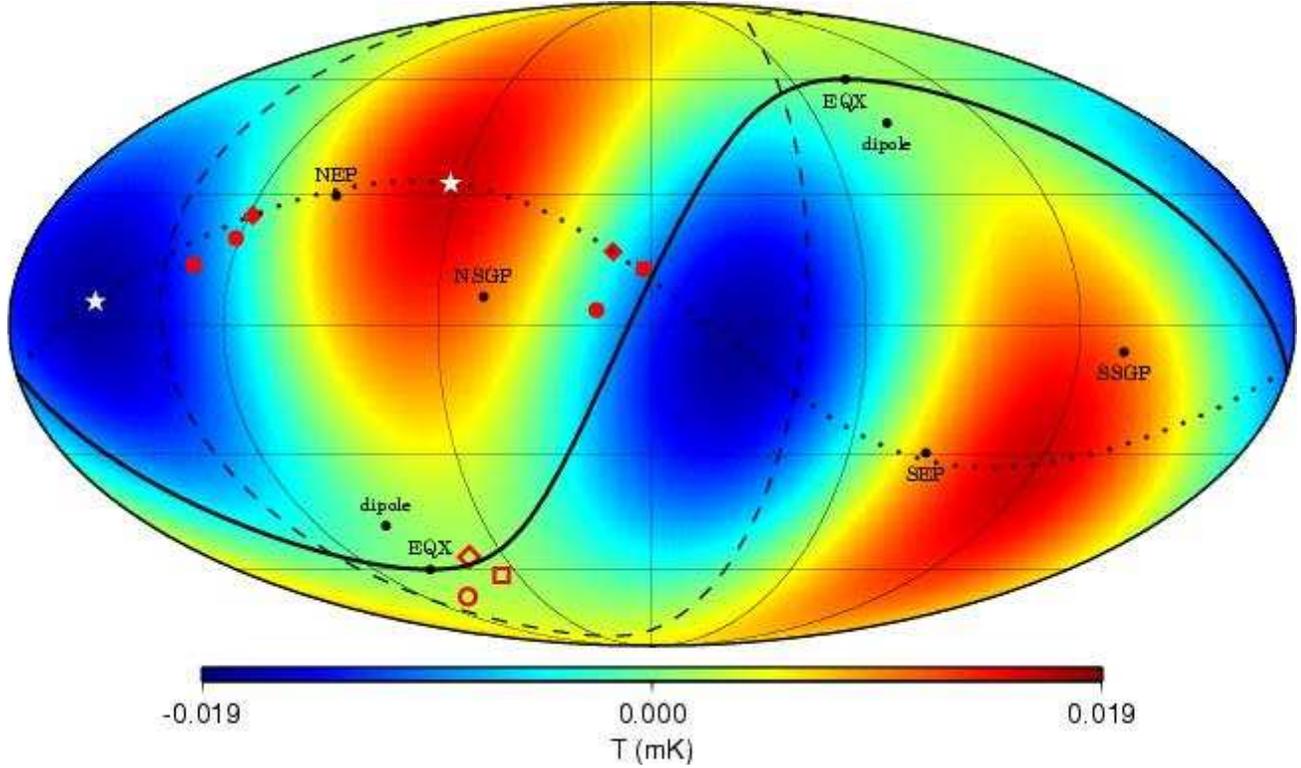}
  \caption{ The $\ell=2$ multipole from the \citet{TOH} cleaned map, presented
    in Galactic coordinates, after correcting for the kinetic quadrupole.  The
    solid line is the ecliptic plane and the dashed line is the supergalactic
    plane.  The directions of the equinoxes (EQX), dipole due to our motion
    through the Universe, north and south ecliptic poles (NEP and SEP) and
    north and south supergalactic poles (NSGP and SSGP) are shown.  The
    multipole vectors are plotted as the solid red (dark gray in gray scale
    version) symbols for each map, ILC (circles), TOH (diamonds), and LILC
    (squares). The open symbols of the same shapes are for the normal vector
    for each map.  The dotted line is the great circle connecting the two
    multipole vectors for this map.  The minimum and maximum temperature
    locations in this multipole are shown as the white stars. The direction
    that maximizes the angular momentum dispersion of any of the maps coincides
    with the respective normal vector as discussed in the text.}
  \label{fig:map:tegmark:2}
\end{figure*}

\subsection{Relation to ``angular momentum dispersion'' axes}
\label{sec:angmom}

The peculiar alignment of the  quadrupole and octopole was first pointed out 
by \citet{deOliveira2004}.
They identified $\hat n_\ell$ as the axis, which, 
when chosen as the fiducial $z$-axis of the coordinate system, maximizes 
the ``angular momentum dispersion'' of each multipole,
\begin{equation}
  \label{eqn:angmom-dispersion}
  (\Delta L)_\ell^2 \equiv \sum_{m=-\ell}^\ell m^2 \vert a_{\ell m}\vert^2.
\end{equation}
By varying the direction of the fiducial $z$-axis over the sky, and recomputing
for each such choice the $a_{\ell m}$ and $(\Delta L)_\ell^2$, they were able
to find the choice of $z$-axis that maximizes $(\Delta L)_\ell^2$.  This
angular momentum axis distills from each multipole a limited amount of
information, reducing the $2\ell$ degrees of freedom in the $\{\hat
v^{(\ell,j)}\}$ to just two.

Interestingly, \citet{deOliveira2004} found that $\hat n_2\cdot \hat n_3=
0.9838$ in the TOH map.  If the $\ell=2$ modes and the $\ell=3$ modes are indeed
statistically independent, then this degree of alignment has only a $1.6\%$
chance of happening accidentally.  (Arguably this probability should be
increased by a factor of two, to $3.2\%$, since we would have been equally
surprised if $\hat n_2$ had been nearly orthogonal to $\hat n_3$.)

We can calculate $\hat n_2$ explicitly for the quadrupole in terms of the
multipole vectors.  
The quadrupole function is 
\begin{equation}
\label{Qofr}
  Q(\vec r) = \left( \hat v^{(2,1)} \cdot \hat e \right)
  \left( \hat v^{(2,2)} \cdot \hat e \right) - \frac13 \hat v^{(2,1)} \cdot
  \hat v^{(2,2)},
\end{equation}
where $\hat e=\vec r/r$.
We now apply to $Q(\vec r)$ the angular momentum operator 
\begin{equation}
  \vec L = -\iimag r \hat e\times \vec\nabla ,
\end{equation}
finding
\begin{equation}
  -\iimag \vec L Q  = (\hat v^{(2,1)} \times \hat e) (\hat v^{(2,2)} \cdot\hat e)
  + (\hat v^{(2,2)} \times \hat e) (\hat v^{(2,1)}\cdot\hat e).
\end{equation}
We want to maximize 
\begin{equation}
\label{eqn:DeltaL2}
  \left(\Delta L\right)_2^2 \equiv \int_{\mathrm{sky}} 
  \vert {\hat n} \cdot {\vec L} Q\vert^2 \dderiv{\hat e}
\end{equation} 
over all possible unit vectors to find ${\hat n}_2$.  The quantity $({\hat
  n}\cdot {\vec L} Q)$ is easily calculated:
\begin{eqnarray}
  \iimag {\hat n}\cdot {\vec L} Q & = & 
  \hat e \cdot \left[ (\hat v^{(2,1)} \times {\hat n}) (\hat
    v^{(2,2)}\cdot{\hat e}) \right. \nonumber \\
  & &
  \quad\left. {} + ( \hat v^{(2,2)} \times {\hat n}) (\hat v^{(2,1)}\cdot{\hat
      e}) \right] .
  \label{eqn:ntwohat}
\end{eqnarray}
The integral in (\ref{eqn:DeltaL2}) is straightforward in any basis.  For an axis 
\begin{equation}
  {\hat n} \equiv \left(\sin\chi\cos\psi,\sin\chi\sin\psi,\cos\chi\right) ,
\end{equation}
in a coordinate system where $\hat v^{(2,1)}$ is identified with the $x$-axis
and $\hat v^{(2,2)}$ is taken to define the $xy$-plane with 
$\hat v^{(2,1)}\cdot\hat v^{(2,2)}\equiv \cos\omega$,
\begin{eqnarray}
  \left(\Delta L\right)_2^2  & = & 
\frac{4\pi}{15} \left[ 
4 - \sin^2\chi \left(2 + \sin^2 \omega + \cos (2\omega-2\psi)
    \right. \right. \nonumber\\
  & & \quad\left. \left. {} + \cos (2\psi)\right) 
  \right] .
\end{eqnarray}

The partial derivative of $\left(\Delta L\right)_2^2$ with respect to
$\chi$ vanishes at $\chi= 0$, $\pi/2$ and $\pi$.  We find that when $\chi=\pi/2$
there are four minima of $\left(\Delta L\right)_2^2$ for 
$\psi = \omega/2 + n \pi/2$, $n = 0,1,2,3$. The directions $\chi=0$ and $\pi$ 
are maxima.  Since $\psi$ is not defined when
$\sin\chi=0$, these are zeroes of all directional derivatives. Thus, the 
``maximum angular dispersion'' is obtained in the direction normal to the 
plane that is defined by the two multipole vectors of the quadrupole, i.e.
$\pm(\hat v^{(2,1)}\times\hat v^{(2,2)})/|\hat v^{(2,1)}\times\hat v^{(2,2)}|$.
The minima are defined by the directions $\pm(\hat v^{(2,1)}\pm\hat v^{(2,2)})/
\sqrt{2}$.

We had previously identified the ``area vectors''
\begin{equation}
 {\vec w}^{(\ell;i,j)}  \equiv {\hat v}^{(\ell,i)} \times {\hat v}^{(\ell,j)} 
\end{equation}
and the corresponding normalized directions $\hat w^{(\ell;i,j)}$
as phenomenologically interesting.  Indeed, most of the interesting statistical
results of \citet{Copi2004} and \citet{Schwarz2004} relate to statistics of
the dot-products of area vectors (or their normalized versions) with one
another or with physical directions on the sky, rather than to the
statistics of the dot-products of the multipole vectors themselves.
We see that for $\ell=2$, the maximum angular momentum dispersion (MAMD) 
axis is parallel to the area vector, i.e.~${\hat n}_2 = \pm {\hat w}^{(2;1,2)}$.

This relation between the area vectors and the MAMD axis cannot extend 
precisely to higher $\ell$.  An octopole, for example, has three multipole 
vectors $\left\{{\hat v}^{(3,i)}\mid i=1,2,3\right\}$ which define three 
distinct planes with area vectors $\left\{{\vec w}^{(3;i,j)} \mid 
i,j=1,2,3; i<j\right\}$.  The octopole function is 
\begin{equation}
  \mathcal{O}(\hat e) =  (\hat v^{(3,1)}\cdot {\hat e})(\hat v^{(3,2)}\cdot {\hat
    e})(\hat v^{(3,3)}\cdot {\hat e}) - \mathcal{T}_3,
  \label{eqn:octopole}
\end{equation}
where the trace term is 
\begin{eqnarray}
  \mathcal{T}_3 & = & \frac{1}{5}\hat e\cdot \left[(\hat v^{(3,1)}\cdot
    \hat v^{(3,2)}) \hat v^{(3,3)} + (\hat v^{(3,2)}\cdot \hat v^{(3,3)})
    \hat v^{(3,1)} \right. \nonumber\\
  & & \quad \left . {} + (\hat v^{(3,3)}\cdot \hat v^{(3,1)}) \hat
    v^{(3,2)} \right] .
  \label{eqn:octopole:trace}
\end{eqnarray}
Once again, the function $({\hat n}\cdot {\vec L} {\cal O})$ is easily
calculated for some arbitrary axis $\hat n$
\begin{eqnarray}
  \iimag {\hat n} \cdot {\vec L} {\cal O} & = & 
  {\hat e}\cdot \left\{
    \left[ 
      (\hat v^{(3,1)}\cdot {\hat e})(\hat v^{(3,2)}\cdot {\hat e})\hat
      v^{(3,3)} \right.\right. \nonumber \\
  & & \left. \left. 
      - \frac15 (\hat v^{(3,1)}\cdot\hat v^{(3,2)}) \hat v^{(3,3)} 
      \right. \right. \nonumber \\ 
  & & \left. \left. + \mbox{\ cyclic\ permutations} \right] \times {\hat n}
      \right\}.
\end{eqnarray}
The MAMD axis, ${\hat n}_3$, is obtained by integrating the square modulus of
this function over the full-sky $\dderiv{\hat e}$, and maximizing with respect to
the choice of axis ${\hat n}$.  Unlike in the quadrupole case, 
we don't find ${\hat n}_3$  to be a simple combination of the 
three octopole vectors $\hat v^{(3,i)}$.  Rather, 
${\hat n}_3$ is some combination of the three directions which is difficult to
determine analytically.

In the specific case where the three octopole vectors lie in a plane,
(corresponding to a so-called ``planar'' octopole) the cross product of any two
are orthogonal to all three.  For a planar octopole, the MAMD axis and the area
vectors of the three octopole planes are all parallel.  In the case of the
observed microwave background, the octopole is relatively planar, for example
for the TOH map:
\begin{eqnarray}
& & \vert{\vec w}^{(3;1,2)} + {\vec w}^{(3;2,3)} + {\vec w}^{(3;3,1)}\vert
  \nonumber \\[0.1cm]
&\simeq & 
  0.8 \left(\vert{\vec w}^{(3;1,2)}\vert + \vert{\vec w}^{(3;2,3)}\vert +
  \vert{\vec w}^{(3;3,1)}\vert\right) .
\end{eqnarray}
Moreover, the three octopole area vectors surround the quadrupole area
vector (see Fig.~\ref{fig:map:tegmark:2+3}), so that the MAMD axis --- which
is some average of the area vectors
--- is even closer to the quadrupole axis.  For these reasons, the MAMD axis
is a moderately good representation of the octopole area vectors.
This explains that the alignment of the quadrupole and
octopole seen by \citet{deOliveira2004} and by us \citep{Schwarz2004} are
indeed the same effect.

\begin{figure*}
  \includegraphics[width=4in,angle=-90]{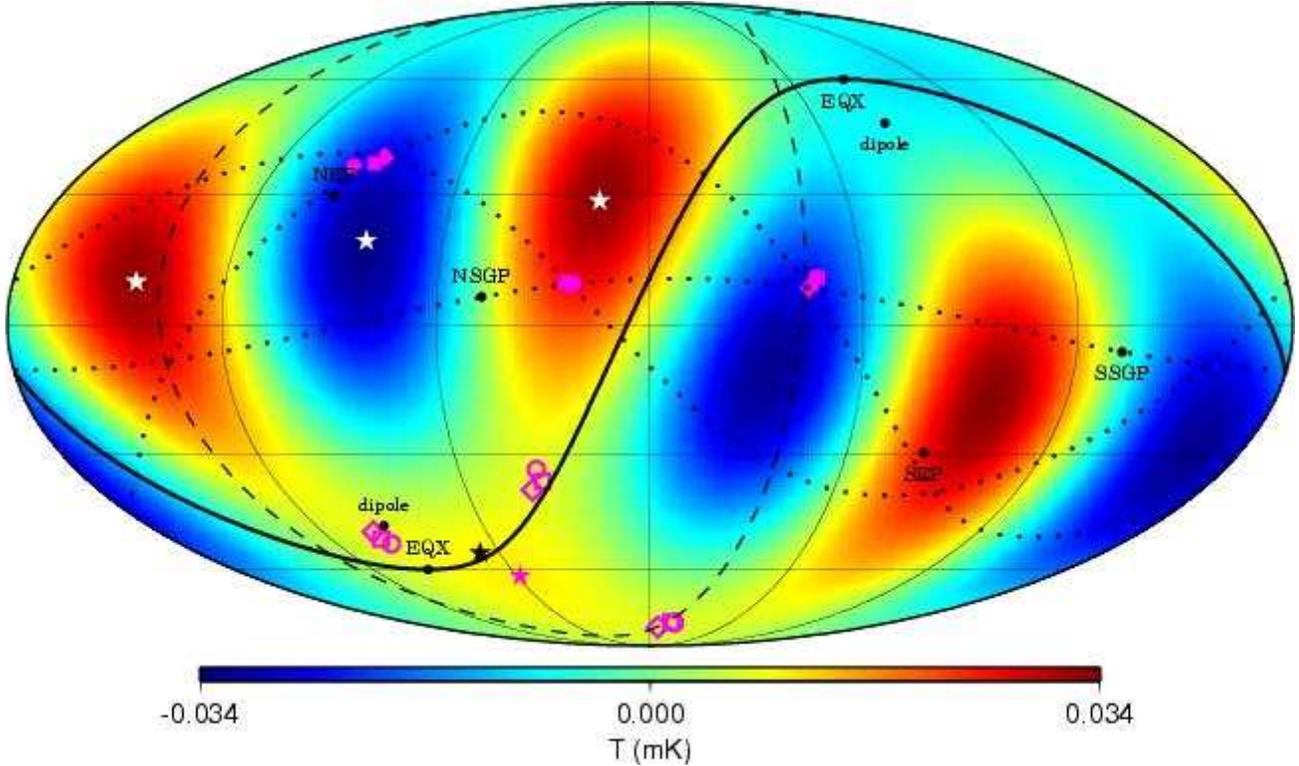}
  \caption{The $\ell=3$ multipole from the \citet{TOH} cleaned map, presented
    in Galactic coordinates.  The solid line is the ecliptic plane and the
    dashed line is the supergalactic plane.  The directions of the equinoxes
    (EQX), dipole due to our motion through the Universe, north and south
    ecliptic poles (NEP and SEP) and north and south supergalactic poles (NSGP
    and SSGP) are shown. The multipole vectors are the solid magenta (medium
    gray in gray scale version) symbols for each map, ILC (circles), TOH
    (diamonds), and LILC (squares).  The open symbols of the same shapes are
    for the normal vectors for each map.  The dotted lines are the great
    circles connecting each pair of multipole vectors for this map.  The light
    gray stars are particular sums of the multipole vectors which are very
    close to the temperature minima and maxima of the multipole.  The solid
    black star shows the direction of the vector that appears in the trace of
    the octopole, $\mathcal{T}_3$~(\ref{eqn:octopole:trace}), of the TOH map.
    The solid magenta (again medium gray in the gray scale version) star is the
    direction to the maximum angular momentum dispersion for the octopole,
    again for the TOH map.}
  \label{fig:map:tegmark:3}
\end{figure*}

\subsection{Relation to minima/maxima directions}
\label{sec:minmax}

The directions toward the minima and maxima of each multipole are simple to
see and at first glance appear to be an equally useful representation
(see \citealt{minmaxdirections}).  This, however, is not the case.

Consider again the quadrupole~(\ref{Qofr}).
The extrema of $Q$ on the sphere occur where the angular momentum is
zero.  That is, they are the solutions of 
\begin{eqnarray}
  0 & = & -\iimag \vec L Q(\vec r ) \nonumber \\
  & = & 
  (\hat v^{(2,1)} \times {\hat e})  (\hat v^{(2,2)}\cdot \hat e) +
  (\hat v^{(2,2)} \times {\hat e})  (\hat v^{(2,1)}\cdot \hat e).
\end{eqnarray}
The  solutions of these equations occur (by inspection) along the directions
\begin{equation}
  \hat e = \hat v^{(2,\pm)} 
  \equiv \frac{\hat v^{(2,1)} \pm \hat v^{(2,2)}}{\vert \hat v^{(2,1)} \pm
    \hat v^{(2,2)}\vert}
  = \frac{\hat v^{(2,1)} \pm \hat v^{(2,2)}} {\sqrt{2(1\pm \hat
      v^{(2,1)}\cdot \hat v^{(2,2)})}}
\end{equation}
It is easily seen that $\hat v^{(2,+)} \cdot \hat v^{(2,-)} = 0$.  Thus we
see that for the quadrupole the maxima and minima are orthogonal to each
other and lie in the same plane as the multipole vectors.  The quadrupole
multipole vectors bracket the two hot spots --- the maxima occur half way
between the two multipole vectors.

The quadrupole multipole vectors $\hat v^{(2,1)}$ and $\hat v^{(2,2)}$ contain
4 pieces of information and thus fully specify the shape of the quadrupole (the
fifth piece of information being the amplitude $A^{(2)}$).  The same is
\textit{not} true of the ${\hat v}^{(2,\pm)}$, the directions to the maxima and
minima.  Since the minima and maxima are orthogonal, these vectors contain only
three pieces of information --- for example, the direction of the first maximum,
and the orientation of the plane of the maxima and minima.  The ratio of the
amplitudes of the two maxima and minima is unspecified.  That is, we do not
know the relative strengths of the maxima and minima.

For higher multipoles the minima/maxima directions, unlike the multipole
vectors, persist in not containing the full information.  The minima and maxima
directions will again lack information about the relative amplitude of the
extrema. There are also additional problems concerning the definition of the
minima/maxima directions: In the case of a pure $Y_{\ell 0}$ mode, there are
degenerate rings of minima/maxima, which do not allow to assign a unique
direction. Moreover, the number of minima and maxima for a fixed value of
$\ell$ is not unique. For a pure mode (spherical harmonic) the number of
extrema depends on $\ell$ and $m$ in general. To see that it is instructive to
remember that $\ell$ corresponds to the total number of nodal lines and $m$
counts the number of meridians that are nodal lines. With this rule in mind one
can easily see that, e.g. $Y_{32}$ has four maxima and four minima, whereas
$Y_{33}$ has three maxima and three minima. For a general multipole there is no
rule on how many minima and maxima we should expect and thus the
minimum/maximum directions are only of limited use.  Conversely, multipole
vectors always contain $2\ell$ pieces of information, including information on
the location, number and relative amplitudes of the minima and maxima.

Therefore, while one might have at first imagined the minima/maxima
directions to be independent, they are actually strongly correlated.  This
weakens the statistical power of tests of the distribution of these
directions.  For these reasons, the statistical properties of the 
minima/maxima directions are not considered further in this work.

\begin{figure*}
  \includegraphics[width=4in,angle=-90]{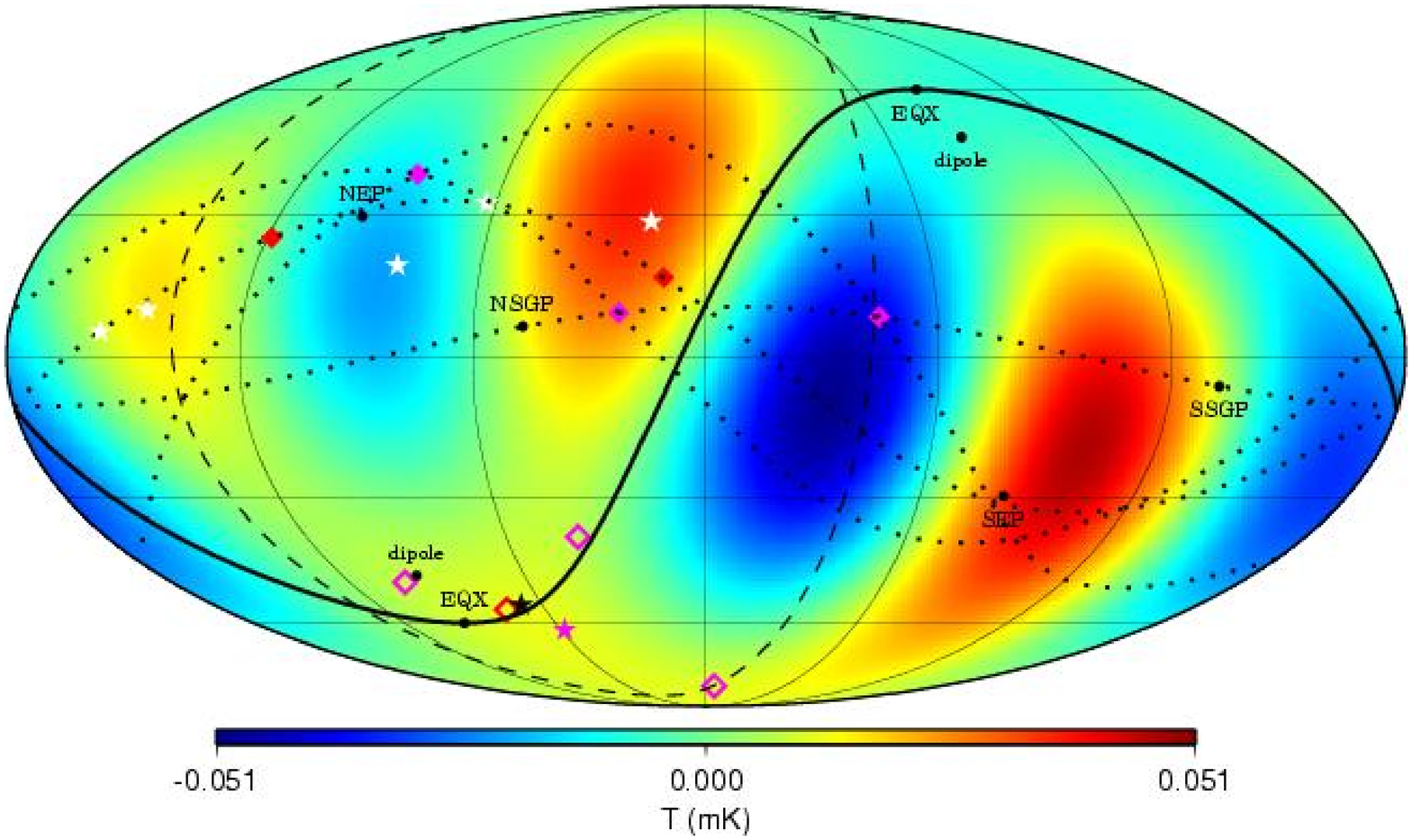}
  \caption{The $\ell=2+3$ multipoles from the \citet{TOH} cleaned map,
    presented in Galactic coordinates.  This is a combination of
    Figs.~\ref{fig:map:tegmark:2} and \ref{fig:map:tegmark:3} with only the
    multipole vectors for the TOH map shown for clarity. The solid line is the
    ecliptic plane and the dashed line is the supergalactic plane.  The
    directions of the equinoxes (EQX), dipole due to our motion through the
    Universe, north and south ecliptic poles (NEP and SEP) and north and south
    supergalactic poles (NSGP and SSGP) are shown.  The $\ell=2$ multipole
    vectors are plotted as the solid red (dark gray in gray scale version)
    diamond and their normal is the open red (dark gray in the gray scale
    version) diamond.  The $\ell=3$ multipole vectors are the solid magenta
    (medium gray in gray scale version) diamonds and their three normals are
    the open magenta (medium gray in the gray scale version) diamonds.  The
    dotted lines are the great circles connecting the multipole vectors for
    this map (one for the quadrupole vectors and three for the octopole
    vectors).  The minimum and maximum temperature locations of the $\ell=2$
    multipole are shown as the white stars.  The light gray stars are
    particular sums of the $\ell=3$ multipole vectors which are very close to
    the temperature minima and maxima of the octopole.  The solid black star
    shows the direction of the vector that appears in the trace of the
    octopole, $\mathcal{T}_3$~(\ref{eqn:octopole:trace}).  The solid magenta
    (again medium gray in the gray scale version) star is the direction to the
    maximum angular momentum dispersion for the octopole, again for the TOH
    map.  }
  \label{fig:map:tegmark:2+3}
\end{figure*}

\subsection{Relation to Land-Magueijo vectors}
\label{sec:LM:vectors}

\citet{Magueijo1995} discussed an alternative approach to the $\ell$-th
multipole, which is well known to be a representation of a symmetric
trace-free tensor of rank $\ell$.  Much as for the multipole vectors, in
this approach one recasts the $\ell$-th multipole as a
$3\times3\times\cdots\times3$ ($\ell$ dimensional) Cartesian tensor, ${\cal
  O}^{i_1\cdots i_\ell}$ (with $i_n=1,2,3$). One then realizes that the
information this Cartesian tensor encodes can be recast as $2\ell-2$
scalars, the $2\ell-2$ independent invariant contractions of the rank
$\ell$ trace-free symmetric tensor, and three degrees of freedom
associated with an orthonormal frame.  (The orientation of the $z$-axis of
the frame is two degrees of freedom.  The orientation of the $x$-axis within
the plane orthogonal to the $z$-axis is the third.  The $y$-axis is then fixed
by orthonormality and the convention of right-handedness.)

\citet{Land2004a} discussed the two scalars and the frame associated with the
microwave background quadrupole.  The two scalars are the power-spectrum and
the bispectrum.  The vectors of the frame are the eigenvectors of the
$3\times3$ Cartesian quadrupole tensor.  This has the advantage that, at least
for the quadrupole, one has clearly separated the issue of non-Gaussianity (the
bispectrum) from that of statistical anisotropy (the frame).  Land and Magueijo
claimed that the frame they found was independent of the multipole vectors
described in \citet{Schwarz2004}.  However, Schwarz and Starkman (private
communication, 2004 as referenced in \citealt{Land2004b}) showed that, in fact,
two of the axes of the frame are parallel to the sum and difference of the
quadrupole multipole vectors $\hat v^{(2,1)}\pm \hat v^{(2,2)}$. The third is,
of course, parallel to their cross-product.

Explicitly, for a quadrupole with multipole vectors $\hat v^{(2,1)}$ and
$\hat v^{(2,2)}$, the  Cartesian tensor representation of the
quadrupole~(\ref{Qofr}) is
\begin{equation}
{\cal Q}_{i j} = 
\frac{{\hat v^{(2,1)}}_i{\hat v^{(2,2)}}_j + {\hat v^{(2,1)}}_j{\hat
    v^{(2,2)}}_i}{2} - \frac{1}{3} {\hat v^{(2,1)}}\cdot{\hat v^{(2,2)}}
\delta_{i j} . 
\end{equation}
 This $3\times3$ matrix can be
diagonalized.  The three eigenvectors making up the Land and Magueijo
orthonormal frame are 
\begin{eqnarray}
  \vec v^{(\mathrm{LM})}_1 & \parallel & \hat v^{(2,1)} \times \hat v^{(2,2)} \nonumber \\
  \vec v^{(\mathrm{LM})}_2 & \parallel & \hat v^{(2,1)} + \hat v^{(2,2)}   \label{eqn:LM} \\
  \vec v^{(\mathrm{LM})}_3 & \parallel & \hat v^{(2,1)} - \hat v^{(2,2)}. \nonumber
\end{eqnarray}
It is noteworthy that $\vec v^{(\mathrm{LM})}_2$ and $\vec
v^{(\mathrm{LM})}_3$ are in the directions of the quadrupole maxima (or
minima), while $\vec v_1^{(\mathrm{LM})}$ is the quadrupole area vector $\vec
w^{(2;1,2)}$.  (The claim by \citet{Land2004a} that these vectors did not
coincide with the quadrupole multipole vectors or area vector of
\citet{Schwarz2004} arose because \citet{Schwarz2004} corrected for the
kinetic quadrupole as discussed in Sec.~\ref{sec:DQ}).

A difficulty that arises with the Land and Magueijo approach is how to go
beyond the quadrupole.  \citet{Land2004b} note that the $2\ell-2$
independent scalars associated with the $\ell$-th multipole are just the
$2\ell-2$ linearly independent combinations of dot-products of the $\ell$
multipole vectors.  The orthonormal frame is then constructed out of the
multipole vectors, as they were for the quadrupole.  However, there are
$\ell$ multipole vectors and so $\ell(\ell-1)$ different ways to
construct the frame (not to mention a much larger number of ways to choose
the set of independent scalars).  To define a unique orthonormal frame
\citet{Land2005b} chose the two vectors that have the two largest values
of 
\begin{equation}
  K_i = \sum_{j\ne i} \left( \hat v_i\cdot\hat v_j \right)^2.
\end{equation}
This is not a unique choice and it is unclear how to fairly choose such a
frame.  When they claim that it is this orthonormal frame that one should use
to test statistical isotropy, we must ask which of the many possible allowed
choices of frame is the correct one.  As \citet{Land2005b} point out the choice
of frame is crucial and their scheme is increasingly sensitive to small
fluctuations in the positions of the multipole vectors as $\ell$ grows.  This
leads to discontinuous noise in the Euler angles of the orthogonal frame making
it difficult to interpret the results.  As \citet{Land2005b} also point out
different orderings of the vectors will be sensitive to different features in
the data.  Without a unique and well understood prescription this approach does
not lead to further understanding.  The multipole vectors, on the other hand
are unique.  It is true, as \citet{Copi2004} and \citet{Schwarz2004} point out,
and as reiterated by \citet{Land2004b}, that they contain information on
\textit{both} statistical isotropy and non-Gaussianity.  Optimal separation of
these two properties remains an open problem.

\subsection{Polydipoles}

The directions of the multipole vectors are somewhat difficult to interpret
physically.  We would argue that in large measure this is due to the
removal of the traces done in constructing the multipole vector expansion
(\ref{eqn:MVE}).  An alternative approach would be to write the general
function $f(\theta,\phi)$ in a slightly different way:
\begin{equation}
  \label{eqn:polydipole}
  f(\theta,\phi) = \sum_{\ell=0}^\infty  {\cal P}_{\ell}(\theta,\phi) \equiv
  \sum_{\ell = 0}^\infty A_{\rm P}^{(\ell)} 
  \prod_{i=1}^\ell ({\hat p}^{(\ell,i)}\cdot {\hat e}).
\end{equation}

The ${\cal P}_\ell$'s, which we term polydipoles, are similar to the multipoles
$f_\ell$ in equation (\ref{eqn:MVE}), except for our failure to remove the
traces in their construction.  But it is this failure which makes them easily
visualized: ${\cal P}_\ell(\theta,\phi)$ is the simple product of $\ell$
dipoles.  Therefore there are $\ell$ great circles on which ${\cal
P}_\ell(\theta,\phi)$ vanishes --- the great circles are normal to each of the
polydipole vectors ${\hat p}^{(\ell,i)}$.  These are clearly related to the
$\ell$ nodal curves of $Y_{\ell m}$.

Unfortunately the polydipole expansion (\ref{eqn:polydipole}) is unstable in
the following important sense.  Suppose that in expanding $f(\theta,\phi)$ we
at first neglect all contributions with angular momentum greater than $L$ and
calculate $f_\ell$ and ${\cal P}_\ell$ for $\ell=1,\ldots,L$.  If we then increase
$L$, the $f_\ell$'s do not change (since the spherical harmonics are an
orthogonal basis) but the ${\cal P}_\ell$ generally do. Thus, the polydipole
expansion is well-defined only if the angular power spectrum falls sufficiently
quickly at large $\ell$.  The precise condition for convergence of the
polydipole expansion has not been determined.

The ``leading'' term in a multipole $f_\ell$ (by which we mean the 
part excluding ${\cal T}_\ell$ in (\ref{eqn:MVE})) is a polydipole,
which we may term the associated polydipole of that multipole.
(Note that the multipole vectors are not however the vectors one would get
in the polydipole expansion, so the $\ell$th polydipole in a polydipole
expansion is not the associated polydipole of the $\ell$th multipole
in a multipole expansion.)  To the extent that it is easier
to visualize the polydipole, it is of some limited interest.

\section{The Strange Properties of the Quadrupole and the Octopole}
\label{sec:quad_oct}

As we have already remarked, the microwave background anisotropies on large
angular scales seem to have several unusual properties.  Most widely known is
that the power in the quadrupole, $C_2$, is substantially less than is expected
from the models that fit the rest of the angular power spectrum (and other)
data.  The power in the octopole is also less than expected, though within
cosmic variance error bars.  By $\ell=4$ the power in the CMB is entirely
consistent with theoretical expectations.  This was first found by COBE
\citep{DMR4_maps} and has now been confirmed by WMAP.  The precise statistical
significance of this deviation is a matter of some dispute
\citep{Efstathiou2003,Slosar2004,Bielewicz2004,O'Dwyer2004}.

It has also  been known since COBE, but had largely been forgotten
until confirmed by WMAP \citep{Spergel2003}, that the two-point angular correlation
function of the microwave background
\begin{equation}
C(\theta) = \langle T(\hat e_1)T(\hat e_2) \rangle 
\end{equation}
(where $\hat e_1 \cdot \hat e_2=\cos\theta$) is nearly zero at angular scales
between about $60\degr$ and $170\degr$.  \citet{Spergel2003} argued that, given
the best fitting $\Lambda$CDM model, this is unlikely at the $0.15$\% level.
What has been under-appreciated, is that this vanishing of $C(\theta)$ is not
merely due to the lack of quadrupole power, but also due to the lack of
octopole power, and maybe even to the ratio of $C_2:C_3:C_4$ \citep{Luminet:2003dx}.

These anomalies relate exclusively to the power in the various multipoles.
More recently, attention has turned to the ``shapes'', ``phase relationships'' 
or ``orientations'' of the multipoles, i.e. to the information contained in 
the multipole vectors.  As remarked above,  several groups of authors have
noticed particular anomalies.  In this section, we will first describe them,
mostly in the language of multipole vectors, before proceeding to try to 
assign them some statistical significance in the next section.

We believe that the observed lack of power at large angles justifies 
singling out the two most responsible multipoles for particular scrutiny.
To do so is no more dubious than the practice of  firefighters to respond
to the house  where the fire alarm is ringing  rather than to all the houses in the 
neighborhood.  That these lowest multipoles represent a physically interesting scale  --- 
the recent scale of the horizon, especially the scale of the horizon at approximately dark
energy domination --- makes it doubly justified to focus our
attention, at least initially, on them.

Table~\ref{tab:vectors} contains the multipole vectors and area vectors for
the quadrupole and octopole that will be discussed below.

\begin{table}
  \caption{Multipole vectors, $\hat v^{(\ell,i)}$, and oriented area vectors,
    $\vec w^{(\ell;i,j)}$, for the quadrupole and octopole in Galactic
    coordinates $(l,b)$.  All vectors are given for the TOH cleaned map after 
    correcting for the kinetic quadrupole and the coordinates are
    consistent with how they are plotted in
    Figs.~\ref{fig:map:tegmark:2}--\ref{fig:map:tegmark:2+3} and
    \ref{fig:add_foregr}. The magnitudes for the oriented area
    vectors are also given.
  }
  \label{tab:vectors}
  \begin{tabular}{lr@{$\fdg$}lr@{$\fdg$}lc}
    \hline
    \multicolumn{1}{c}{Vector} & \multicolumn{2}{c}{$l$} &
    \multicolumn{2}{c}{$b$} & Magnitude \\ \hline
    $\hat v^{(2,1)}$ & $118$&$9$ & $25$&$1$ & --- \\
    $\hat v^{(2,2)}$ & $11$&$2$ & $16$&$6$ & --- \\
    $\vec w^{(2;1,2)}$ & $74$&$3$ & $-56$&$6$ & 0.990 \\
    $\hat v^{(3,1)}$ & $86$&$9$ & $39$&$3$ & --- \\
    $\hat v^{(3,2)}$ & $22$&$6$ & $9$&$2$ & --- \\
    $\hat v^{(3,3)}$ & $-44$&$9$ & $8$&$2$ & --- \\
    $\vec w^{(3;1,2)}$ & $101$&$6$ & $-49$&$8$ & 0.902 \\
    $\vec w^{(3;2,3)}$ & $-6$&$3$ & $-79$&$5$ & 0.918 \\
    $\vec w^{(3;3,1)}$ & $38$&$4$ & $-38$&$9$ & 0.907 \\
    \hline
  \end{tabular}
\end{table}

\subsection{The queerness of the quadrupole}
Figure \ref{fig:map:tegmark:2} shows the $\ell=2$ multipole from the
\citet{TOH} cleaned map after subtraction of the kinetic quadrupole.  The
solid line is the ecliptic plane and the dashed line is the supergalactic
plane.  The directions of the equinoxes (EQX), dipole due to our motion through
the Universe, north and south ecliptic poles (NEP and SEP) and north and south
supergalactic poles (NSGP and SSGP) are shown.  The multipole vectors are
plotted as the solid red symbols for each map (see figure caption), while the
open symbols of the same shapes are for the normal vector for each map.  The
dotted line is the great circle connecting the two multipole vectors for this
map.  The minimum and maximum locations of the temperature in this multipole
are shown as the white stars.  The following observations can be made about
the quadrupole:
\begin{enumerate}
\item
The great circle defined by the quadrupole multipole vectors $\hat v^{(2,1)}$
and $\hat v^{(2,2)}$ passes through the NEP and SEP.  This is especially true
for the TOH and LILC maps, with some slight deviation for the ILC map.  We can
rephrase this to say that the (normalized) quadrupole area vector $\hat
w^{(2;1,2)}$ lies on the ecliptic plane.
\item
The axis of this great circle (i.e., $\hat w^{(2;1,2)}$) is aligned with 
both the dipole and the equinoxes.
\end{enumerate}

\subsection{The oddness of the octopole}
Figure \ref{fig:map:tegmark:3} shows the $\ell=3$ multipole from the
\citet{TOH} cleaned map.  Many of the features are the same as for the
quadrupole map (see Fig.~\ref{fig:map:tegmark:2}).  Note that there are now 3
multipole vectors (closed symbols) and 3 normal vectors (open symbols) plotted.
Also there are 3 dotted lines showing the great circles connecting each pair of
multipole vectors for this map.  Several perplexing observations can also be
made about the properties of the octopole.
\begin{enumerate}
\item The octopole has three multipole vectors: $\hat v^{(3,1)}$, $\hat
  v^{(3,2)}$ and $\hat v^{(3,3)}$.   One of these, $\hat v^{(3,1)}$, lies
  quite near the ecliptic poles.  Therefore, the two great circles defined
  by the pairs $(\hat v^{(3,1)},\hat v^{(3,2)})$ and the pair
  $(\hat v^{(3,1)},\hat v^{(3,3)})$
  each nearly pass through the ecliptic poles.   
  But in fact, one of these great circles passes much closer to the poles
  than the position of $\hat v^{(3,1)}$.
  The associated area vectors, $\vec {w}^{(3;1,2)}$ and $\vec{w}^{(3;3,1)}$,
  therefore lie on (or nearly on) the ecliptic.  
\item The third pair of multipole vectors,
  $(\hat v^{(3,2)},\hat v^{(3,3)})$ define a great circle that includes the
  supergalactic poles.  This pair also lies within $9\degr$ of the Galactic plane.  
  The associated area vector $\vec {w}^{(3;2,3)}$, therefore lies on the supergalactic plane
  and only $10\fdg5$ from the Galactic poles.
\item The octopole is noticeably planar, but not overwhelmingly
  so.  Namely, the area vectors of the octopole cluster only somewhat compared to
  expectations.  
\item The three maxima and three minima are very nearly at $\pm \hat
  v^{(3,1)}\pm \hat v^{(3,2)}\pm \hat v^{(3,3)}$, even though these are not
  generically where the angular momentum is zero.
  (Note that there are 8 such choices but only 6 are either minima or maxima direction.)
  This contrasts with the quadrupole, where the extrema can be
  shown analytically to be at $\pm \hat v^{(2,1)}\pm \hat v^{(2,2)}$ (see
  section~\ref{sec:minmax}).
\item The vector from the octopole trace, $\mathcal{T}_3$, (see
  Eqn.~\ref{eqn:octopole:trace}) lies on the ecliptic plane.
  The importance of this vector is unclear.
  Note that $\mathcal{T}_3$ is proportional to the difference between the
  octopole and the polydipole, $\mathcal{P}_3$.
\end{enumerate}

\subsection{The remarkable relation of the quadrupole and the octopole}

Figure \ref{fig:map:tegmark:2+3} shows the sum of $\ell=2$ and $\ell=3$
in the TOH cleaned map, after kinetic quadrupole subtraction.  
We can see that, beyond the separate oddities of
$\ell=2$ and $\ell=3$, there are additional unexpected relationships
between the quadrupole and octopole:
\begin{enumerate}
\item  The quadrupole normal vector $\hat w^{(2;1,2)}$ (open red symbols) --- 
or equivalently the quadrupole MAMD axis $\hat n_2$ --- is aligned with the octopole 
MAMD axis $\hat n_3$ (magenta star).  It is also aligned with the three 
octopole area vectors (open magenta symbols), especially with one  of them. 
\item The ecliptic carefully traces a zero of the combined map, 
doing so almost perfectly over the entire hemisphere centered on the Galactic center,
and even relatively well over the antipodal hemisphere.
\item Two of the extrema south of the ecliptic are clearly stronger than any north of the
ecliptic.   The weakest southern extremum is essentially equal in power to the strongest
northern extremum.  (It is slightly stronger in the LILC map, slightly weaker in the
TOH and ILC maps.)
\end{enumerate}

\section{Statistics}
\label{sec:stat}
We now revisit the statistics used to quantify the alignment of the various
vectors with each other and with the Solar System. In \citet{Schwarz2004}, we
considered the dot-products, $A_i$, of the three octopole area vectors with the
quadrupole area vector, and the mutual dot-products, $D_i$, of the three
octopole normal vectors (area vectors normalized to unit length) with the
quadrupole normal vector.  We found that both the $A_i$ and the $D_i$ are
unusually large.  Note that we can take as a convention that $A_i\geq0$ since
each vector is ambiguous in sign.  The various dot-products are shown in Table
\ref{tab:oldstat}.  As we emphasized in \citet{Schwarz2004}, all of the
aforementioned alignments are statistically significant at 99.8\% C.L.\ or
higher.

To compute all probabilities we compare statistics applied to WMAP maps to that
applied to Monte Carlo simulations. Unless otherwise noted, Monte Carlo
simulations are comprised of 100,000 realizations of Gaussian random, statistically isotropic 
maps with WMAP's (inhomogeneous) pixel noise.

	\begin{table}
	  \caption{Values of various vector dot-products for the TOH
	    cleaned map, corrected for the kinetic quadrupole. 
	    We show the values of the dot-products of the
	    three octopole area vectors with the quadrupole area vector ($A_i$,
	    $i=1,2,3$); the dot-products of the three octopole normal vectors
	    with the quadrupole normal vector ($D_i$, $i=1,2,3$; the normals
	    are the unit area vectors); and the dot-products of the four
	    quadrupole and octopole area and normal vectors with the 
	    north ecliptic pole (NEP),
	    the north Galactic pole (NGP), the north supergalactic pole (NSGP),
	    the dipole and the equinox.} 
	  \label{tab:oldstat}
	  \begin{tabular}{lcccc}
	    \hline
	    & \multicolumn{4}{c} {Value of the dot-product}\\
	    Test & ${\vec w}^{(2;1,2)}$ & ${\vec w}^{(3;1,2)}$ 
	         & ${\vec w}^{(3;1,3)}$ & ${\vec w}^{(3;2,3)}$ \\
	    \hline
	    $A_i$                  & --- & 0.851 & 0.783 & 0.762 \\
	    $\vec{w}\cdot\mathrm{NEP}$ & 0.027 & 0.161 & 0.041 & 0.481 \\
	    $\vec{w}\cdot\mathrm{NGP}$ & 0.827 & 0.688 & 0.570 & 0.903 \\
	    $\vec{w}\cdot\mathrm{NSGP}$ & 0.392 & 0.262 & 0.630 & 0.0011 \\
	    $\vec{w}\cdot\mathrm{dipole}$ & 0.974 & 0.883 & 0.755 & 0.674 \\
	    $\vec{w}\cdot\mathrm{equinox}$ & 0.968 & 0.886 & 0.681 & 0.766 \\
	    \hline
	    $D_i$                  & --- & 0.953 & 0.872 & 0.838 \\
	    $\hat{w}\cdot\mathrm{NEP}$ & 0.027 & 0.179 & 0.045 & 0.523 \\
	    $\hat{w}\cdot\mathrm{NGP}$ & 0.835 & 0.763 & 0.629 & 0.983 \\
	    $\hat{w}\cdot\mathrm{NSGP}$ & 0.396 & 0.291 & 0.694 & 0.0012 \\
	    $\hat{w}\cdot\mathrm{dipole}$ & 0.984 & 0.979 & 0.832 & 0.733 \\
	    $\hat{w}\cdot\mathrm{equinox}$ & 0.978 & 0.982 & 0.751 & 0.834 \\
	    \hline
	  \end{tabular}
	\end{table}

\subsection{Requirements for robust statistics}

In previous work \citep{Schwarz2004}, we used a particular set of statistics, for example,
for a  set of vectors that have dot-products $\{A_i \mid i=1,\ldots,n\}$  
(with one another, or with  a particular physical direction on the sky) 
with ``unusually'' high values.  We asked what is the
probability that a random Monte Carlo map has the highest dot-product higher than $A_1$, 
the second highest one higher than $A_2$, the third highest one higher than
$A_3$, and so on down to the $n\mbox{-}th$ such dot product.
For several cases that number was, as reported, very small. 
For example, for the dot products between the quadrupole area vector $\vec w^{(2;1,2)}$,
and each of the three octopole area vectors $\left\{\vec w^{(3;1,2)},\vec
  w^{(3;2,3)},\vec w^{(3;3,1)}\right\}$,
only 21 out of 100,000 MC maps
(for the TOH DQ-corrected map) satisfied the criterion, i.e. only 21 had
$A_1$ larger than the TOH value of $A_1$,
$A_2$ larger than the TOH value of $A_2$ and
$A_3$ larger than the TOH value of $A_3$.

One can ask if this statistic preferentially returns a small probability
even if there is really none to be found. This is clearly not the case --- the vector
dot-products could have been either large, small or ``average''; we noticed they
were large and found an easy to understand statistic that quantified the effect. 
Weeks  (private communication) however has pointed out that
the above-described statistics for the $A_i$ do not define an ordering relation
on the set of possible $A_i$; 
they therefore implicitly incorporate  some {\it a posteriori} knowledge.
One would therefore like to confirm this result with different, independent 
statistics.

\subsection{S and T statistics --- definitions}
\label{sec:ST:def}

To quantify the various alignments we found, it is desirable to choose the
statistics in such a way that the {\it a posteriori} knowledge of the
particular nature of the alignments is not used to find unjustly small
probabilities.  With that in mind we define and discuss two statistics, $S$ and
$T$, which do define ordering relations.  The first of these was briefly
mentioned and used in \citet{Schwarz2004} as suggested by Weeks.

Two natural choices of statistics which define ordering relations on the three
dot-products $A_i$, each lying in the interval $[0,1]$, are:
\begin{eqnarray}
S &\equiv& {1\over 3} \left (A_1 + A_2 + A_3\right ) \\[0.1cm]
{\rm and}\nonumber \\
T &\equiv& 1-{1\over 3}\left [(1-A_1)^2 + (1-A_2)^2 + (1-A_3)^2\right ].
\end{eqnarray}
Both $S$ and $T$ can be viewed as the suitably defined
``distance'' to the vertex $(A_1, A_2, A_3)=(0, 0, 0)$.
A third  obvious choice, $(A_1^2+A_2^2+A_3^2)/3$, is just $2S-T$.
Of course many other choices exist, involving higher powers of $A_i$.

One could also ask about the probability that, for example, two out
of three normals are aligned, and so we generalize the definitions to
\begin{eqnarray}
S^{(n, m)} &\equiv& {1\over m}\sum_{i=1}^m A_i\\[0.1cm]
T^{(n, m)} &\equiv& 1-{1\over m}\sum_{i=1}^m (1-A_i)^2, 
\end{eqnarray}
where the $A_i$ are ordered from largest to smallest
Note
that, if the $A_i$ are large (near $1$), both $S^{(n, m)}$ and $T^{(n, m)}$
will be large (near $1$).

One could further generalize these statistics to
  arbitrary weighting by different alignments; 
for example 
\begin{equation}
  S^{(n;\vec\alpha)}\equiv \left. \sum_{i=1}^{n} \alpha_i A_i \right/
  \sum_{i=1}^n \alpha_i
\end{equation}
with $0\leq\alpha_i\leq 1$.  Hereafter we consider only the values $\alpha_i=0$
or $1$. Finally, there is nothing special about the area vector products $A_i$
and we apply the statistics $S^{(n, m)}$ to dot-products of the vector normals,
$D_i$, and also to dot-products of the normals with specific directions or
planes in the sky (ecliptic plane, NGP, supergalactic plane, dipole and
equinox) that were discussed in \citet{Schwarz2004}.  When comparing to planes,
we order the dot products (taken with the plane's axis), from smallest to
largest.

\subsection{S and T statistics --- quadrupole and octopole}
\label{sec:ST:QO}

\begin{figure*}
  \includegraphics[width=2.0in,angle=-90]{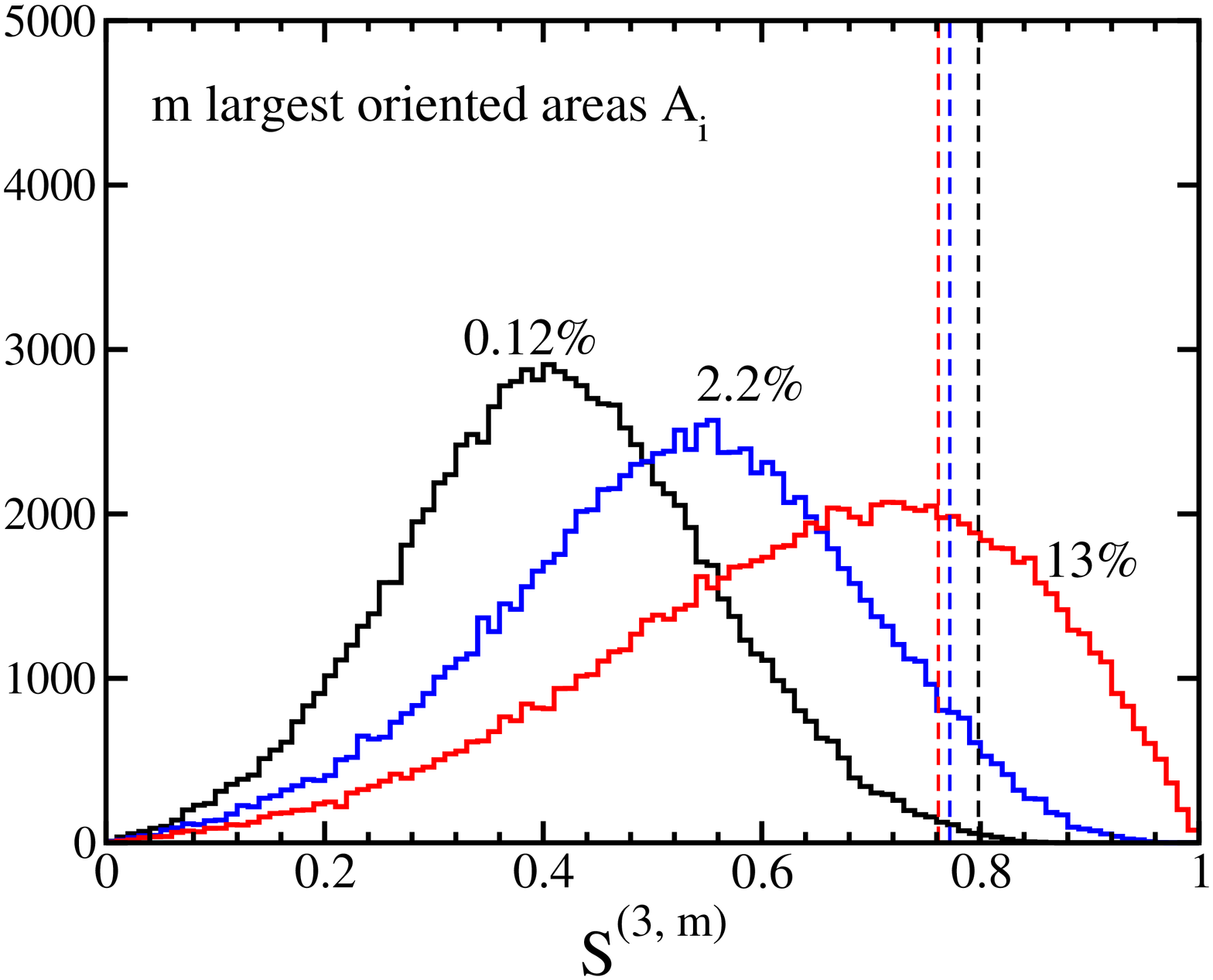}\hspace{0.2cm}
  \includegraphics[width=2.0in,angle=-90]{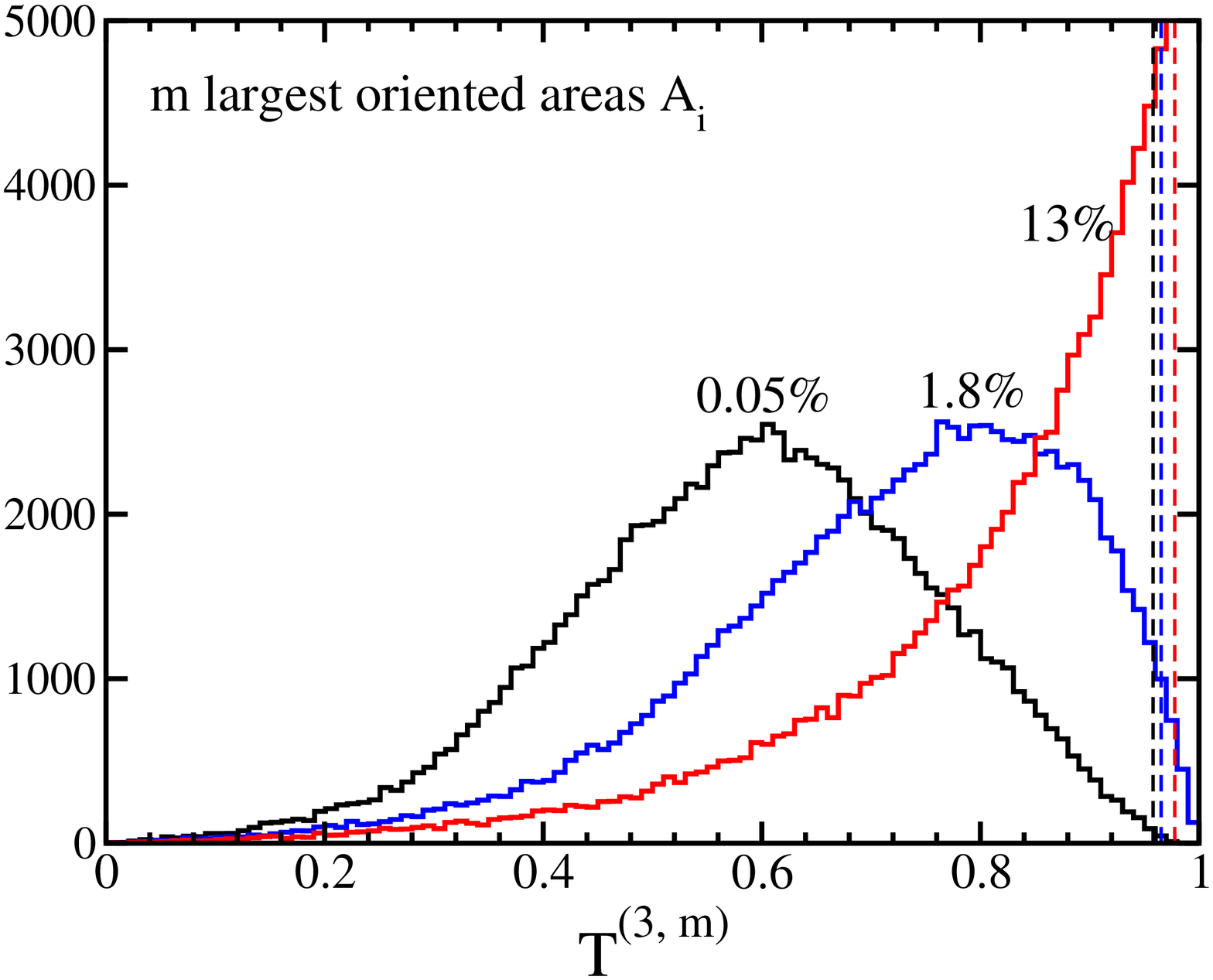}\\[-0.2cm]
  \includegraphics[width=2.0in,angle=-90]{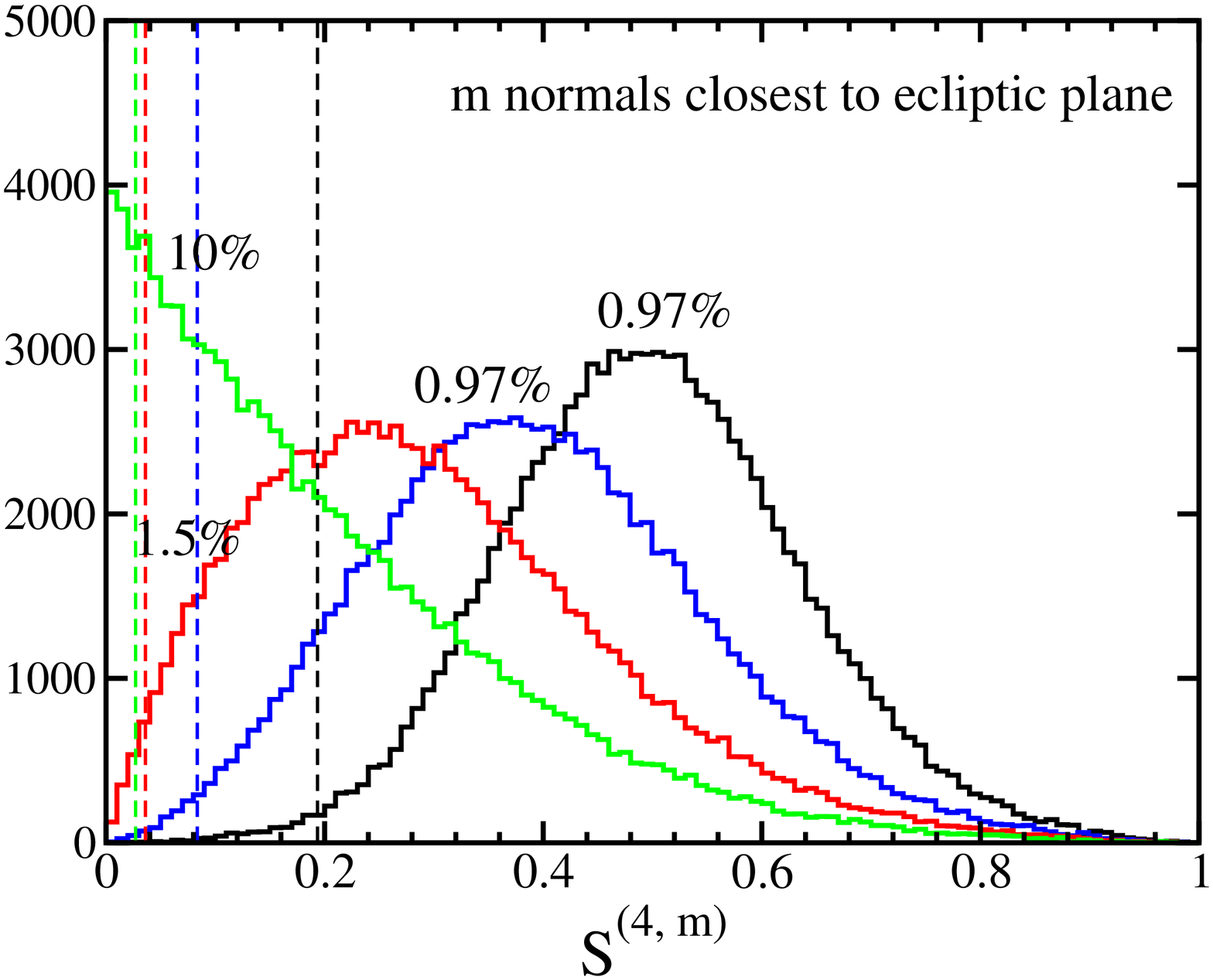}\hspace{0.2cm}
  \includegraphics[width=2.0in,angle=-90]{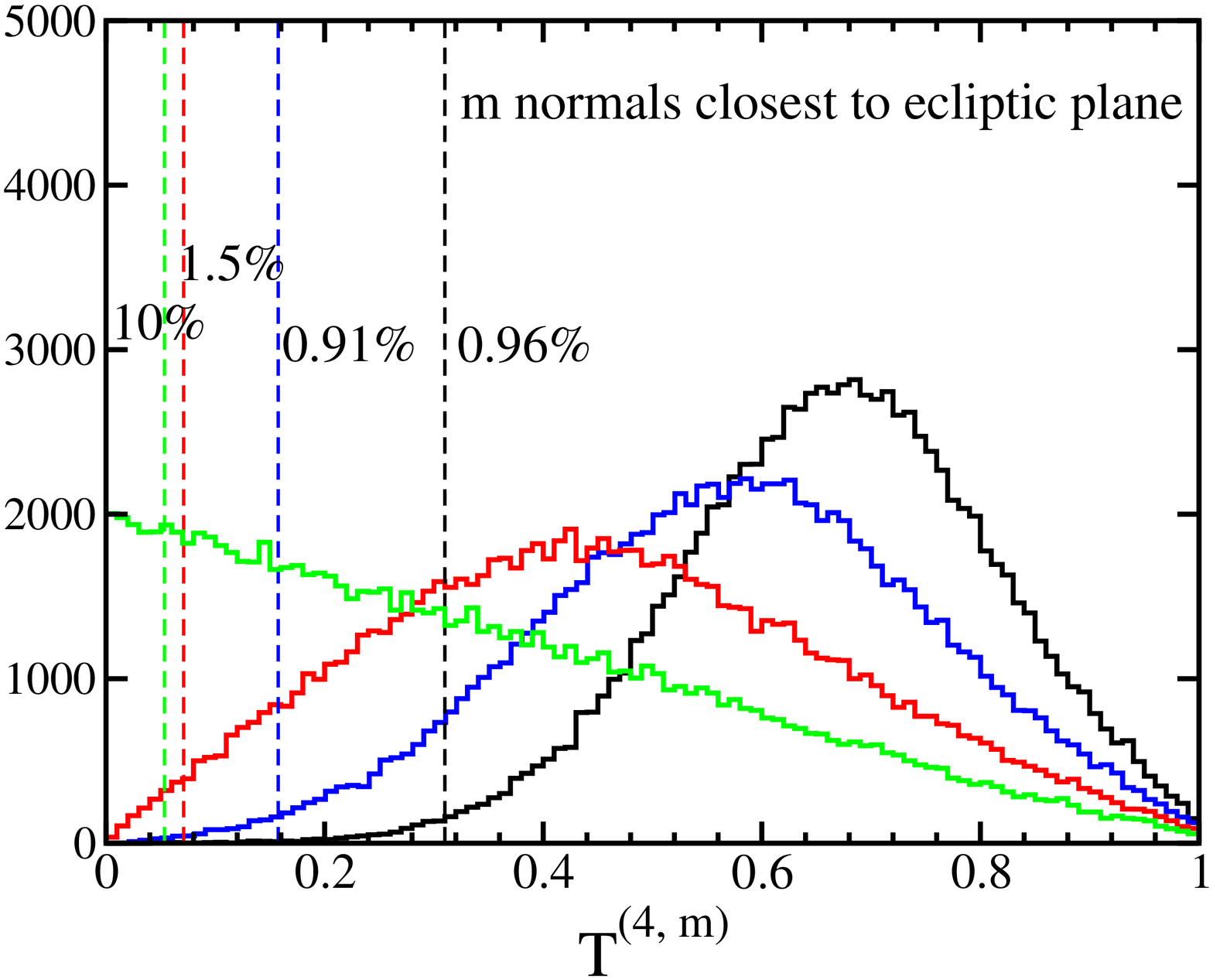}\\[-0.2cm]
  \includegraphics[width=2.0in,angle=-90]{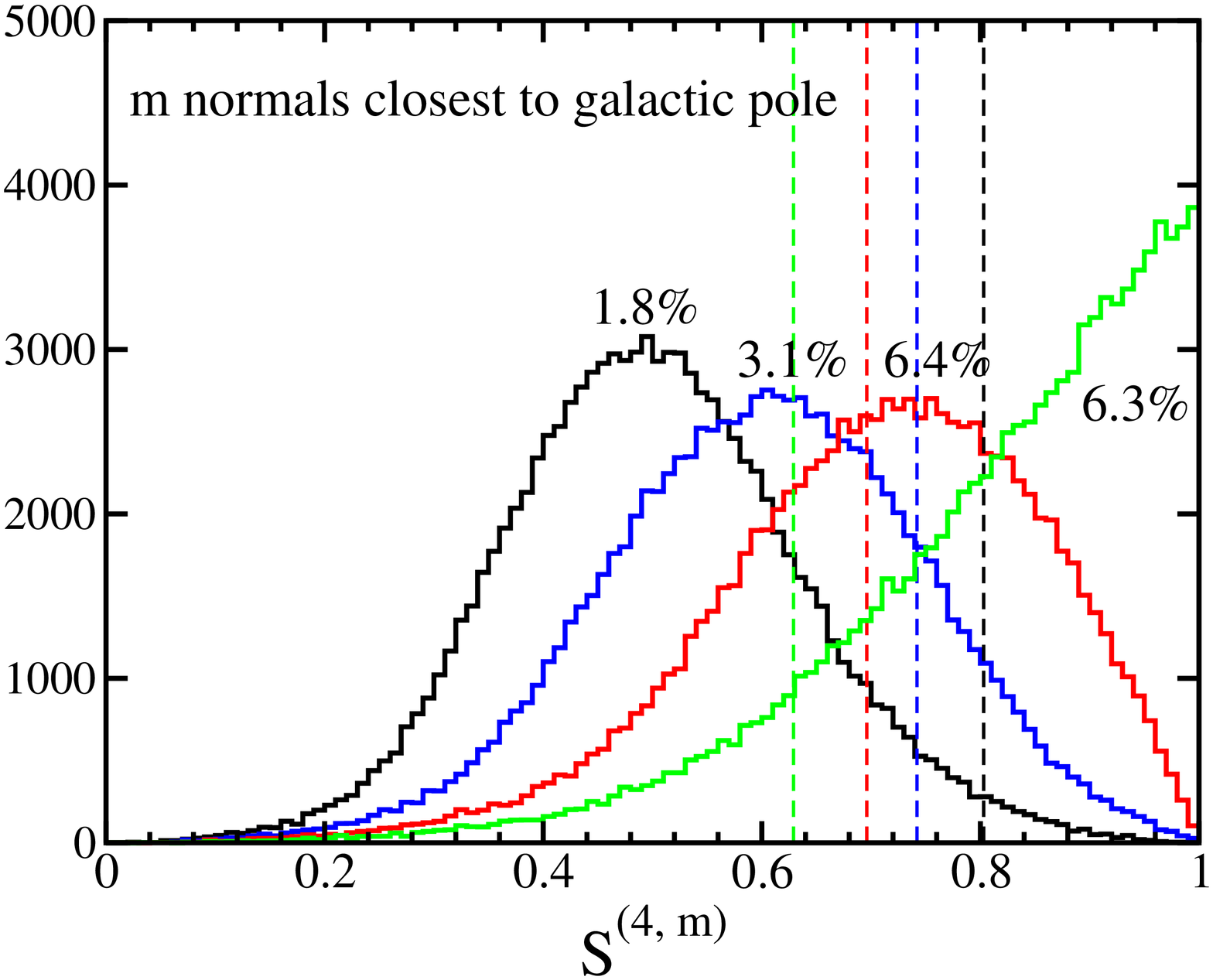}\hspace{0.2cm}
  \includegraphics[width=2.0in,angle=-90]{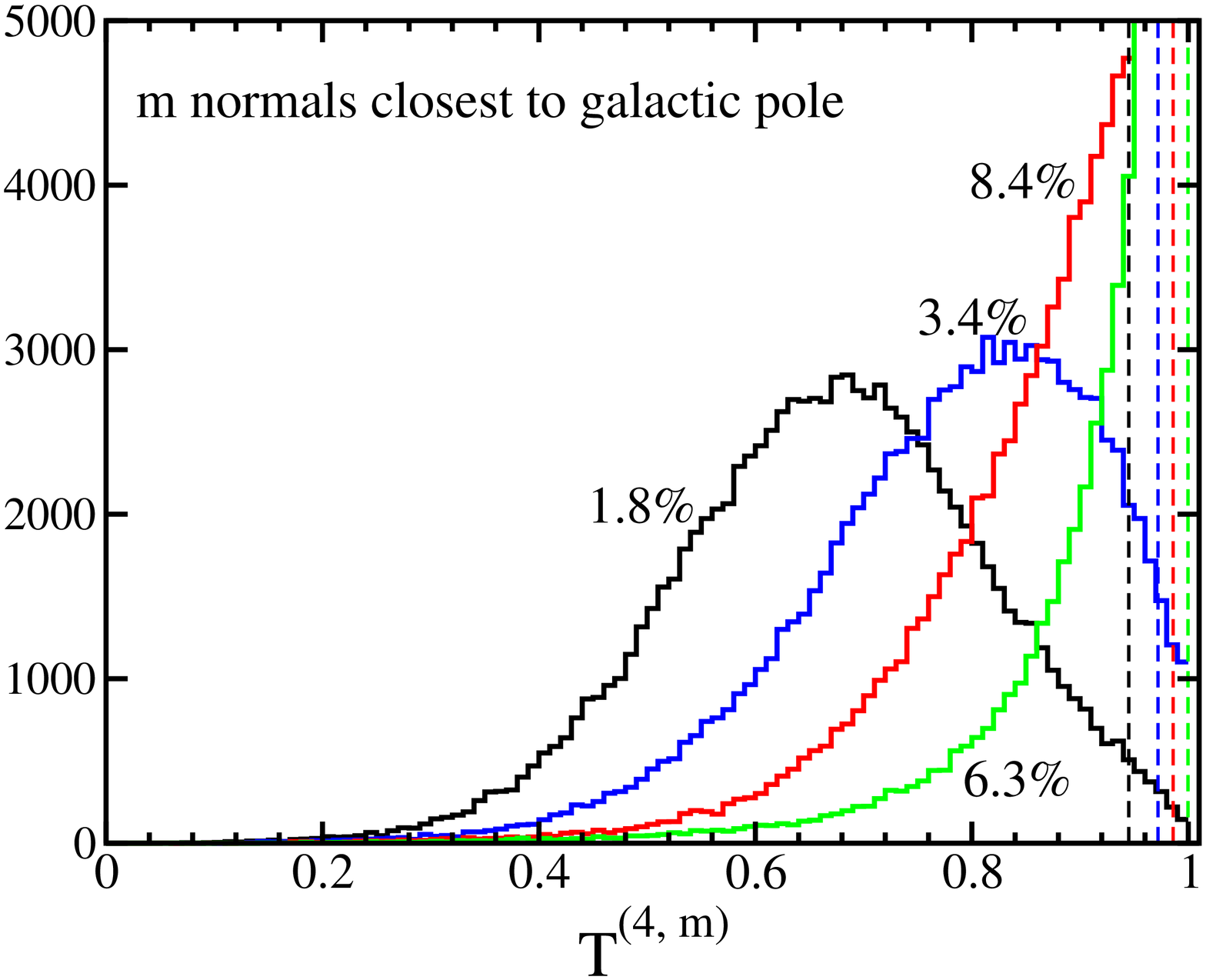}\\[-0.2cm]
  \includegraphics[width=2.0in,angle=-90]{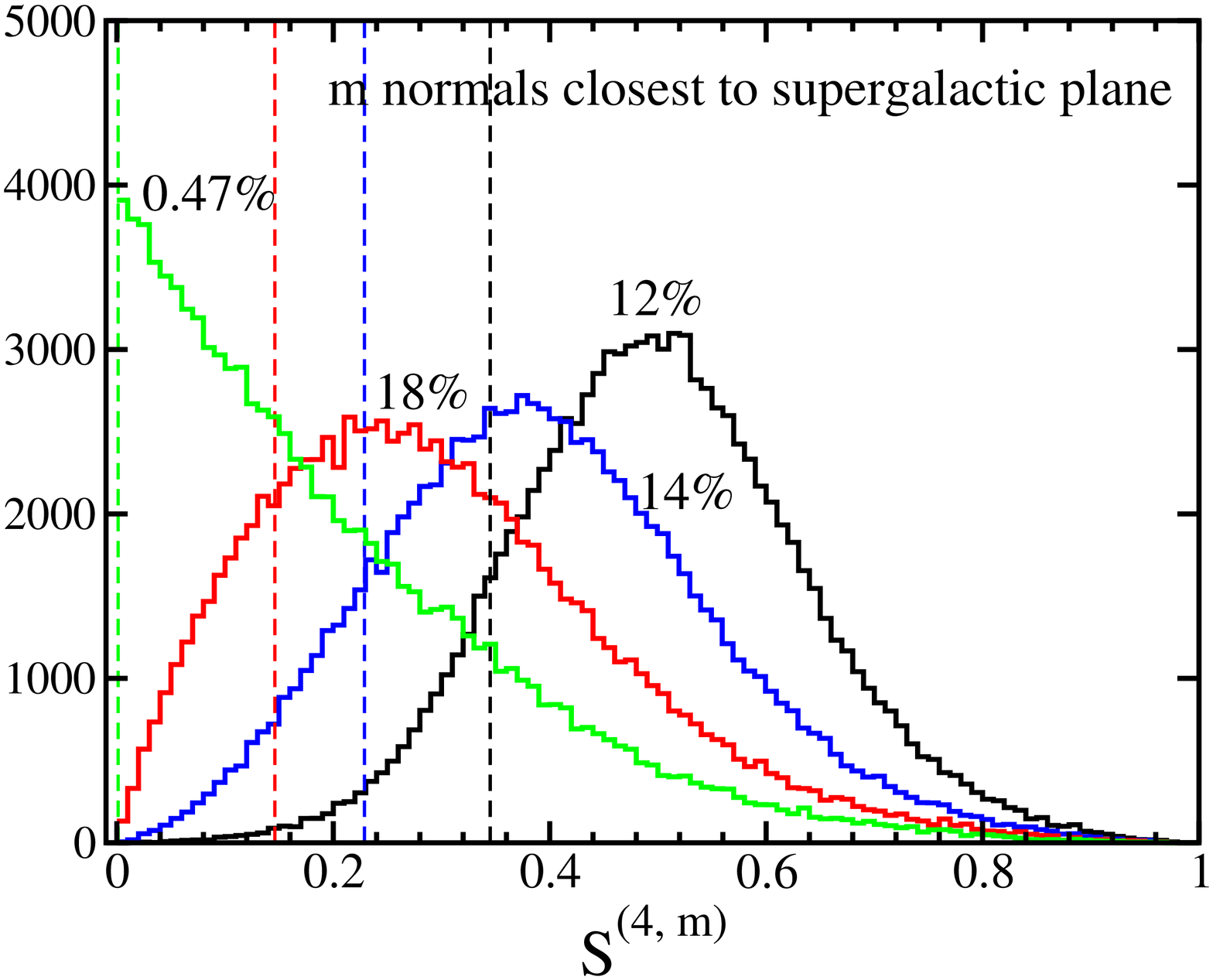}\hspace{0.2cm}
  \includegraphics[width=2.0in,angle=-90]{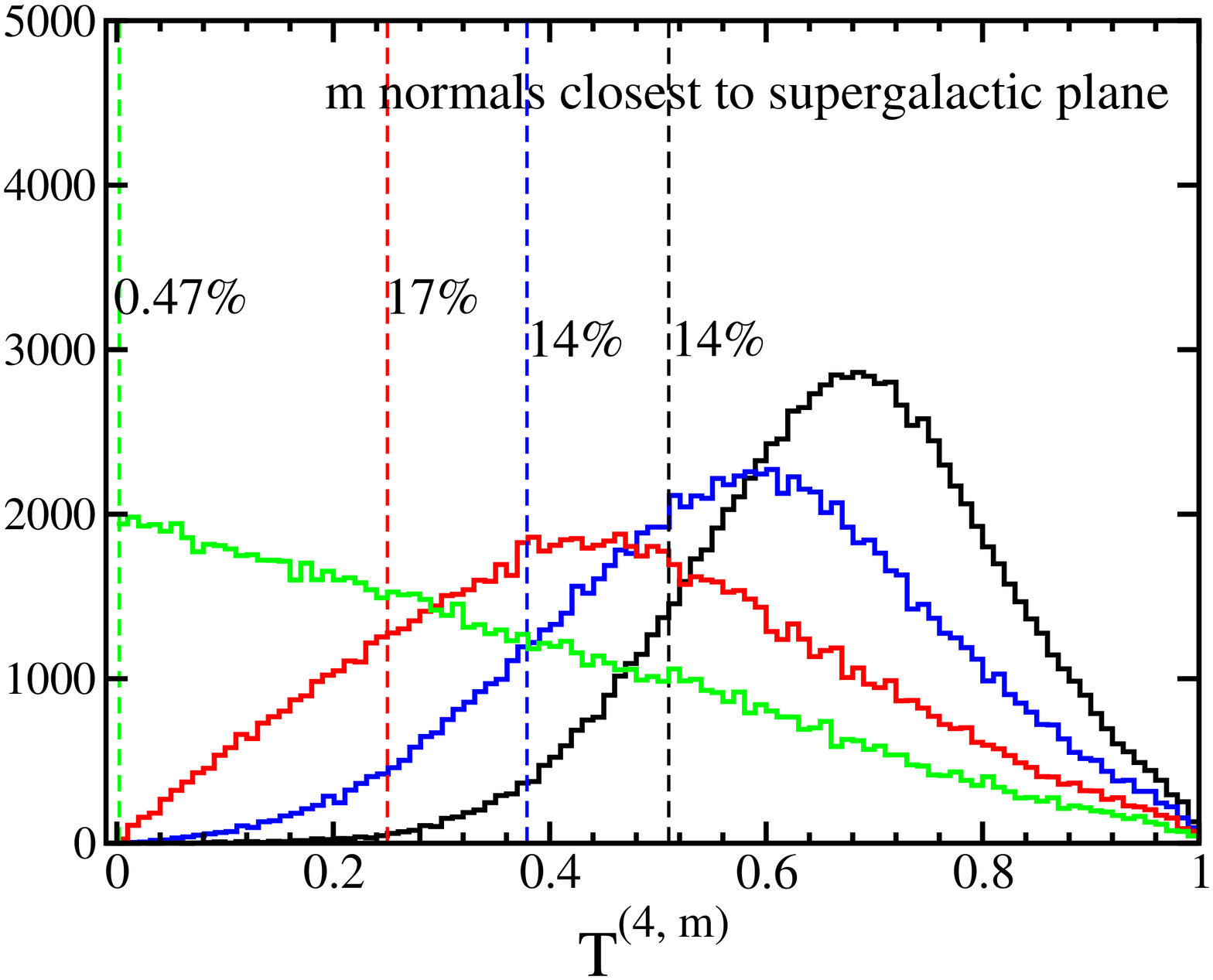}
\caption{ Histograms of $S^{(n, m)}$ statistics (left column) and $T^{(n, m)}$
statistics (right column) for Gaussian random, statistically isotropic Monte
Carlo maps with $m=1\ldots n$ of the most aligned area vectors considered
separately in each panel.  First row panels show the $n=3$ mutual dot-products
of quadrupole and octopole area vectors $A_i$, second row shows dot-products of
the $n=4$ normals with the ecliptic plane, third row shows dot-products of the
$n=4$ normals with the north Galactic pole while the fourth row shows
dot-products of the $n=4$ normals with the supergalactic plane.  Specific
values for the WMAP (TOH DQ-corrected map), for each $m$, are shown with
vertical lines.  The numbers show the percentage of MC maps that have a more
extreme value (i.e.\ larger mutual dot-products $A_i$, larger products with the
NGP, or smaller dot-products with the NEP and NSGP).  In other words, the
numbers show the extremeness of each vertical line's value in the corresponding
histogram. Note that the statistical significance is strongest when all vectors
are considered (that is, when $m=n$), except for the supergalactic plane where
a single octopole normal is only $0\fdg07$ away from this plane while the other
three are not particularly unusual.  }
\label{fig:S_T_stat}
\end{figure*}

Figure \ref{fig:S_T_stat} shows histograms of $S^{(n, m)}$ statistics (left
column) and $T^{(n, m)}$ statistics (right column) for Gaussian random,
statistically isotropic MC maps, as well as values for the TOH DQ-corrected
map.  The statistics shown are $S^{(3, m)}$ and $T^{(3, m)}$ for the intrinsic
alignment of the octopole area vectors with the quadrupole area vector, and
$S^{(4, m)}$ and $T^{(4, m)}$ for the alignment of normals with the ecliptic
plane, Galactic poles, and supergalactic plane.  The dipole and equinox
alignments are similar to the ecliptic plane and are not shown.

Note several interesting features in Fig.~\ref{fig:S_T_stat}. First, the
probabilities for the algebraically-related values of $S^{(n, 1)}$ and $T^{(n,
1)}$ are by definition identical, since they measure the extremeness of a
single parameter.  Second, evidence for alignment of the quadrupole and octopole
and for alignment of multipoles with the ecliptic plane (and similarly with
the dipole and 
equinoxes) is strongest when all dot-products are considered --- i.e. when
$m=n$.  This gives us further confidence that these probabilities are generally
not dominated by one or two unusual alignments, but rather when alignment of
all four normals, either mutual or with the specified direction, are
considered. 

We note that the alignment of normals with either the supergalactic plane or
the Galactic poles is dominated by a single normal, $\hat w^{(3; 2, 3)}$. In
the case of the supergalactic plane, this normal is only $0\fdg07$ away from
the plane (in the TOH map), while the other three are not particularly close to
the plane at all; see Table \ref{tab:oldstat}, Figure \ref{fig:map:tegmark:2}
and \citet{Schwarz2004}.  Therefore $S^{(4,1)} < S^{(4,m)}$ for $m\geq 2$.
Although this normal is still within approximately $1\fdg5$ in the ILC and LILC
maps, this is still sufficient to raise the probability of $S^{(4,1)}$ to
approximately $10$\%.  The fact that only the $S^{(4,1)}$ statistic and only
the TOH map show small probabilities suggests that the supergalactic
correlation is a statistical fluctuation.

The same normal $\hat w^{(3; 2, 3)}$ is approximately $10\degr$ from the
Galactic pole. This is largely responsible for the measured correlation of the
quadrupole and octopole with the Galactic pole, although it is true that
$S^{(4,4)} < S^{(4,1)}$ for this case.  We discuss this further in Sec.~\ref{sec:ecvsgal}.

\begin{table*}
  \caption{Table of probabilities (in percent) for the seven different tests
    using the TOH, LILC and ILC maps.  We show the results based on both
    DQ-corrected and uncorrected maps. The statistics $S^{(n,n)}$, computed for
    each microwave background map and compared to 100,000 realizations of
    Gaussian random, statistically isotropic maps with WMAP's pixel noise, has
    been used to compute all probabilities. Except for the test $D_i$, all
    tests are based on the dot-products of the area vectors, $\vec{w}$, of the
    quadrupole and octopole.}
  \label{tab:S_stat_WMAP}
  \begin{tabular}[b]{lcccccc}
    \hline
    Test & TOH DQ-corr & LILC DQ-corr & ILC DQ-corr 
    & TOH uncorr & LILC uncorr & ILC uncorr \\
    \hline
    $A_i$                  & 0.117 & 0.602 & 0.289 & 0.582 & 2.622 & 0.713\\
    $D_i$                  & 1.246 & 1.309 & 2.240 & 1.262 & 1.309 & 2.567\\
    ecliptic plane    & 1.425 & 1.480 & 2.006 & 1.228 & 1.735 & 2.724\\
    NGP               & 0.734 & 0.940 & 0.508 & 0.909 & 1.265 & 0.497\\
    SG plane  & 14.4  & 13.4  & 8.9   & 11.6  & 10.2  & 6.5 \\
    dipole  & 0.045 & 0.214 & 0.110 & 0.093 & 0.431 & 0.207\\
    equinox & 0.031 & 0.167 & 0.055 & 0.064 & 0.315 & 0.080\\
    \hline
  \end{tabular}
\end{table*}

In the absence of any model, $S^{(n,n)}$ and $T^{(n,n)}$
statistics seem the fairest choice, and the one we adopt.
These statistics treat planes and directions identically (i.e. the ordering of the
relevant dot-products does not matter).
Table \ref{tab:S_stat_WMAP} shows the probabilities for the seven different tests
applied on the TOH, LILC and ILC maps using the $S^{(4,4)}$ statistic.  We show the
results based on both DQ-corrected and DQ-uncorrected maps.

Finally, there has been some confusion regarding constraints from the
application of the $S$ statistic to the normalized, $D_i$, and
unnormalized, $A_i$, area vectors.  It has been claimed that the $S$
statistic applied to the area vectors $A_i$ is both not a measure
of alignment between the octopole and quadrupole and
not a very robust statistic. It was argued that one should only
use the $D_i$. The heuristic argument is that $S$ applied to the $A_i$ is not
actually a measure of alignment since the $A_i$ incorporate information
about the lengths of the area vectors, not just their directions (see, for
example, \citealt{Weeks}).  However, what this means is that the $S$
statistic applied to the $A_i$ weights the contribution of each plane
according to how well the two associated multipole vectors define that
plane --- the closer they are to orthogonal, the more well defined the plane
is, and the more heavily the plane is weighted; the more nearly parallel
they are, the less well defined the plane is and the smaller its weighting
in the statistic.  This weighting thus seems entirely intuitive and
appropriate.

The confusion noted above has arisen from inconsistently applying the test
to maps, some with the kinetic quadrupole correction applied and others
without it.  The specific concern about the robustness of this statistical
test was that the $S$ statistic applied to the $A_i$ of the LILC map was
apparently much less unlikely than the $S$ statistic applied to the $A_i$
of the TOH map \citep{Weeks}.  However, this result compared the
Doppler-corrected TOH map to the uncorrected LILC map.  The former has
probability around $0.1\%$, whereas the latter has probability
$2.6\%$.  When we Doppler-correct the LILC map (as one should), we find that
the probability falls to around $0.6\%$, comparable to that of the TOH
map~(see Table~\ref{tab:S_stat_WMAP}).
Also the Doppler corrected ILC map yields a similar $0.2\%$ probability.
Far from challenging the robustness of the $S$ statistic for the area
vectors, the probabilities derived from different maps support it.

\subsection{The evidence for ecliptic (and other) alignment}
\label{sec:ecvsgal}

The values in Table \ref{tab:S_stat_WMAP} show alignment of the area vectors of
the quadrupole and octopole with each other and with some specific physical
directions.  Also shown in Table~\ref{tab:S_stat_WMAP} is the statistical
evidence for a correlation with the dipole and equinox direction (at larger
$99.7\%$C.L.  in all three maps), with the NGP (at larger $99\%$C.L.) and the
ecliptic plane (at larger $98\%$C.L.). As discussed in Sec.~\ref{sec:ST:QO}, 
our use of the $S^{(4,4)}$ statistic (instead of the $S^{(4,1)}$ statistic)
shows that the correlation with the supergalactic plane is not significant.

Previously we have claimed that there is evidence for ecliptic alignment,
but not for Galactic alignment. Yet Table~\ref{tab:S_stat_WMAP} shows a
slightly higher significance for the Galactic than the ecliptic alignment.
In this Section we demonstrate why the ecliptic correlation is significant and
the Galactic one is not. Simultaneously, we consider the suggestion by
\citet{Bielewicz2005} that the only important correlation is between the
quadrupole and octopole area vectors themselves.  We show that actually 
there is  a $>98$\% C.L. further alignment with the ecliptic plane (and argue
that this alignment is in fact $>99.9$\% C.L. unlikely), but that the
additional alignment with  the Galaxy is not significant.

\begin{figure}
  \includegraphics[width=2.8in,angle=-90]{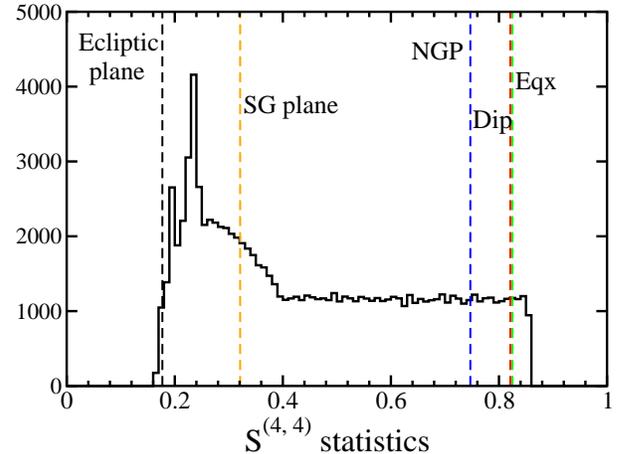}\hspace{-0.5cm}
\caption{Histogram of the $S^{(4, 4)}$ statistics applied to the TOH map
quadrupole and octopole area vectors and a fixed direction or plane on the sky,
where the area vectors have been rotated together in a random direction $10^5$
times.  Vertical lines show the $S$ statistics of the actual area vectors
applied to the ecliptic plane, NGP, supergalactic plane, dipole and equinox
directions (Table \ref{tab:given23} shows the actual product percentile ranks
among the random rotations for all three full-sky maps). This Figure and
Table~\ref{tab:given23} show that, even {\it given} the relative location of
the quadrupole-octopole area vectors (i.e.\ their mutual alignment), the
ecliptic plane, dipole and equinox alignments are unlikely at the $\ga 95\%$
C.L.\ while the NGP and supergalactic plane alignments are not.  }
\label{fig:given23}
\end{figure}

To begin we take the correlations in the quadrupole and octopole (their area
vectors) to be fixed as measured.  We then compute 100,000 rotations of the TOH
quadrupole and octopole on the sky and compute the $S^{(4,4)}$ statistic for
each direction in each of these rotated microwave background skies (see
\citet{Bielewicz2005} for a similar approach).  The results are shown in
Fig.~\ref{fig:given23}.  The histogram is the distribution of the $S^{(4,4)}$
we get from these TOH rotated skies and the dashed vertical lines are the
values of $S^{(4,4)}$ for each of the ecliptic plane, NGP, supergalactic plane,
dipole and equinox axes.  In Table \ref{tab:given23}, we list the percentiles
of the values of $S^{(4,4)}$ for these five physical directions for all three
of the TOH, the ILC and the LILC maps.  (The distributions for the ILC and LILC
maps are quite similar to the TOH distribution.)

The results are striking. The percentile for the ecliptic plane is between
$0.2$\% (LILC) and $1.7$\% (ILC). The percentile for the Galactic pole is
between $87$\% and $90$\%, so that the two-sided probability is only between
$74$\% and $80$\%. (We justify below why a two-sided probability is not
appropriate for the ecliptic plane alignment.) This shows that, given the
observed shapes and alignment of the quadrupole and octopole, the evidence from
the area vectors for additional correlation with the ecliptic is at least 10
times stronger than for additional correlation with the Galaxy.  There is also
mild evidence for additional correlation with the dipole or equinox at
approximately the 95\% C.L.\ (approximately the 90\% C.L.\ when we take two-sided
probabilities).

Qualitatively we can understand from inspection of
Fig.~\ref{fig:map:tegmark:2+3} why the quadrupole and octopole normals are so
much better correlated with the ecliptic than with the Galactic pole.  These
four normals essentially surround the ecliptic, therefore it is relatively hard
to be more correlated; similar is true, though to a lesser extent for the
dipole and equinox directions.  On the other hand just one of the normals comes
somewhat close to the Galactic pole, while the other three are further away.
The situation for the dipole and equinox is intermediate to these two.

It is also very important to notice that {\em the quadrupole area vector 
does not contain all the information about the quadrupole}. That means the 
information that a zero of the sum of quadrupole and octopole traces the 
ecliptic for about $1/3$ of the sky is not contained in our $S^{(4,4)}$ 
statistics. Moreover, {\em the $S^{(4,4)}$ statistics does not make use of 
all the information contained in the octopole area (normal) vectors}.
To see this consider a somewhat idealized sky
which is dominated by $Y_{33}$ in some frame. (The is nearly true for the
$\ell=2+3$ sky.)  In this case, all the $\ell=3$ multipole vectors lie approximately
in a single plane, and the three normal vectors are closely aligned,
and define a small circle. Any rotation around the axis defined by 
the centre of that small circle will give rise to the same results in our 
statistical tests, because this rotation merely leads to a rotation of 
the $Y_{33}$ minima and maxima within the octopole plane. The three minima and 
three maxima are separated by three nodal lines, which are great circles. For 
one of those great circles to be the ecliptic, the axis of rotation (the 
direction defined by the three normal vectors) would have to lie on the 
ecliptic. This is what we have nearly found to be the case in the data.

However, placing the normal vectors on the ecliptic plane does not itself
guarantee that the ecliptic plane is one of the null contours between extrema.
The freedom to rotate all the multipole vectors in their plane remains.
With $60\degr$ between extrema, the chance of the ecliptic plane being a
null-contour within the observed tolerance of about $3\fdg5$ is about
$6$\%.  Since this rotational freedom is entirely independent of the
alignment of the area vectors, we can multiply the probability of the
area-vector alignment by the probability of the rotational alignment to
obtain at most $0.1$\% (ILC), and as little as $0.01$\% (LILC).  (It is not
appropriate to use two-sided probabilities, because if the normal vectors
had been aligned with the ecliptic poles rather than the ecliptic plane,
then the ecliptic plane could not have been a null contour between
extrema.)  The ecliptic plane only traces the null-contour over one third
of the sky due to the fact that the sky isn't a pure $Y_{33}$ mode.
However, the ecliptic plane also doesn't pass between just any two extrema
but instead splits the weaker extrema in the north from the stronger
extrema in the north.  Thus these two effects approximately cancel each
other leaving us with the probabilities estimated above.  Though this is
an estimate and a more detailed statistical analysis is warranted,
the probability of the quadrupole and octopole being this aligned with
the solar system is unlikely at greater than $99.9$\% C.L.

We note that it has been suggested that one should reduce the significance
of this discovery by some large number of possible ``physical great
circles'' with which we could have noticed a correlation --- that our focus
on the ecliptic is purely (and by implication fatally) {\it a posteriori}
\citep{Bielewicz2005,SS2004}.  To the contrary, we would argue that
given the experiment there are precisely two great circles and their axes with which one
{\em must} look for correlations --- the ecliptic and the Galactic equator.
The former because WMAP orbits the sun deep within the solar system, and
correlations with the ecliptic could be a sign of either systematic errors
or a local foreground; the latter because the Galaxy is an important
foreground source, and correlations to the Galactic equator would be a sign
of residual Galaxy contamination (cf.\ Sec.~\ref{sec:foregr}).

\begin{table}
  \caption{Percentile ranks of the quadrupole-octopole area vector dot-products
with specific directions, among the $10^5$ such products where the area vectors
have been rotated together in a random direction.  We show dot-products for the
TOH, LILC and ILC area vectors, and the NEP, NGP, NSGP, dipole and equinox
directions.  This Table shows that, even {\it given} the relative location of 
the quadrupole-octopole area vectors (i.e.\ their mutual alignment), the ecliptic plane,
dipole and equinox alignments are unlikely at the $\ga 95\%$ C.L.\ while 
the NGP and supergalactic ones are not.  }
  \label{tab:given23}
  \begin{tabular}[b]{lccc}
    \hline
    Test & TOH DQ-corr & LILC DQ-corr & ILC DQ-corr \\
    \hline
    ecliptic plane    & 1.0   & 0.2  & 1.7\\
    NGP               & 87    & 88   & 90\\
    SG plane          & 34    & 33   & 25 \\
    dipole            & 95.6  & 93.8 & 94.5 \\
    equinox           & 96.1  & 94.4 & 96.4\\
    \hline
  \end{tabular}
\end{table}

\subsection{Single multipole alignment test for $\ell \leq 50$}
\label{sec:single_ell}

So far we have devoted attention to alignments of the area vectors defined by
the quadrupole and octopole. We would now like to investigate alignment in
higher multipoles. To simplify the analysis, we consider a single multipole
$\ell$ at a time.  We then ask whether the multipole in question is unusually
planar. Note however that this test is by no means exhaustive in finding
unusual correlations --- for example, we are not considering {\it pairs} of
multipoles, as we did for the quadrupole and octopole. In fact, the octopole,
when considered alone, is not unusually planar as first shown by \citet{TOH}
and further illustrated using our tests below. Nevertheless, considering the
more complicated tests applied to higher multipoles is not efficient at this
time without a specific suspected theory or model in mind. Moreover, we are
curious to investigate whether the ``bites'' observed on the angular power
spectrum at a handful of multipoles \citep{Hinshaw2003} are somehow correlated
to the multipole vector alignments in those multipoles. Therefore, we proceed
with testing the single multipole alignment of area vectors.

For a fixed multipole $\ell$ we have $N_\ell= \ell(\ell-1)/2$ area (headless)
vectors $\vec{w}_i$.  We would like to find the plane that is defined by these
vectors. We define the plane as the one whose normal, $\hat{n}$, has the
largest dot product with the sum of the area vectors. Since $\vec{w}_i \cdot
\hat{n}$ is defined only up to a sign, $\vec{w}_i$ is headless, we take the
absolute value of each dot product. Therefore, we need to find $\hat{n}$ that
maximizes
\begin{equation}
\mathcal{S}\equiv \frac 1 {N_{\ell}} 
\sum_{i=1}^{N_\ell} \left |\vec{w}_i \cdot \hat{n}\right |.
\label{eq:S_single_mult}
\end{equation}
An exact procedure to find the best aligned direction is non-trivial due to 
the presence of the absolute value. Instead of writing an absolute value we 
replace $\vec{w}_i$ by $s_i \vec{w}_i$, where $s_i$ is $\pm 1$, such that 
there are only positive terms in the sum. The problem is that we cannot know 
the signs before we know the best aligned directions and thus we have to try 
out all the possibilities. There are $N_\ell$ area 
vectors, and thus there are $2^{N_\ell}$ possible choices for the set 
$\{s_i\}$. As is easily seen, for large $\ell$ this becomes computationally 
more expensive than a numerical search for the best aligned 
direction. For any fixed set of signs $\{s_i\}$, the best aligned direction 
$\hat{n}$ expressed in spherical coordinates, $(\theta,\phi)$, is given by
\begin{eqnarray}
  \tan \phi &=& \left. \left(\sum s_i w_{i,y}\right) \right/
  \left(\sum s_i w_{i,x}\right)
  \\
  \tan \theta &=& \left. \left[\sum s_i \left(w_{i,x} \cos \phi + w_{i,y}
        \sin \phi\right)\right] \right/ \left(\sum s_i w_{i,z}\right), 
  \nonumber
\end{eqnarray}
where $w_{i,x}$ denotes the $x$ component of $\vec{w}_i$, \textit{etc}. 

In practice, we used an iterative solution to the problem:
pick some trial direction $\hat{n}$ so that we can immediately compute
$\mathcal{S}$. Repeat many times for $\hat{n}$ randomly chosen on a unit 
sphere, and simply find one that has maximal $\mathcal{S}$. We 
find that the search converges after a few thousand trials for the 
direction $\hat{n}$.

We compute $\mathcal{S}_{\rm WMAP}$ for any given multipole $\ell$ in the TOH
map, and then compare it to the value of $\mathcal{S}_{\rm MC}^{(i)}$, at that
multipole, computed in a large number of Monte Carlo realizations of a Gaussian
random, statistically isotropic sky with WMAP's pixel noise added. We then
rank-order $\mathcal{S}_{\rm WMAP}$ among the $\mathcal{S}_{MC}^{(i)}$. A low
rank indicates that the WMAP areas are roughly aligned, or equivalently, that
the multipole vectors are planar and most of the power lies in the
corresponding plane. Conversely, a high rank would imply that the multipole in
question is non-planar relative to a Gaussian random, statistically isotropic
expectation. We use 10,000 Monte Carlo realizations at $l\leq 20$ and at any
$l$ that has a very low or high rank in order to obtain sufficient statistics;
at all other $\ell$ we use 1000 realizations.

\begin{table}
  \caption{Ranks of the $\mathcal{S}$ statistics, R, given as percentiles. Low
    ranks indicate multipoles that are planar (i.e.\ their area vectors are
    aligned), while high ranks indicate multipoles that are non-planar. The
    statistic $\mathcal{S}$, defined in Eqn.~\ref{eq:S_single_mult}, is applied
    to the TOH cleaned map and compared with Gaussian random, statistically
    isotropic skies to obtain the ranks. We used 10,000 realizations of such
    skies for $\ell\leq 20$ and for $\ell=44$, and 1000 realizations for all
    other multipoles.  Values within 5 percentiles of 0 or 100 are shown in
    bold.  }
  \begin{tabular}{cc|cc|cc|cc|cc}
    \hline
    \multicolumn{10}{c}
    {
      Single-$\ell$ Ranks of the $\mathcal{S}$ Statistic (\%) }\\
    $\ell$ & R & $\ell$ & R & $\ell$ & R & $\ell$ & R & $\ell$ & R\\
    \hline
   2+3 & {\bf 0.35} & 11 & 42  	   & 21 & 32 	   & 31 & 33 	   & 41 & 26 \\
    2  & ---        & 12 & 19  	   & 22 & 48  	   & 32 & 46 	   & 42 & {\bf 3} \\
    3  & 7     	    & 13 & {\bf 3} & 23 & 20  	   & 33 & 65 	   & 43 & 86 \\
    4  & 75         & 14 & {\bf 5} & 24 & 15  	   & 34 & 11 	   & 44 & {\bf 99.7} \\
    5  & {\bf 99.7} & 15 & 62  	   & 25 & 48  	   & 35 & 13 	   & 45 & 35 \\
    6  & {\bf 4}    & 16 & {\bf 4} & 26 & 46  	   & 36 & {\bf 96} & 46 & 31 \\
    7  & 45    	    & 17 & {\bf 3} & 27 & {\bf 96} & 37 & 89 	   & 47 & 49 \\
    8  & 77    	    & 18 & 18  	   & 28 & 80  	   & 38 & 47 	   & 48 & 27 \\
    9  & {\bf 95}   & 19 & 19  	   & 29 & 58  	   & 39 & 86 	   & 49 & 91 \\
    10 & 65    	    & 20 & 29  	   & 30 & {\bf 4}  & 40 & 12 	   & 50 & 19  \\
    \hline
  \end{tabular}
\label{tab:single_ell}
\end{table}

Table \ref{tab:single_ell} shows the ranks of the $\mathcal{S}$ statistic for
$3\leq \ell\leq 50$, given as percentages\footnote{The quadrupole alone has
only one area vector which picks out a unique direction $\hat{n}$ and therefore
cannot be used with the $\mathcal{S}$ statistics as defined in
Eqn.~(\ref{eq:S_single_mult}).}.  For example, 25\% would indicate that 25\% of
Monte Carlo maps are more planar and 75\% are less planar at that $\ell$. Note
that, while the octopole alone is planar only at the 92\% C.L., the quadrupole
and octopole together are planar at 99.65\% C.L. The planarity of the
quadrupole and octopole is therefore very significant, in agreement with other
tests (see e.g.\ $A_i$ and $D_i$ in Table \ref{tab:S_stat_WMAP}).  The
$\ell=5$, in contrast, is non-planar and only 0.3\% of Monte Carlo realizations
exhibited lower planarity.  What has been called the ``sphericity'' of $\ell=5$
was first pointed out by \citet{LILC}; their test found it unusual at the
5-10\% level.

Visual inspection of table \ref{tab:single_ell} suggests that there
is an excess of both high and low values.  Simple attempts to quantify this 
do indeed find such anomalies at  between $95$\% C.L.\ and $99$\% C.L\@.
We might therefore conclude that the alignment test as defined in
Eqn.~(\ref{eq:S_single_mult}) gives strong hints of something unusual 
at $4\leq \ell\leq 50$ in the TOH cleaned map, but without further evidence, 
the case is not sufficiently strong to stand on its own for any bold claims.
Moreover, we find that at $\ell\geq 8$, the values of the $\mathcal{S}$ 
statistic differ substantially among the TOH, ILC and LILC maps.

Apart from looking at the $\mathcal{S}$ statistics, we also inspected the 
best aligned directions $\hat n$ (the direction that maximize $\mathcal{S}$). 
We would expect that only the best aligned directions of planar multipoles have
a well defined meaning. For $\ell < 8$ only $\ell = 6$ is singled out by 
our statistics. We find the corresponding vector at $(l,b) = (152\fdg 4,
50\fdg 3)$, which is $46\fdg 2$ from the ecliptic pole. Among the higher 
best aligned directions, $\ell = 21$ and $\ell = 44$ are within 
$2\fdg 6$ and $9\fdg 0$, respectively, of the dipole. All other vectors are 
more than $10\degr$ from any physical direction studied in this work.

In addition to the $\mathcal{S}$ statistic described above, we also applied 
Bingham's
statistic test of isotropy \citep{Fisher, Morgan}. Let us assume we have $N$
unit vectors with components $(x_i, y_i, z_i)$ $(i=1, \ldots, N)$ and that we
want to check whether they are distributed isotropically.  We construct the
orientation matrix
\begin{equation}
\mat{T} = \frac{1}{N}
\sum^N_{i=1}
\left( \begin{array}{ccc}
x_i x_i & x_i y_i & x_i z_i \\
y_i x_i & y_i y_i & y_i z_i \\
z_i x_i & z_i y_i & z_i z_i
\end{array} \right) \,,
\end{equation}
which is real and symmetric with unit trace, so that that the
sum of its eigenvalues $e_k$ ($k=1,2,3$) is unity. For an
isotropic distribution all three eigenvalues should
be equal to $1/3$ to within statistical fluctuations. Bingham's modified
statistic ${\cal B}^{\star}$~\citep{Bingham} is defined as
\begin{eqnarray}
{\cal B}^{\star} &=& {\cal B} \left( 1 - \frac{1}{N}\left[\frac{47}{84} +
\frac{13}{147}{\cal B} + \frac{5}{5292}{\cal B}^2 \right]\right) \,, 
{\,\,\,\,\rm where} \nonumber \\
{\cal B} &=& \frac{15N}{2}\sum^3_{k=1}\left( e_k - \frac{1}{3} \right)^2.
\end{eqnarray}
For isotropically distributed vectors and $N\gg 1$, ${\cal B}^{\star}$ is
distributed as $\chi^2_5$. Here we compare WMAP's value of ${\cal B}^{\star}$
for a given multipole to Monte Carlo simulations directly, and therefore
do not require assuming $N\gg 1$.

Bingham's statistic results for $2\leq \ell\leq 50$ are broadly consistent
with those for the $\mathcal{S}$ statistic shown in
Table~\ref{tab:single_ell} (and hence we do not show them separately).  We
find that $\ell=5$ is non-planar at the $99.8\%$ C.L.\ but, apart from that,
other multipoles are neither planar nor non-planar at a level not expected from
such a statistical sample. Finally, note that Bingham's statistic,
being general and coordinate independent, does not use all information 
and is typically not as strong as the coordinate dependent tests.

\subsection{Higher multipole angular momentum vectors}
\label{sec:angmom:statistic}

The angular momentum dispersion~(\ref{eqn:angmom-dispersion}) can be
maximized for all multipoles and thus serve as a statistic.  That is we can
find the axis $\hat n_\ell$ around which $(\Delta L)_\ell^2$ is maximized.
The spherical harmonics provide an irreducible representation of the
rotation group in three dimensions and transform as $\vec Y_\ell' = \vec
Y_\ell^{T} \mat{D}^{(\ell)}$ where $\vec Y_\ell$ is a vector of the
$\ell$-th multipole spherical harmonics ($2\ell+1$ components) and
$\mat{D}^{(\ell)}$ is a rotation of this multipole.  Since a scalar
function (such as the temperature) is invariant under rotations the $\alm$
must transform as $\vec a_\ell' = \mat{D}^\dagger \vec a_\ell$ under
rotations.  The rotations can be parametrized in terms of the Euler angles
$\alpha$, $\beta$, $\gamma$ in the $zyz$ representation as
\begin{equation}
  \mat{D}(\alpha\,\beta\,\gamma) = \exp\left( \frac{i\alpha}{\hbar}L_z
  \right) \exp\left( \frac{i\beta}{\hbar}L_y \right) \exp\left(
    \frac{i\gamma}{\hbar}L_z \right)
\end{equation}
where $L_y$ and $L_z$ are the $y$ and $z$ components of the angular
momentum operator, respectively.  The discussion here follows
\citet{Edmonds}.  For an alternative representation see appendix A of
\citet{deOliveira2004}.

To perform the maximization it is convenient to use the matrix representation
for the rotations
\begin{equation}
  D^{(\ell)}_{m' m} (\alpha\,\beta\,\gamma) = e^{im'\gamma} d^{(\ell)}_{m'
    m} e^{im\alpha},
\end{equation}
where
\begin{eqnarray}
  d^{(\ell)}_{m' m}\!\!\!\! &=&\!\!\!\! \sum_k \frac{(-1)^{\ell-m'-k} \sqrt{(\ell+m')!
      (\ell-m')! (\ell+m)! (\ell-m)!}}{k! (l-m'-k)! (l-m-k)! (m+m'+k)!}
  \nonumber \\
  & & \quad {} \times 
  \left(\cos\frac{\beta}2\right)^{2k+m'+m}
  \left(\sin\frac{\beta}2\right)^{2\ell-2k-m'-m}.
\end{eqnarray}
In this representation, the angular momentum
dispersion~(\ref{eqn:angmom-dispersion}) becomes
\begin{eqnarray}
  (\Delta L)^2_\ell \!\!\!\! &=&\!\!\!\! \sum_{m',m''} 
  a_{\ell m'}^* a_{\ell m''} e^{i(m'-m'')\gamma}
  \sum_m m^2 d^{(\ell)}_{m' m}(\beta)
    d^{(\ell)}_{m'' m}(\beta) \nonumber \\
  & \equiv & \sum_{m',m''} H^{(\ell)}_{m' m''}(\gamma) G^{(\ell)}_{m''
    m'}(\beta) \nonumber \\
  & = & \trace\left( \mat{H}^{(\ell)}(\gamma) \mat{G}^{(\ell)}(\beta) \right).
\end{eqnarray}
Notice that this expression separates into a term that depends only on
$\gamma$ and the $\alm$, $\mat{H}^{(\ell)}(\gamma)$, and a term that only
depends on $\beta$, $\mat{G}^{(\ell)}(\beta)$.  To extremize this function
we take derivatives of $(\Delta L)^2_\ell$ with respect to $\beta$ and
$\gamma$ which also separates. It is easy to show that
\begin{equation}
  \partial_\gamma H^{(\ell)}_{m' m''} (\gamma) = i(m'-m'')H^{(\ell)}_{m'
    m''} (\gamma)
\end{equation}
and that both $G^{(\ell)}_{m'' m'} (\beta)$ and $\partial_\beta
G^{(\ell)}_{m'' m'} (\beta)$ can be calculated quickly and efficiently (see
\citet{Edmonds} for details).  These rotation angles are related to
standard Galactic coordinates via
\begin{equation}
  (l,b) = (\gamma-180\degr, 90\degr-\beta).
\end{equation}
Finally, we will find it convenient to work with the normalized angular
momentum dispersion (the $t$ statistic of \citet{deOliveira2004})
\begin{equation}
  (\Delta \tilde L)^2_\ell \equiv (\Delta L)^2_\ell \left/ \ell^2 \sum_m
  \left| \alm \right|^2\right. .
\end{equation}
The normalized dispersion takes a value between $(\ell+1)/3\ell$ and one.

\begin{table}
  \caption{The maximum angular momentum dispersion results for the TOH
    cleaned map.  Shown in the table is the multipole number, $\ell$, the 
    Galactic
    coordinates of the direction in which the axis around which the angular
    momentum dispersion is maximized, $(l,b)$, the value of the normalized
    angular momentum dispersion around this axis, $(\Delta\tilde L)_\ell^2$,
    and the percentage of Monte Carlos that had a maximum angular momentum
    dispersion larger than the value found from the TOH map. Values within
    5 percentiles of 0 and 100 are shown in bold.  See the text
    for details.}
  \label{tab:angmom-dispersion}
  \begin{tabular}{cr@{$\fdg$}lr@{$\fdg$}lcc}
    \hline
    $\ell$ & \multicolumn{2}{c}{$l$} & \multicolumn{2}{c}{$b$}
    & $(\Delta\tilde L)_\ell^2$ & MC Larger \\
    \hline
    $2\mbox{+}3$ & $-112$&$7$ & $59$&$7$ & $0.962$ & $\mathbf{\hphantom{0}0.37\%}$ \\
    $2$ & $105$&$7$ & $56$&$6$ & $0.993$ & --- \\
    $3$ & $121$&$6$ & $62$&$0$ & $0.942$ & $11.24\%$ \\
    $4$ & $-106$&$0$ & $36$&$3$ & $0.637$ & $75.93\%$ \\
    $5$ & $-170$&$0$ & $24$&$0$ & $0.484$ & $\mathbf{99.44\%}$ \\
    $6$ & $-160$&$5$ & $44$&$1$ & $0.817$ & $\mathbf{\hphantom{0}3.00\%}$ \\
    $7$ & $-119$&$7$ & $54$&$8$ & $0.585$ & $58.13\%$ \\
    $8$ & $149$&$7$ & $20$&$1$ & $0.546$ & $62.28\%$ \\
    $9$ & $162$&$1$ & $77$&$9$ & $0.504$ & $80.69\%$ \\
    $10$ & $153$&$8$ & $11$&$1$ & $0.511$ & $70.17\%$ \\
    $11$ & $-76$&$9$ & $12$&$2$ & $0.540$ & $45.15\%$ \\
    $12$ & $46$&$3$ & $37$&$2$ & $0.565$ & $28.75\%$ \\
    $13$ & $-71$&$7$ & $41$&$7$ & $0.651$ & $\mathbf{\hphantom{0}2.73\%}$ \\
    $14$ & $59$&$3$ & $9$&$3$ & $0.636$ & $\mathbf{\hphantom{0}2.60\%}$  \\
    $15$ & $-81$&$3$ & $31$&$1$ & $0.503$ & $49.27\%$ \\
    $16$ & $137$&$9$ & $78$&$7$ & $0.632$ & $\mathbf{\hphantom{0}1.82\%}$ \\
    $17$ & $-163$&$5$ & $30$&$2$ & $0.593$ & $\mathbf{\hphantom{0}4.70\%}$ \\
    $18$ & $152$&$8$ & $19$&$0$ & $0.540$ & $14.85\%$ \\
    $19$ & $-121$&$2$ & $60$&$3$ & $0.530$ & $16.20\%$ \\
    $20$ & $124$&$2$ & $20$&$9$ & $0.500$ & $28.78\%$ \\
    \hline
  \end{tabular}

\end{table}

We have maximized the normalized angular momentum dispersion for $\ell=2$
to $20$ for the TOH cleaned map.  The ILC and LILC maps give similar
results.  The results are shown in table~\ref{tab:angmom-dispersion}. 
For each multipole we provide the direction in Galactic coordinates,
$(l,b)$, for the axis around with the angular momentum dispersion is
maximized and the value of the maximum angular momentum dispersion.  We
have also performed 10,000 Monte Carlo simulations of Gaussian random,
statistically isotropic skies and found the maximum angular momentum
dispersion for each one of them.  The final column in the table gives the
percentage of MC skies that had an angular momentum dispersion larger than
that for the WMAP data.  We have also done the same procedure for a joint
fit to the quadrupole and octopole ($\ell=2\mbox{+}3$).  That is, we find
the single axis that maximizes both multipoles.  

We find that the octopole is planar, but only at 89\% C.L\@. 
The strong correlation
between the quadrupole and octopole is seen by the fact that less than $0.4\%$ of
all Gaussian random, statistically isotropic skies have the quadrupole and
octopole this well aligned.  We again confirm the ``sphericity'' of
$\ell=5$ first pointed out by \citet{LILC}.  We find the angular
momentum dispersion to be very low --- all but $0.56\%$ of the MC skies have a
value larger than the WMAP data.  They have also suggested that $\ell=6$ is
planar; by this test it is somewhat planar with only $3\%$ of MC skies
being more planar than the data.  We find a total of $5$ multipoles that
are somewhat planar (less than $5\%$ of MC skies having a larger angular
momentum dispersion), those being $\ell=6$, $13$, $14$, $16$, and $17$.  We
only find the $\ell=5$ multipole to be particularly non-planar.

Setting aside the $\ell=2+3$ result, we see that $6$ of the $18$ angular
momentum dispersions are in either the top or bottom 5 percentile.  The
probability of having $6$ or more of the $18$ so anomalously high or low is
$0.6$\%.  We also see that of these $6$, all but $\ell=5$ are in the top 5
percentile.  The probability of having $5$ or more of the $18$ so anomalously
high is $0.15$\%.

Inspecting the directions of maximum angular momentum dispersion we find that 
only the $\ell=4$ direction is close to one of the physical directions
under consideration: its distance to the ecliptic pole is $10\fdg 3$.
Note that this confirms the qualitative impression from looking at the 
$\ell = 4$ map (see \citet{Schwarz2004}) that this mode has its minima and 
maxima aligned with the ecliptic plane.
It is also interesting to note that the directions given in 
table~\ref{tab:angmom-dispersion}, especially for the planar multipoles, are 
consistent with the ones found as the best aligned directions.

These results are another suggestion that the higher $\ell$ multipoles are not
statistically isotropic.  Reassuringly, comparison of Tables
\ref{tab:single_ell} and \ref{tab:angmom-dispersion} shows the same $5$
multipoles which had a high angular momentum dispersion also exhibited
comparably low ranks of the $\mathcal{S}$ statistic, while $\ell=5$ showed a
high rank of $\mathcal{S}$.  This difference in range of $\ell$ in
Secs.~\ref{sec:angmom:statistic} and \ref{sec:single_ell} was purely a result
of computational limitations.

\subsection{``Shape'' statistic}\label{sec:shape}

The angular momentum dispersion searches for planarity through a weighted
average that favors modes with $m=\ell$.  \citet{Land2005a} have suggested
the use of the ``shape'' statistic which finds the preferred axis and the
preferred $m$ for each multipole.  The statistic is defined as
\begin{equation}
  r_\ell = \max_{m,\hat n} r^{(\ell)}_m
  \label{eqn:shape:statistic}
\end{equation}
where
\begin{equation}
  r^{(\ell)}_m \equiv \left( 2 - \delta_{m,0} \right) \left| \alm \right|^2
  \left/ \sum_m \left| \alm \right|^2 \right. .
\end{equation}
and $\hat n$ is the $z$-axis of the coordinate system in which the $\alm$ are
computed.  Note that the angular momentum dispersion is a weighted sum of these
terms,
\begin{equation}
  (\Delta L)^2_\ell = \frac1{\ell^2}\sum_{m=0}^\ell m^2 r^{(\ell)}_m .
\end{equation}

The maximization of the shape statistic~(\ref{eqn:shape:statistic}) follows
the same formalism as for the angular momentum dispersion and will not be
discussed further (see Sec.~\ref{sec:angmom:statistic}).  We have
performed this maximization and confirm the results of \citet{Land2005a}.
In particular, we find that the surface defined by $r_\ell$ is complicated
with many local maxima.  The results are quite sensitive to the data and
are not consistent among the three full-sky maps made from the WMAP data
(see Fig.\ 2 of \citet{Land2005a}).  Unfortunately this sensitivity is not
understood in terms of features of the data.  That is, the variability in
the results cannot be understood in terms of features such as non-Gaussianity
or a violation of statistical isotropy. The sensitivity of the shape
statistic is related to the difficulty in uniquely defining
the Land-Magueijo vectors (\citealt{Land2005b}, also see
Sec.~\ref{sec:LM:vectors}). For these reasons, the shape
statistic does not serve as a robust statistic for separating nor
understanding Gaussianity versus statistical isotropy.

\section{Foregrounds}\label{sec:foregr}

So far, we have taken into account the effects of noise in the full-sky cleaned
maps by including the WMAP pixel noise into our Monte Carlos. We now explore
the effect of the foregrounds on the quadrupole-octopole anomaly.

While it has repeatedly been emphasized that there might be residual
foreground contamination left in the cleaned maps, it seems that such a
contamination should lead to \textit{Galactic} and not \textit{ecliptic}
correlations. Here we explicitly show this with a quantitative analysis.
We slowly add the measured WMAP foreground contaminations to WMAP full-sky
maps and monitor how the directions defined by the quadrupole and octopole change.

\begin{figure*}
\includegraphics[width=3.5in,angle=-90]{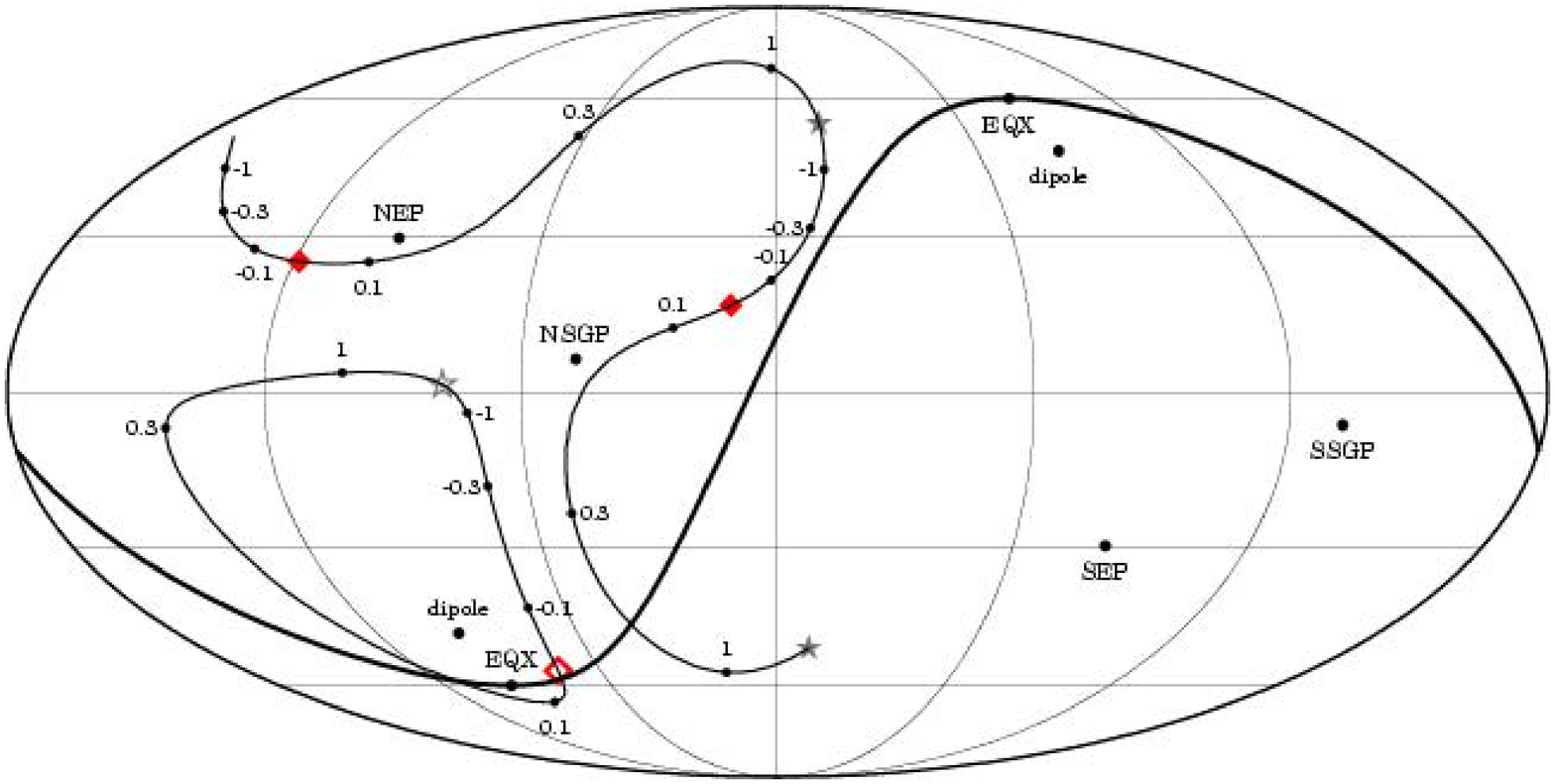}\\[0.2cm]
\includegraphics[width=3.5in,angle=-90]{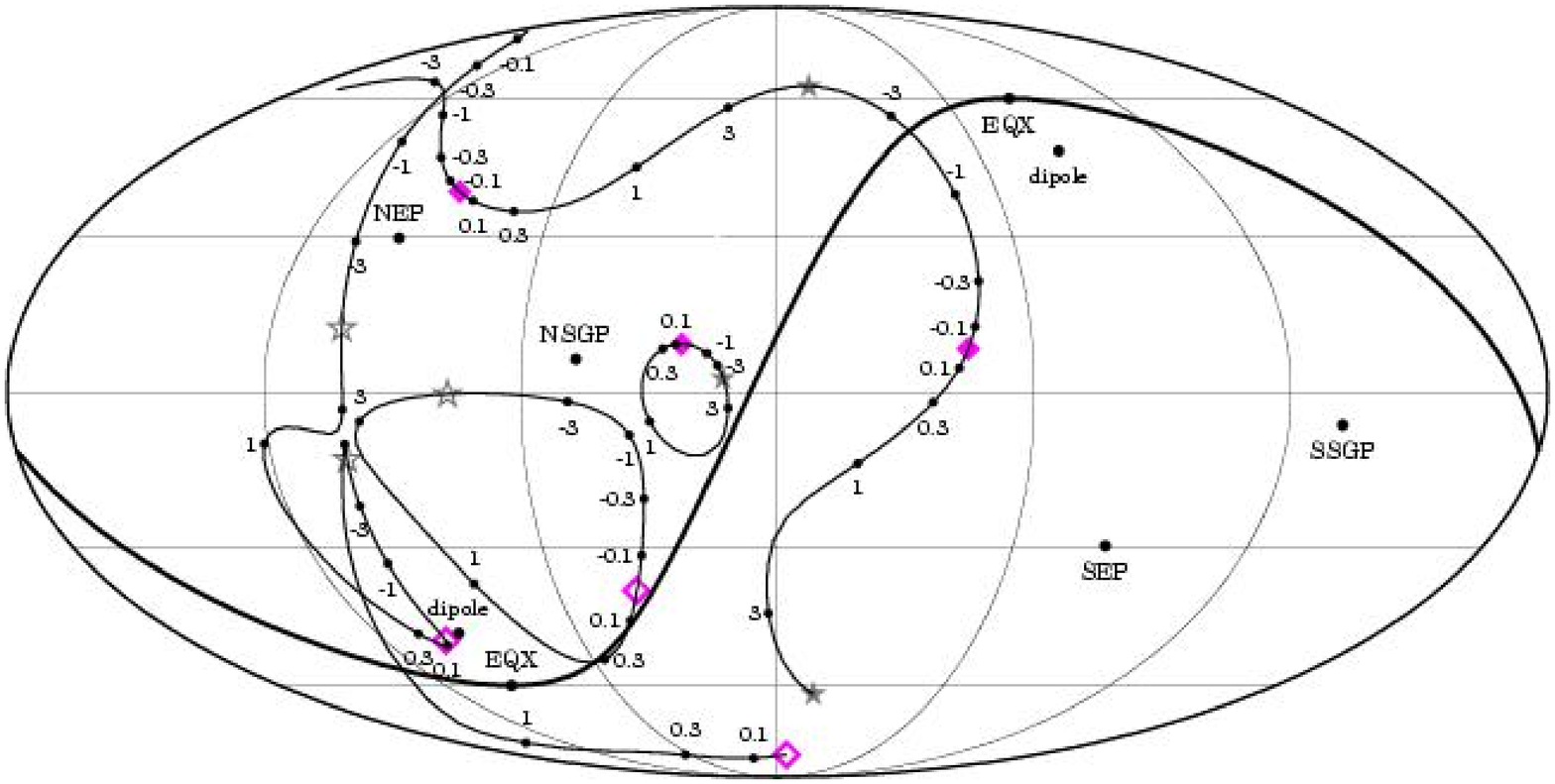}
\caption{Trajectories of the multipole vectors and their normals as increasing
amounts of the foreground is added to or subtracted from the microwave background map.  We have
use the synchrotron V-band foreground and the TOH cleaned map; the coefficient
$c$ shows the rms contribution of the foreground to the microwave background
(see Eqns.~\ref{eq:add_foregr} and \ref{eq:add_foregr_var}). The solid
diamond symbols show the zero-foreground locations of the multipole vectors
while the solid stars refer to their pure-foreground locations. Similarly, the
open diamond and star symbols refer to the beginning and end points of the
normal vectors. On each trajectory we label a few values of the coefficient
$c$. The top panel shows $\ell=2$ (two vector and one normal) and the bottom
panel shows $\ell=3$ (three vectors and three normals).  Not all trajectories
end on plotted symbols.  For these trajectories they end on the multipole
vector (or normal) that is the negative of the plotted vector at the start
of the trajectory.  Note that, in the
large-foreground limit, the quadrupole vectors move near the $z$-axis and the
normal into the Galactic plane, while for the octopole all three normals become
close to the Galactic disk at $90^\circ$ from the Galactic center. Therefore,
as expected {\it Galactic foregrounds lead to Galactic, and not ecliptic,
correlations of the quadrupole and octopole}.}
\label{fig:add_foregr}
\end{figure*}

Let $T_{\rm CMB}(\hat n)$ be the cleaned-map microwave background temperature in some direction,
and $T_{\rm FOR}(\hat n)$ the temperature from one of the three basic
foreground maps (thermal dust, free-free emission or synchrotron emission)
provided by the WMAP team. We form the total contaminated map as
\begin{equation}
T_{\rm tot}(\hat n)=T_{\rm CMB}(\hat n)+ c\, T_{\rm FOR}(\hat n) 
\sqrt{\var(T_{\rm CMB})\over \var (T_{\rm FOR})}.
\label{eq:add_foregr}
\end{equation}
where $c$ is the foreground fraction. Note that the second term has been
normalized so that 
\begin{eqnarray}
\var (T_{\rm tot})
&=&\var (T_{\rm CMB}) + c^2\,\var (T_{\rm FOR})
{\var (T_{\rm CMB})\over \var(T_{\rm FOR})}\nonumber \\[0.1cm]
&=& \var (T_{\rm CMB}) (1+ c^2)
\label{eq:add_foregr_var}
\end{eqnarray}
so that the rms contribution to the total rms temperature from the added
foreground is a factor, $c$, relative to the contribution of the cleaned
microwave background map.  (For reference, the constant $\sqrt{\var (T_{\rm
CMB})/\var(T_{\rm FOR})}$ is of order unity in all cases we consider.)  In
\citet{Copi2004}, we have performed tests with $c \leq 0.2$ and found no
significant changes to the results in that paper. Note that the contribution in
power of known foregrounds to the microwave background, after removing the
foregrounds, is estimated to be less than a percent in the V and W bands
\citep{Bennett_foregr}. However this estimate is for the multipole range $2\leq
\ell\leq 100$ while the foreground contamination is most significant at low
multipoles (see Fig.~10 in \citealt{Bennett_foregr}). Therefore it is
reasonable to assume that the contribution of residual foregrounds at large
angular scales is less than about 10\% in power, or $c\la 0.3$.

We have added increasing amounts of foreground, corresponding to $c$ taking
values from zero (no foreground) to $\pm$ 1000 (essentially pure foreground) and
recomputed the multipole vectors and their normals.  Fig.~\ref{fig:add_foregr}
shows the trajectories of the multipole vectors and their normals as increasing
amount of the foreground (V-band synchrotron map produced by WMAP) is added to
the microwave background map.  The solid diamond symbols show the
zero-foreground locations of the multipole vectors while the solid stars 
refer to their pure-foreground locations. Similarly, the empty diamond and 
star symbols refer to the zero- and pure-foreground normal vectors.  On each 
trajectory we label a few positive and negative 
values of the coefficient $c$. 
The top panel of Fig.~\ref{fig:add_foregr} shows $\ell=2$ (two vector and one 
normal) and the bottom panel shows $\ell=3$ (three vectors and three normals). 

In the pure-foreground limit, the quadrupole vectors move near the $z$-axis and
their normal into the Galactic plane as is expected for an almost pure $Y_{20}$
mode (see the discussion in Sec.~\ref{sec:full-sky}). In the same limit, two of
the octopole multipole vectors move close to the Galactic poles and one close
to the Galactic center.  Consequently, all three normals become close to the
Galactic disk at $90^\circ$ from the Galactic center. This is indeed the
signature of the expected $\mathrm{Re}(Y_{31})$ domination in the foreground
octopole. Therefore, both the vectors and their normals clearly migrate from
locations correlated with the ecliptic and other directions discussed in
\citet{Schwarz2004} and in this paper to locations specified by the Galactic
foreground emission. Further, note that appreciable admixture of the foreground
($|c|\ga 0.1$ for the quadrupole and $|c|\ga 0.3$ for the octopole) is
necessary for this migration to become apparent by eye.  This confirms the
argument in Sec.~\ref{sec:full-sky} that the quadrupole foreground is the most
critical one.

The reader might notice that two of the octopole multipole vectors are within
$10\degr$ of the Galactic plane, or, equivalently, one of the octopole normals
is about $10\degr$ from the Galactic poles.  One might ask if that could be a
sign of residual Galactic contamination of the cleaned full-sky maps.  However,
as seen in figure \ref{fig:given23} and table \ref{tab:given23}, given the
observed pattern of quadrupole and octopole area vectors, the alignment of
these area vectors with the NGP is significant at $<90$\% C.L.  (compared to a
99\% C.L. correlation with the ecliptic plane).  Moreveor, since a correlation
of the normal with the Galactic plane would have been even at least as
noteworthy, this 90\% C.L. figure should be reduced to 80\%.

We also observe that when adding foreground, one of the two multipole vectors
close to the Galactic plane moves far away from that plane.  The other
multipole vector close to the Galactic centre does not move very far (about
$10\degr$). But, a $10\degr$ alignment of one of the cleaned-map multipole
vectors with one of the foreground multipole vectors is not statistically
significant by any statistical test applied in this work.  Even less so when we
consider that we would remarked as well on a similar alignment of a cleaned-map
area vector with a foreground area vector.

Finally, we do not expect more than one foreground multipole vector to lie near
the Galactic plane since the dominant fairground mode of the octopole is indeed
$\mathrm{Re}(Y_{31})$. The second biggest foreground mode is
$\mathrm{Re}(Y_{33})$, which has its three multipole vectors in the Galactic
plane at $l = 30\degr$, $90\degr$ and $150\degr$. But, the cleaned full-sky map
does not resemble that pattern either. We are not able to identify a Galactic
contamination of the cleaned full-sky maps and thus see no evidence to question
more significant ecliptic correlations found in the previous sections.

We have checked that the results are qualitatively unchanged if we use the
W-band synchrotron map, or the V-band free-free and dust foreground maps
instead of the V-band synchrotron map.  We have done some further testing,
recomputing the $S$ statistics applied to different alignments but now with the
foreground incrementally added to both the microwave background and Monte Carlo
maps. We found that the results are consistent with those inferred from
Fig.~\ref{fig:add_foregr} and indicate that large admixture of the known
Galactic foreground would not cause the alignments found in
\citet{Schwarz2004}.

Results of this investigation are therefore in agreement with the intuitive
expectation: {\it Galactic foregrounds lead to Galactic, and neither ecliptic
nor dipole, correlations of the quadrupole and octopole}. In fact, it is 
difficult to see how any known foreground that has most of its power in the 
Galactic plane can lead to the solar system correlations that we find. 

\section{Effects of cut skies}
\label{sec:cutsky}

As mentioned in \citet{Copi2004}, Galaxy cuts of a few degrees or larger
introduce significant uncertainties to the reconstructed full-sky multipole
vectors.  This is precisely why we used the cleaned full-sky maps --- sky cuts
of 20 degrees or so would simply lead to large uncertainties in our
statistical tests. Nevertheless, we would like to look at the issue of sky cuts
in more detail; in particular, we would like to explore how correlation
significance varies as we introduce a sky cut.

\begin{figure*}
  \includegraphics[width=1.7in,angle=-90]{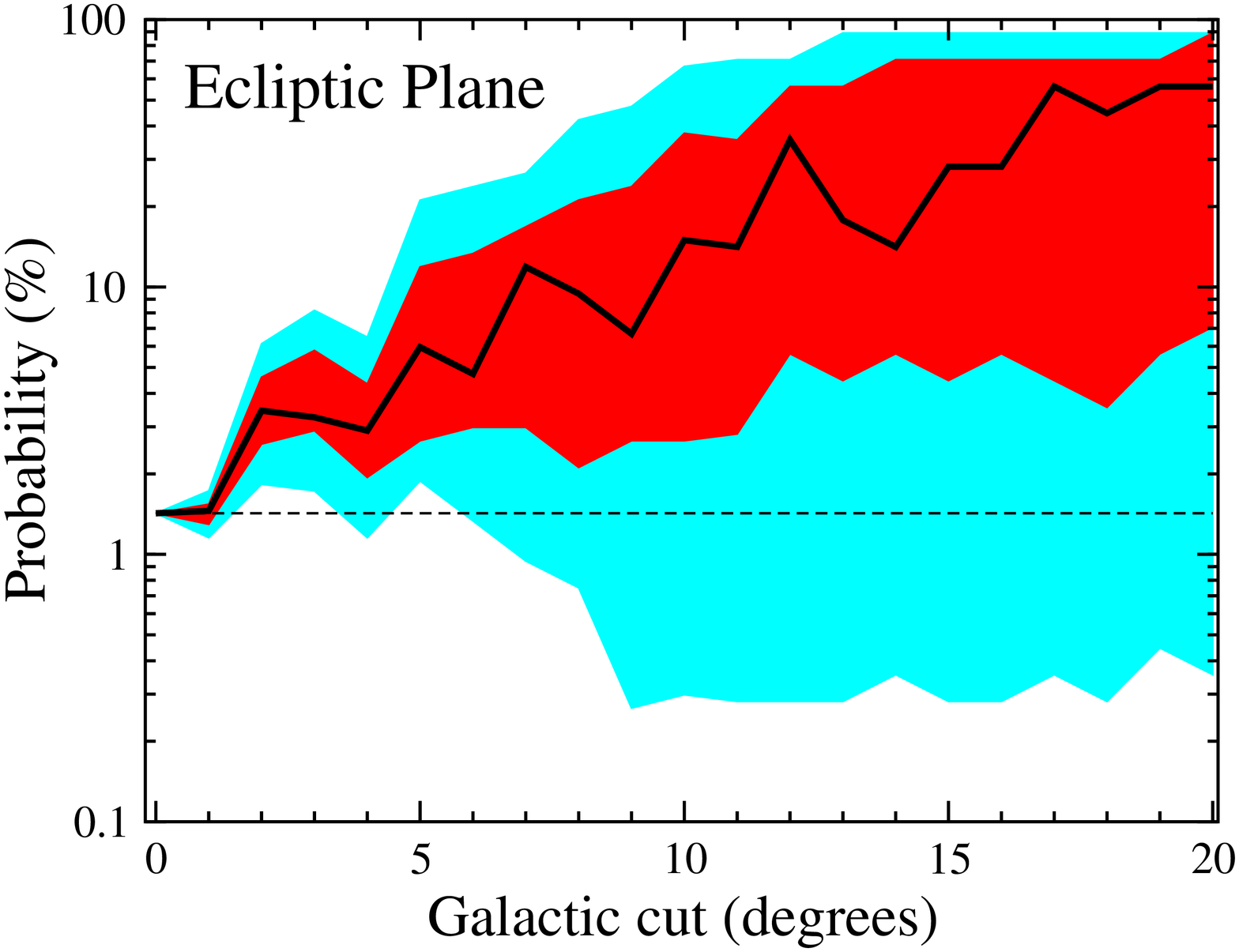}\hspace{-0.2cm}
  \includegraphics[width=1.7in,angle=-90]{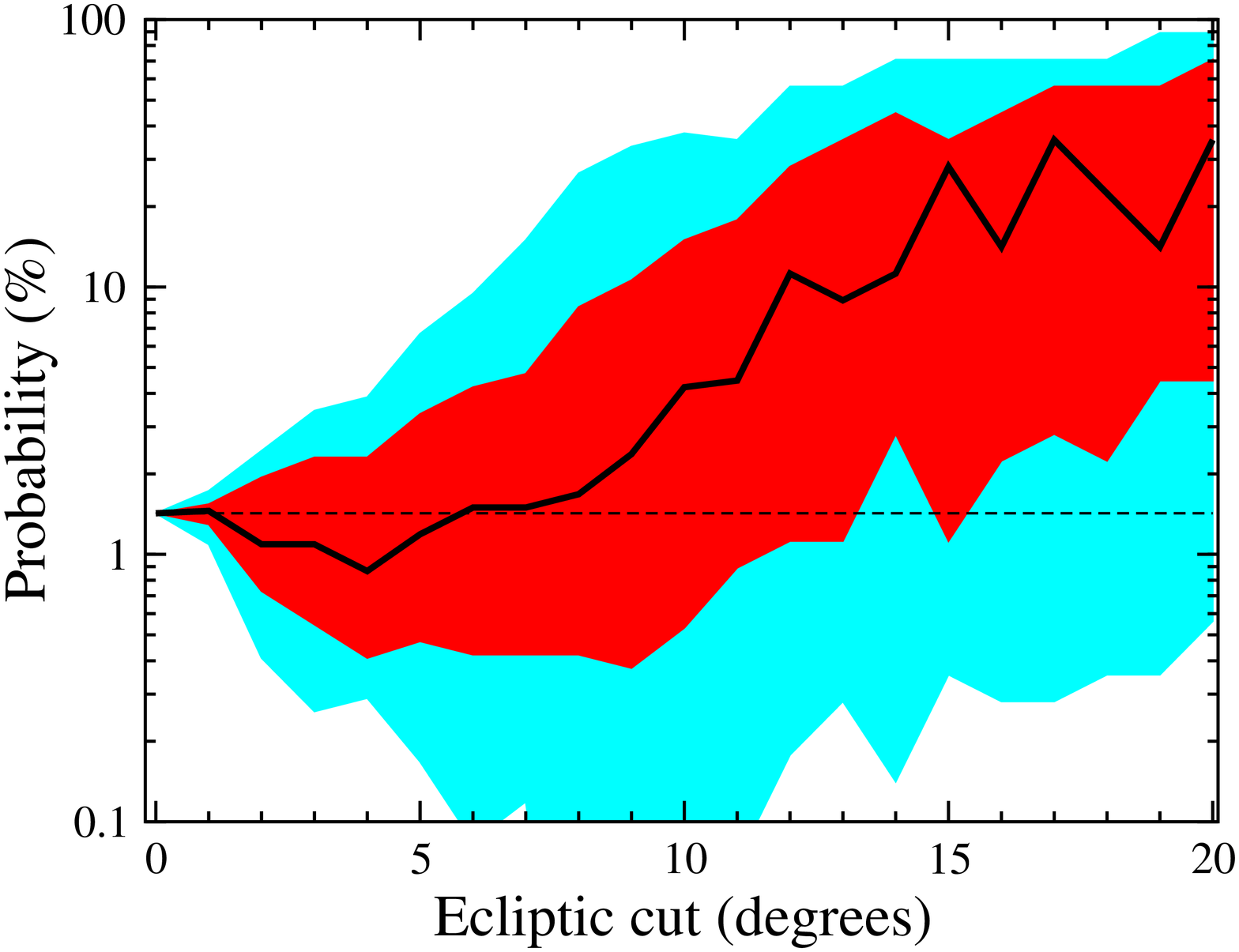}\hspace{-0.2cm}
  \includegraphics[width=1.7in,angle=-90]{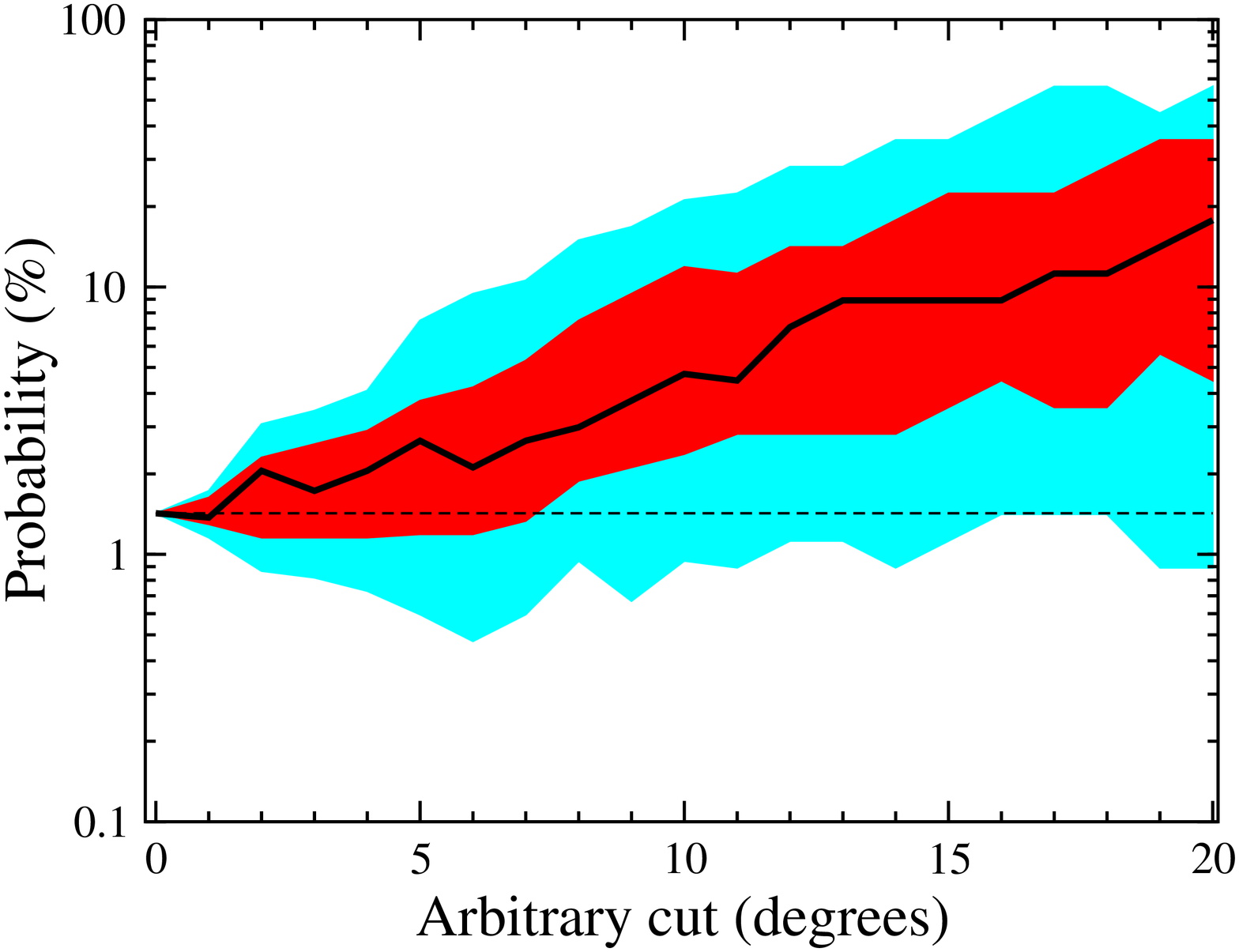}\\
  \includegraphics[width=1.7in,angle=-90]{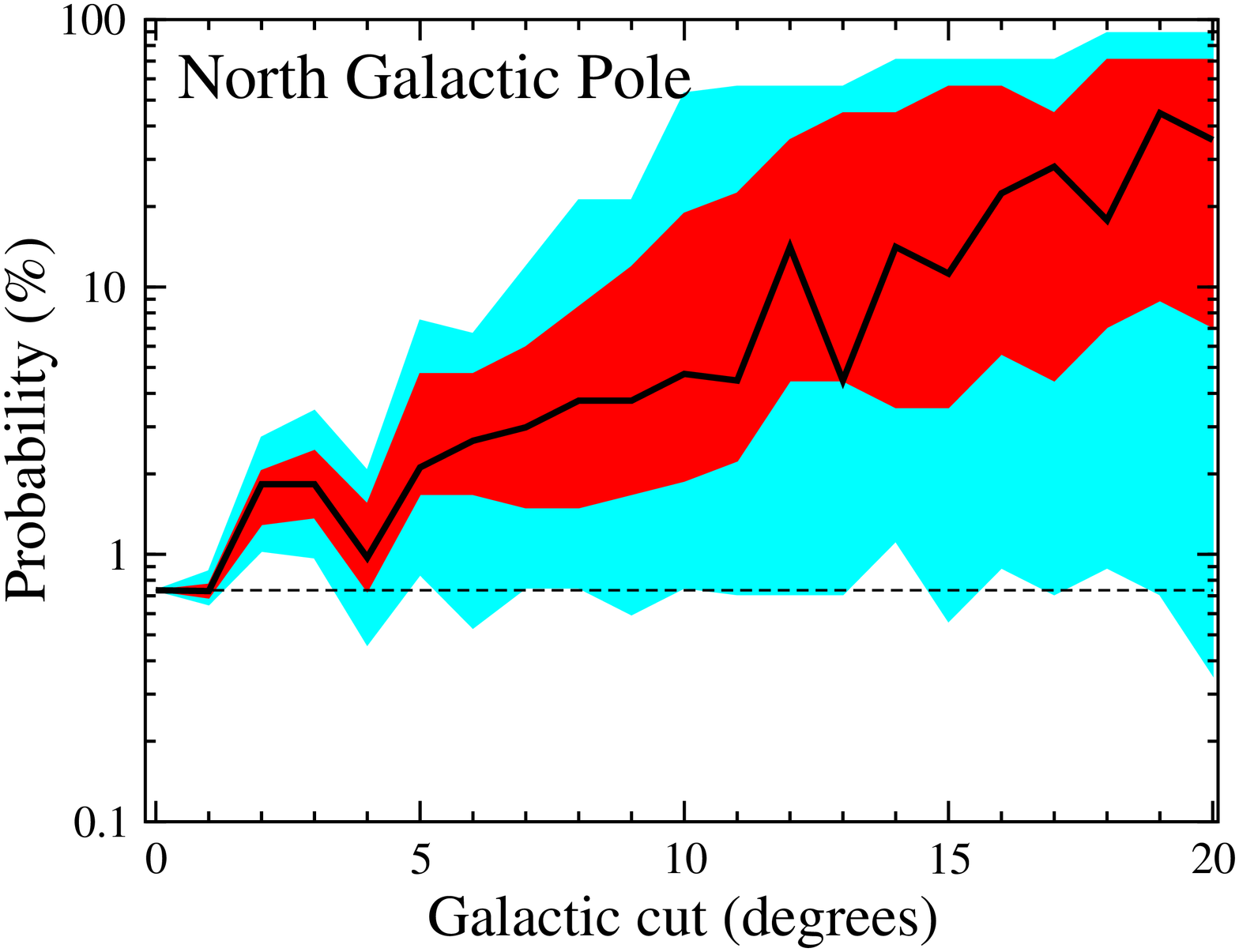}\hspace{-0.2cm}
  \includegraphics[width=1.7in,angle=-90]{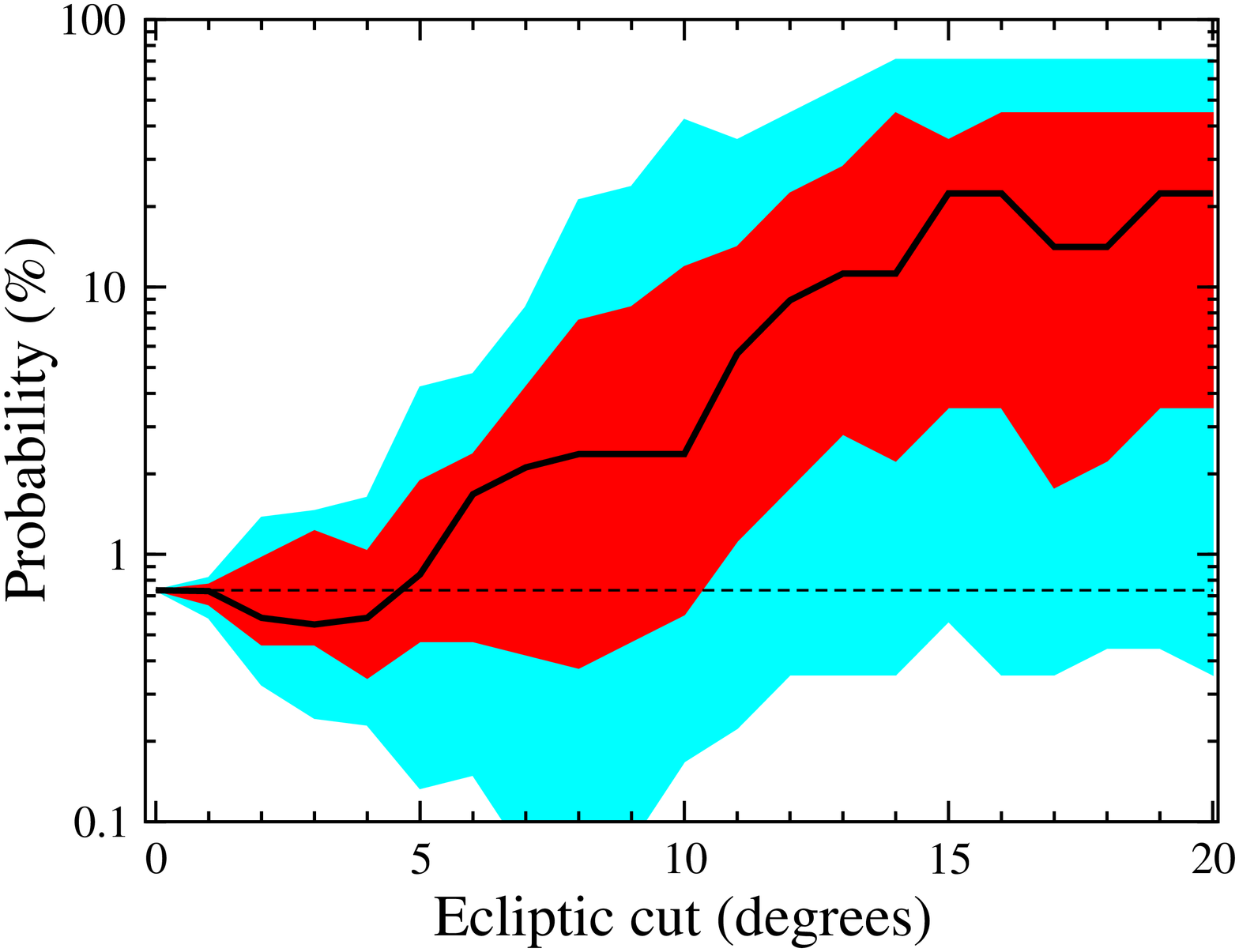}\hspace{-0.2cm}
  \includegraphics[width=1.7in,angle=-90]{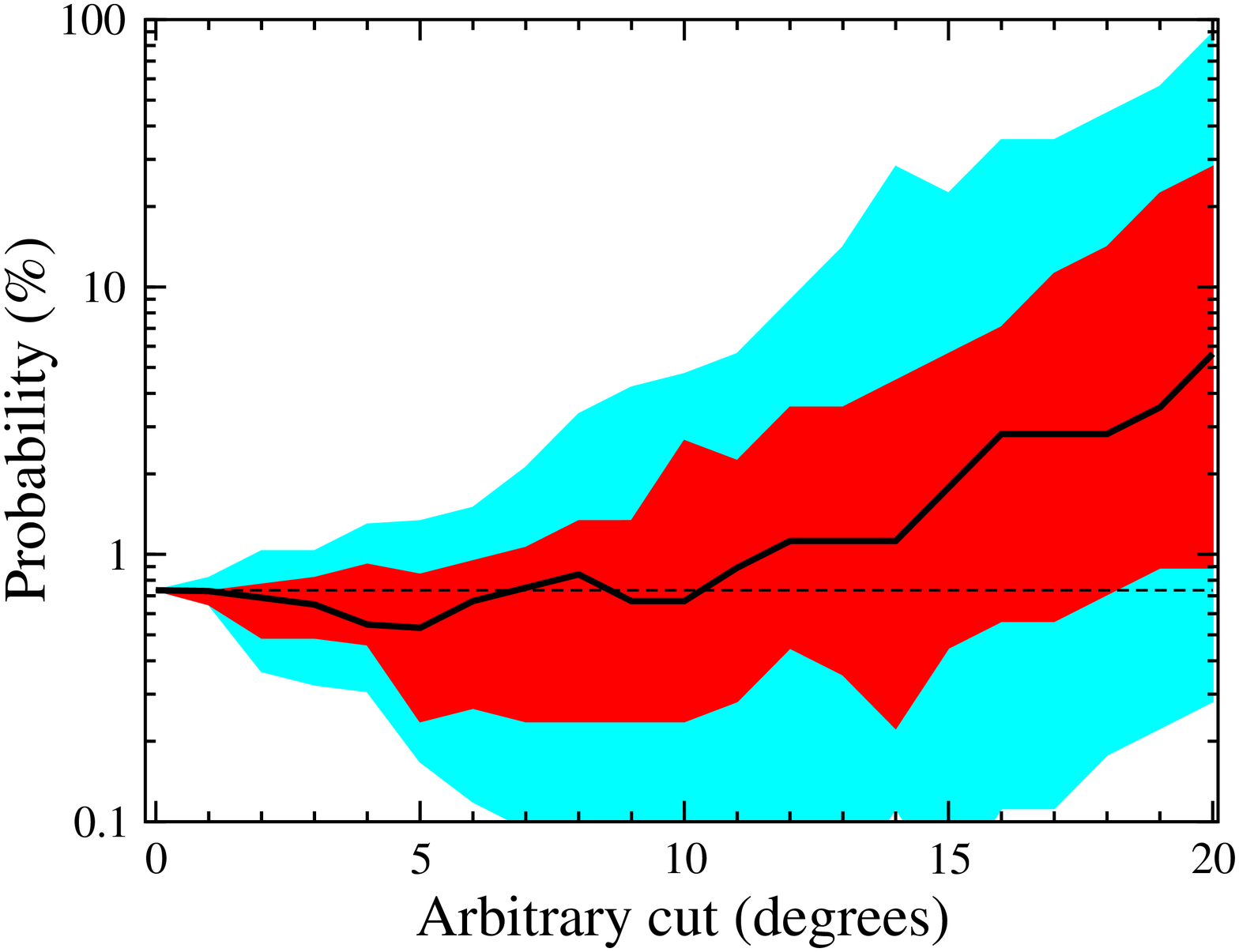}\\
  \includegraphics[width=1.7in,angle=-90]{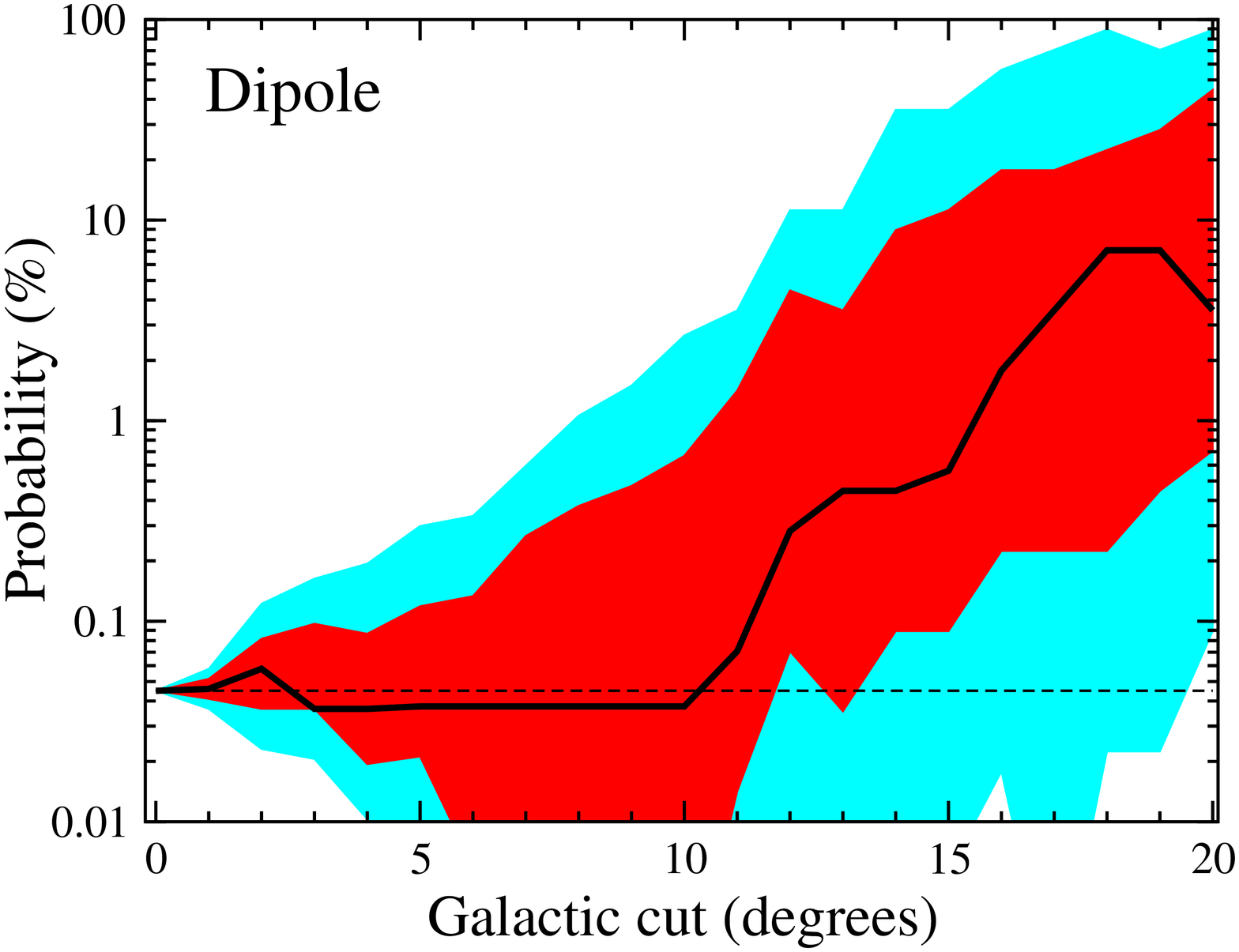}\hspace{-0.2cm}
  \includegraphics[width=1.7in,angle=-90]{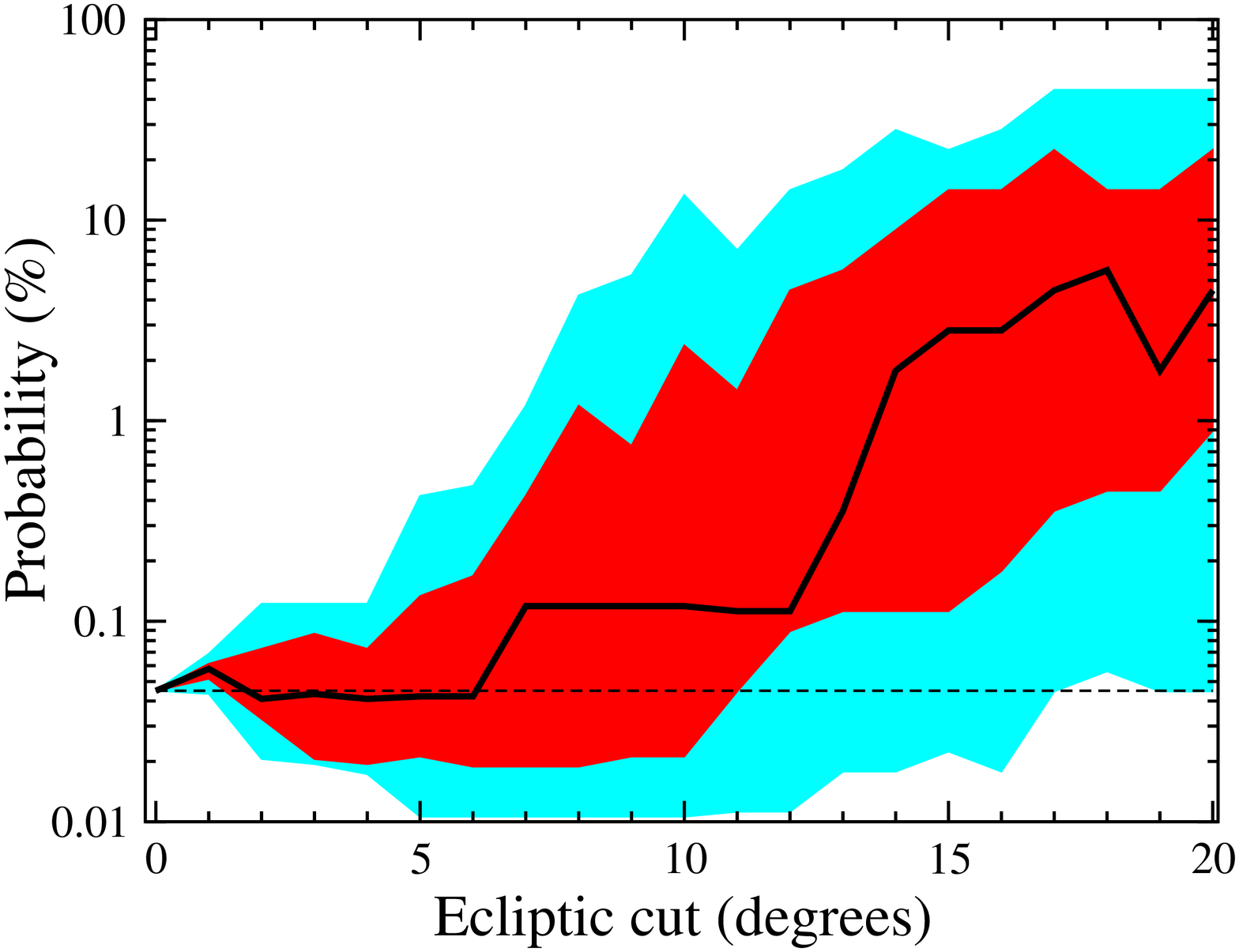}\hspace{-0.2cm}
  \includegraphics[width=1.7in,angle=-90]{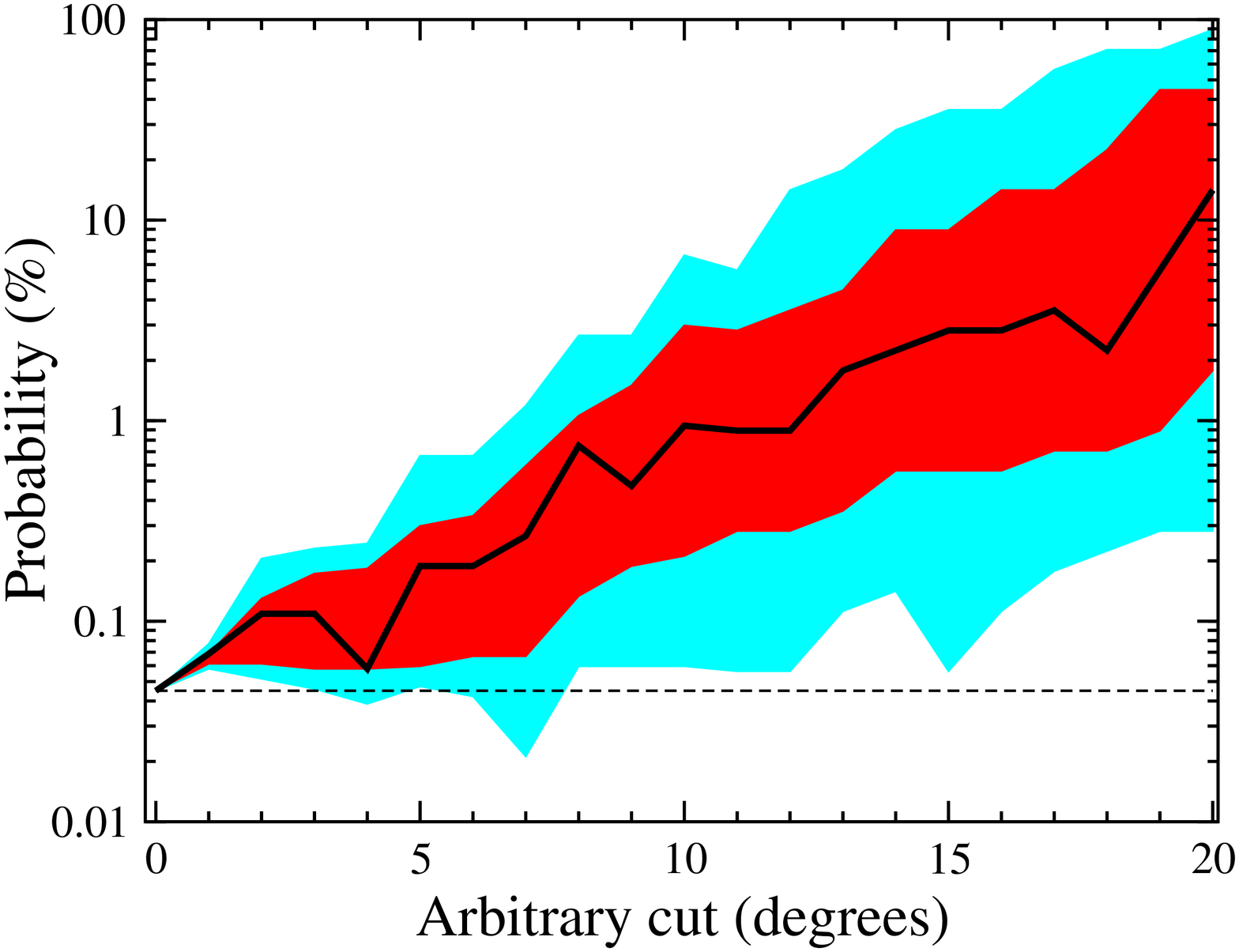}\\
  \includegraphics[width=1.7in,angle=-90]{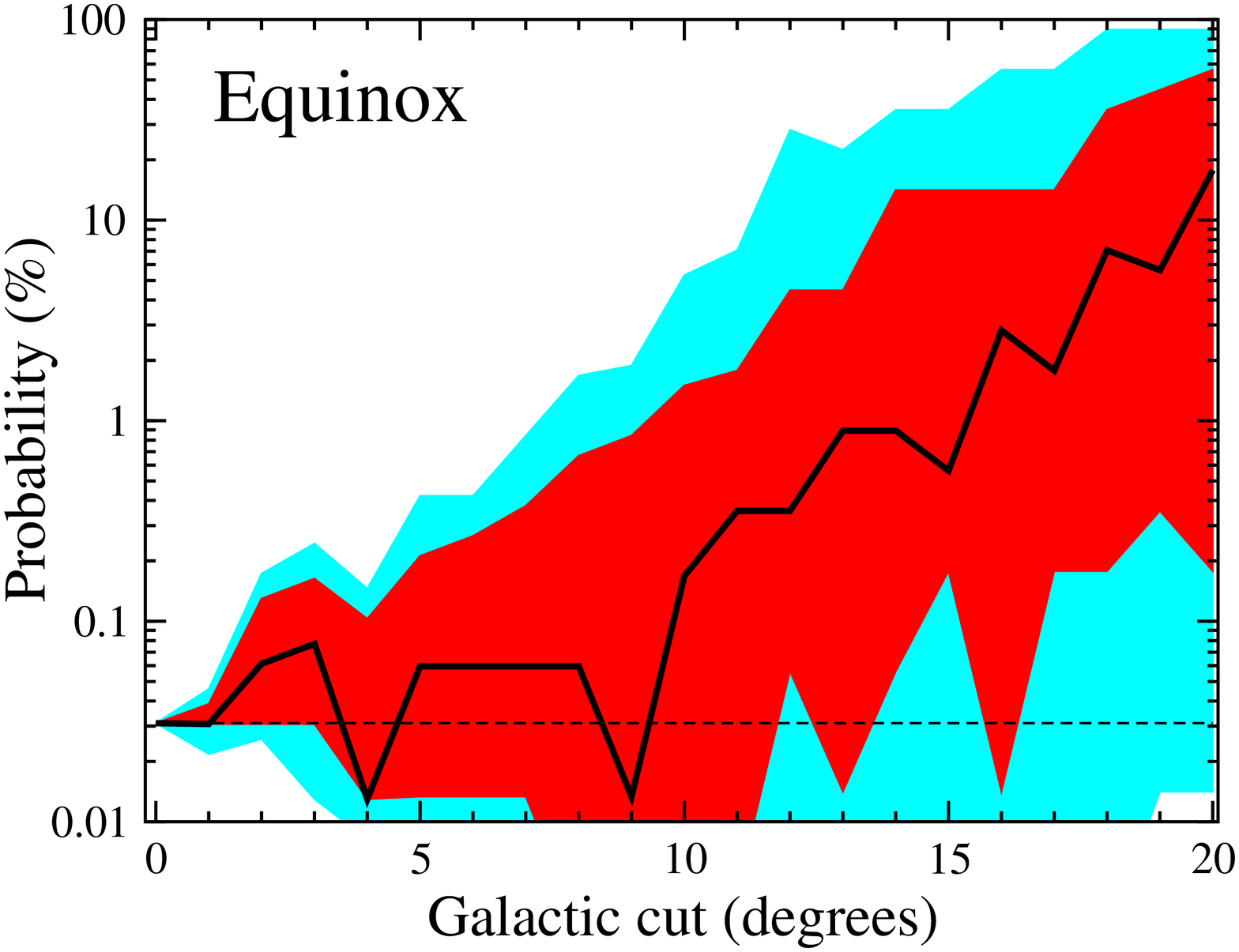}\hspace{-0.2cm}
  \includegraphics[width=1.7in,angle=-90]{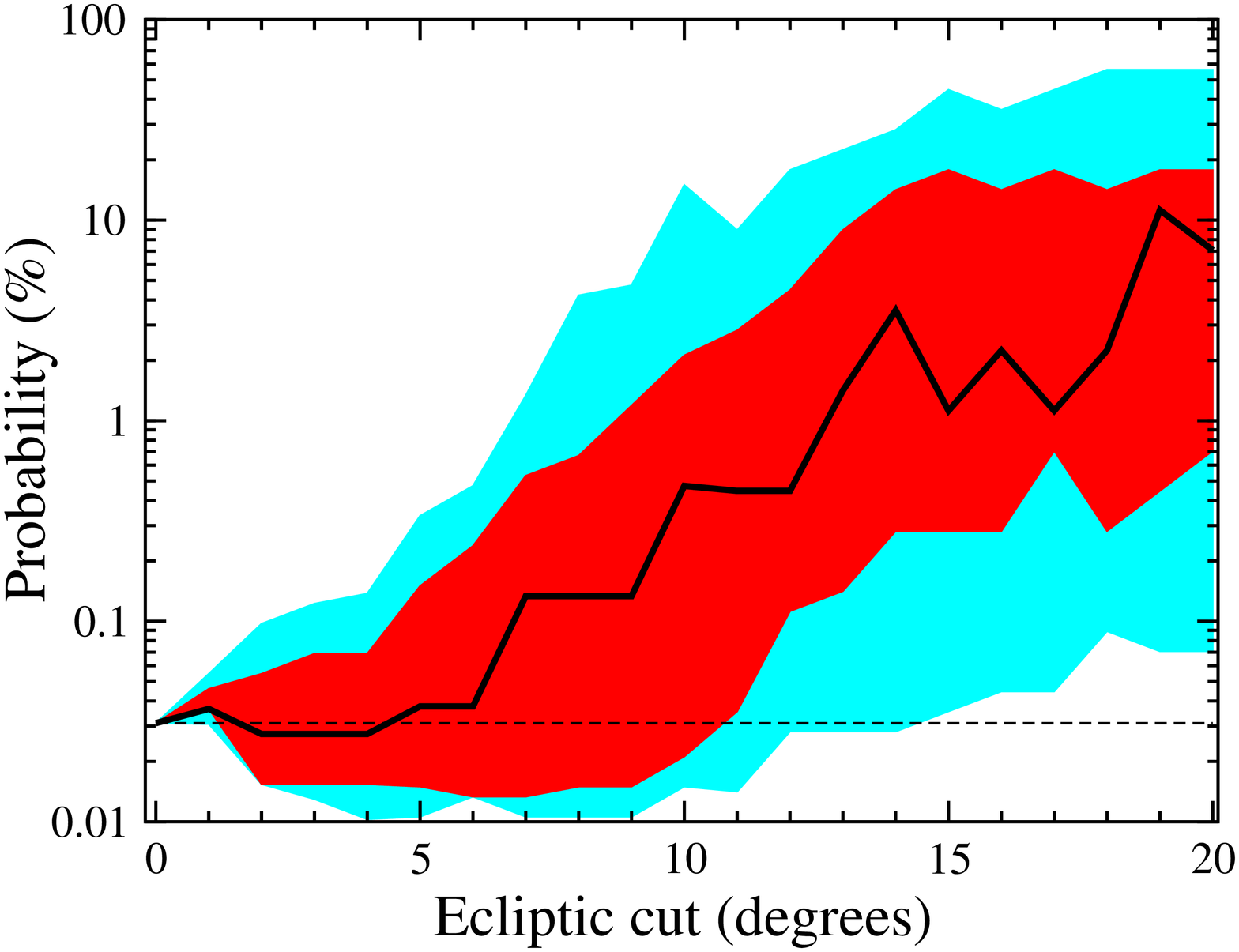}\hspace{-0.2cm}
  \includegraphics[width=1.7in,angle=-90]{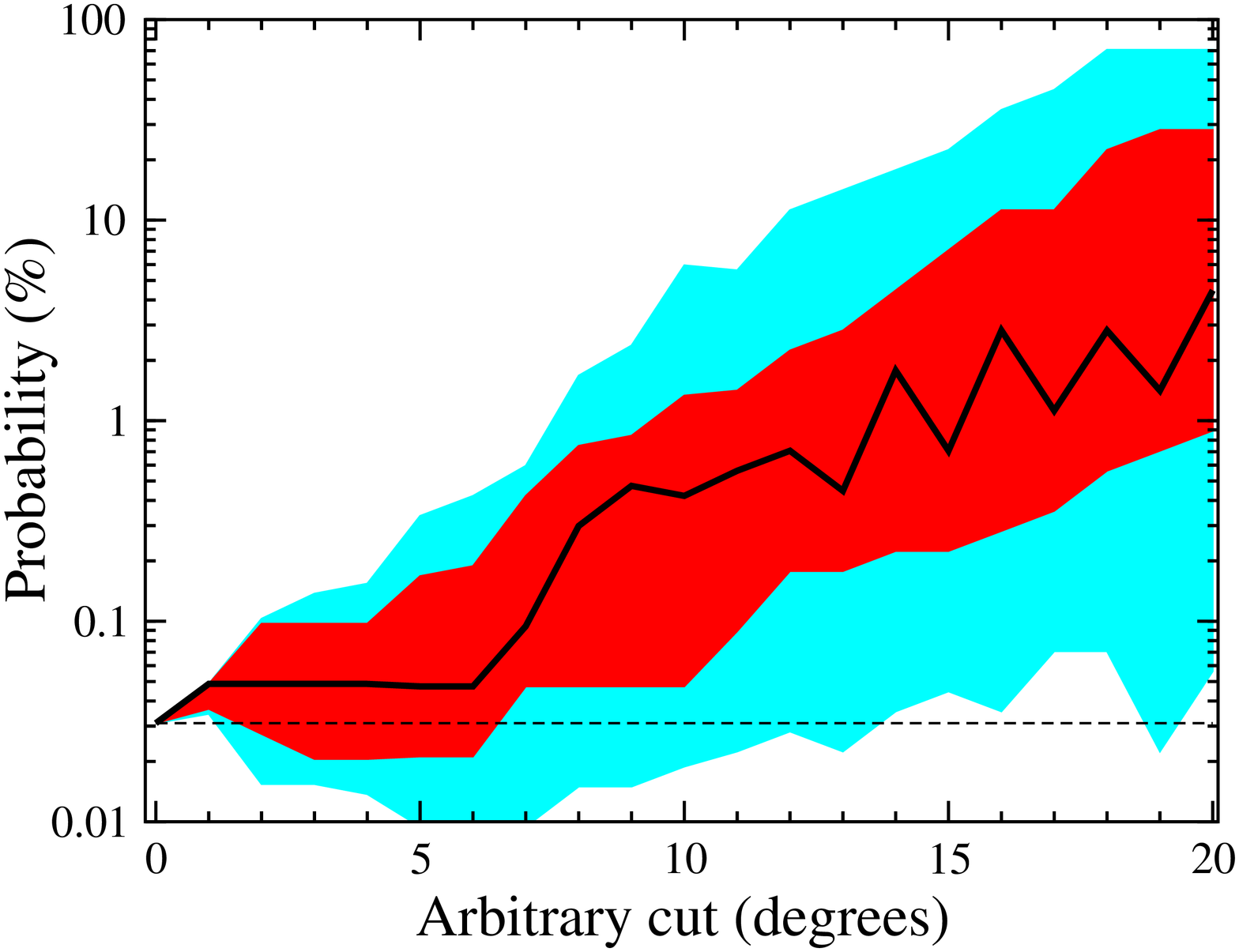}
  \caption{Quadrupole-octopole probabilities for the TOH DQ-corrected map for
    an increasingly larger isolatitude cut of $\pm$(degrees shown), performed
    symmetrically around the Galactic plane (left column), ecliptic plane
    (central column) or an arbitrarily chosen plane (right column). We consider
    the $S$ statistic probabilities applied to the ecliptic plane,
    north Galactic pole, dipole and the equinoxes (first to fourth row
    respectively). The solid line is the mean value, while the dark and light
    regions represent 68\% C.L.\ and 95\% C.L.\ regions, respectively, from 1000
    realizations of reconstructed $a_{\ell m}$ coefficients that take into
    account the noise in the reconstruction process. The dashed line denotes
    the probability obtained from the full-sky map, corresponding to the case
    of zero cut.  }
  \label{fig:cutsky}
\end{figure*}

We are interested in finding the full-sky (true) $\alm$ which are derived
from the full-sky temperature distribution and denoted by
$\alm^t$.  In general, we cannot view the full sky as the Galaxy
obscures our field of view and must be cut out.  There are well know
techniques for relating the cut sky decomposition, $\alm^c$, to
the true sky decomposition \citep{cutsky:wandelt,cutsky:mortlock}. We briefly
discuss the key facts here.  The decompositions are related by
\begin{equation}
  \alm^c = \sum_{\ell' m'} W_{\ell\ell',mm'} a_{\ell' m'}^t
\end{equation}
where
\begin{equation}
  W_{\ell\ell',mm'} \equiv \int_{S_{\rm cut}} Y_{\ell' m'}^* (\Omega)
  Y_{\ell m}
  (\Omega) \dderiv\Omega
\end{equation}
and $S_{\rm cut}$ is the cut sphere.  There are fast, stable recursion
relations for calculating these $W_{\ell\ell',mm'}$
\citep{cutsky:wandelt,cutsky:mortlock}. In this work, we restrict ourselves
to longitudinal cuts symmetric across the Galactic ($xy$) plane.  In this
case, $m=m'$ and $\mat{W}$ is a symmetric matrix.  For notational convenience
we will drop the $m$ index and keep in mind that the subsequent equations
hold independently for each $m$.  Thus we write
\begin{equation}
  a_\ell^c = \sum_{\ell'} W_{\ell\ell'} a_{\ell'}^t.
\end{equation}

Since information is lost in the cut, $\mat{W}$ is not an invertible matrix.
We can, however, replace $\mat{W}$ with an invertible matrix $\mat{\tilde
  W}$ constructed from $\mat{W}$ by removing the rows and columns with
small eigenvalues.  That is, $\mat{\tilde W} \equiv \mat{\tilde V}
\mat{\tilde \lambda} \mat{\tilde V}^T$ where $\mat{\tilde\lambda}$ is the
diagonal matrix of eigenvalues such that $\tilde\lambda_j = \lambda_j$ if
$\lambda_j > \lambda_{\rm threshold}$ and $\tilde\lambda_j =
\tilde\lambda_j^{-1} = 0$ otherwise.  A threshold of $\lambda_{\rm
  threshold} < 0.1$ is typically sufficient and is used in this analysis.
An estimate, $\tilde a_\ell^t$  for the true decomposition is
\begin{equation}
  \tilde a_\ell^t = \sum_{\ell'} \tilde W_{\ell\ell'}^{-1} a_{\ell'}^c.
\end{equation}
We can prevent leakage of power from non-cosmological monopole and dipole
modes by projecting out these modes using a partial Householder
transformation (see Appendix C of \citealt{cutsky:mortlock} for details).

There is an error in this approximation which is evident from the fact that
\begin{equation}
  \tilde{\vec a}^t = \mat{\tilde W}^{-1} \vec a^c = \mat{\tilde W}^{-1}
  \mat{W} \vec a^t 
\end{equation}
and $\mat{\tilde W}^{-1} \mat{W} \ne \mat{I}$ due to the loss of
information in the cut.  Our error in the approximation is found to be
\begin{eqnarray}
  \left\langle \left|\vec a^t-\tilde{\vec a}^t\right|^2\right\rangle &=&
  \left\langle \left(\vec a^t-\tilde{\vec a}^t\right)^* \left(\vec
      a^t-\tilde{\vec a}^t\right)^T \right\rangle
  \\
  &=& \left(\mat{I}-\mat{\tilde W}^{-1}\mat{W}\right) \left\langle (\vec
    a^t)^* (\vec a^t)^T
  \right\rangle
  \left(\mat{I}-\mat{\tilde W}^{-1}\mat{W}\right)^T.
  \nonumber
\end{eqnarray}
Here $\langle (a^t_\ell)^* a^t_\ell\rangle = {\cal C}_\ell$ and this error
can be readily calculated.

Fig.~\ref{fig:cutsky} shows the quadrupole-octopole correlation probabilities
with the sky cut between zero and $\pm$20 degrees performed along the Galactic
plane (left column), ecliptic plane (central column) or an arbitrarily chosen
plane\footnote{The ``arbitrary'' plane is chosen as one obtained by rotating
the map in Galactic coordinates by $+45$ degrees around the $z$-axis and then
by $-60$ degrees around the new $x$-axis. The resulting map has neither the
ecliptic nor the Galactic plane located along the equator.} (right column). We
consider the $S$ statistic probabilities applied to the ecliptic plane,
north Galactic pole, dipole and the equinoxes (first to fourth row
respectively). The solid line is the mean value, while the dark and light
regions represent 68\% C.L.\ and 95\% C.L.\ regions, respectively, from 1000
realizations of reconstructed $a_{\ell m}$ coefficients that take into account
the noise in the reconstruction process. While the increasing cut clearly
increases error in the vector reconstruction and therefore uncertainty in the
final probability, it is clear that in essentially all cases the probabilities
remain consistent with the full-sky values at 95\% C.L., and in most cases at the
68\% C.L.\ for cuts up to 10 degrees.

This figure clearly shows that sky cuts of a few degrees or larger introduce
significant uncertainty in the extracted multipole vectors and their normals,
leading to increased error in all alignment tests.  Nevertheless, the cut-sky
alignments are consistent with their full-sky values even for relatively large
cuts. Note that the shift of the mean value of the alignments (black curves in
the panels of Fig.~\ref{fig:cutsky}) to less significant values, as the cut is
increased, is entirely expected: an unlikely event, in the presence of noise in
the data, becomes less unlikely because any perturbation will shift the
multipole and area vectors away from their aligned locations. While the results
of this exercise are in good agreement with those found by \citet{SS2004} and
\citet{Bielewicz2005}, unlike these authors, we emphasize that the cut sky is
always expected to lead to shift in the alignment values and to increased
errors (see again Fig.~\ref{fig:cutsky}).

\section{Comparisons with COBE} \label{sec:COBE}

Since the alignments we are studying are on very large scales (i.e. quadrupole
and octopole scales), it is natural to ask whether they can be seen in COBE-DMR
maps \citep{DMR4_maps}.  COBE angular resolution is about $7\degr$, which is more than
sufficient for this test.  However, full-sky maps produced by the COBE team
are very noisy (G. Hinshaw, private communication) while, as discussed in
Sec.~\ref{sec:cutsky}, using the cut sky maps produces too much uncertainty in the vectors
for Galaxy cuts larger than a few degrees.

Fortunately, Markov chain Monte Carlo (MCMC) Gibbs sampler realizations of the
full-sky COBE maps have been produced.  Using their Global Estimation Method,
\citet{Wandelt:GEM} generate 10,000 realizations of the COBE sky consistent
with DMR measurements and expected foregrounds.  Following their own
conservative approach, we drop the first 1000 maps which might be affected by
the burn-in of the MCMC, and choose every 200th map from the remaining 9000.
The resulting 45 maps are essentially uncorrelated and their analysis is
computationally undemanding. We then compute the $S$ statistic for each COBE
map and rank-order it relative to 100,000 Monte Carlo realizations of the
Gaussian random, statistically isotropic sky. The COBE derived values for the
statistic $S$ is represented by histograms, while a WMAP full-sky map is
represented by a single value.  We then ask whether the two are consistent.
Note that {\it we have corrected all COBE maps for the DQ} using the procedure
described in Sec.~\ref{sec:DQ}.

\begin{figure*}
  \includegraphics[width=1.7in,angle=-90]{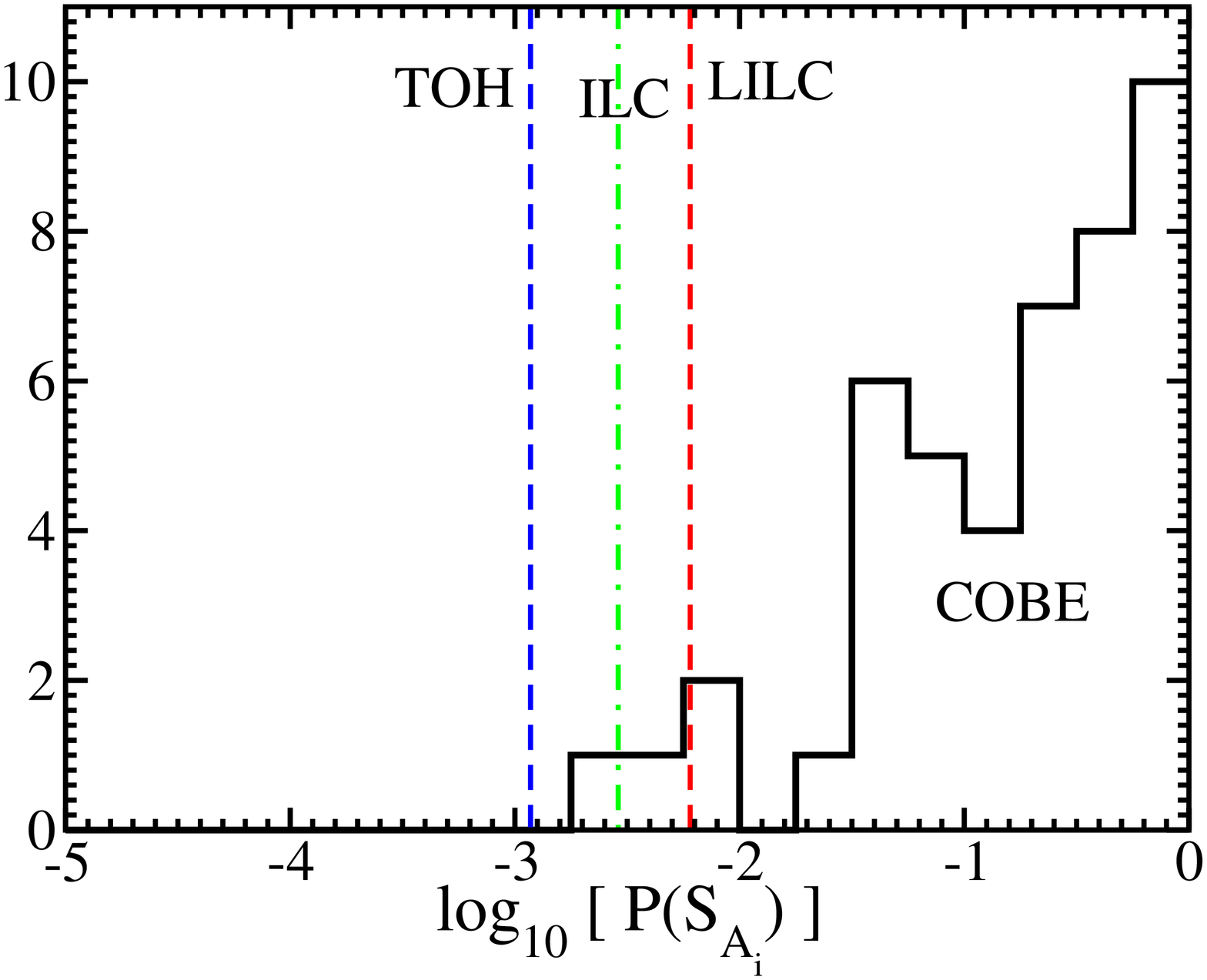}\hspace{-0.2cm}
  \includegraphics[width=1.7in,angle=-90]{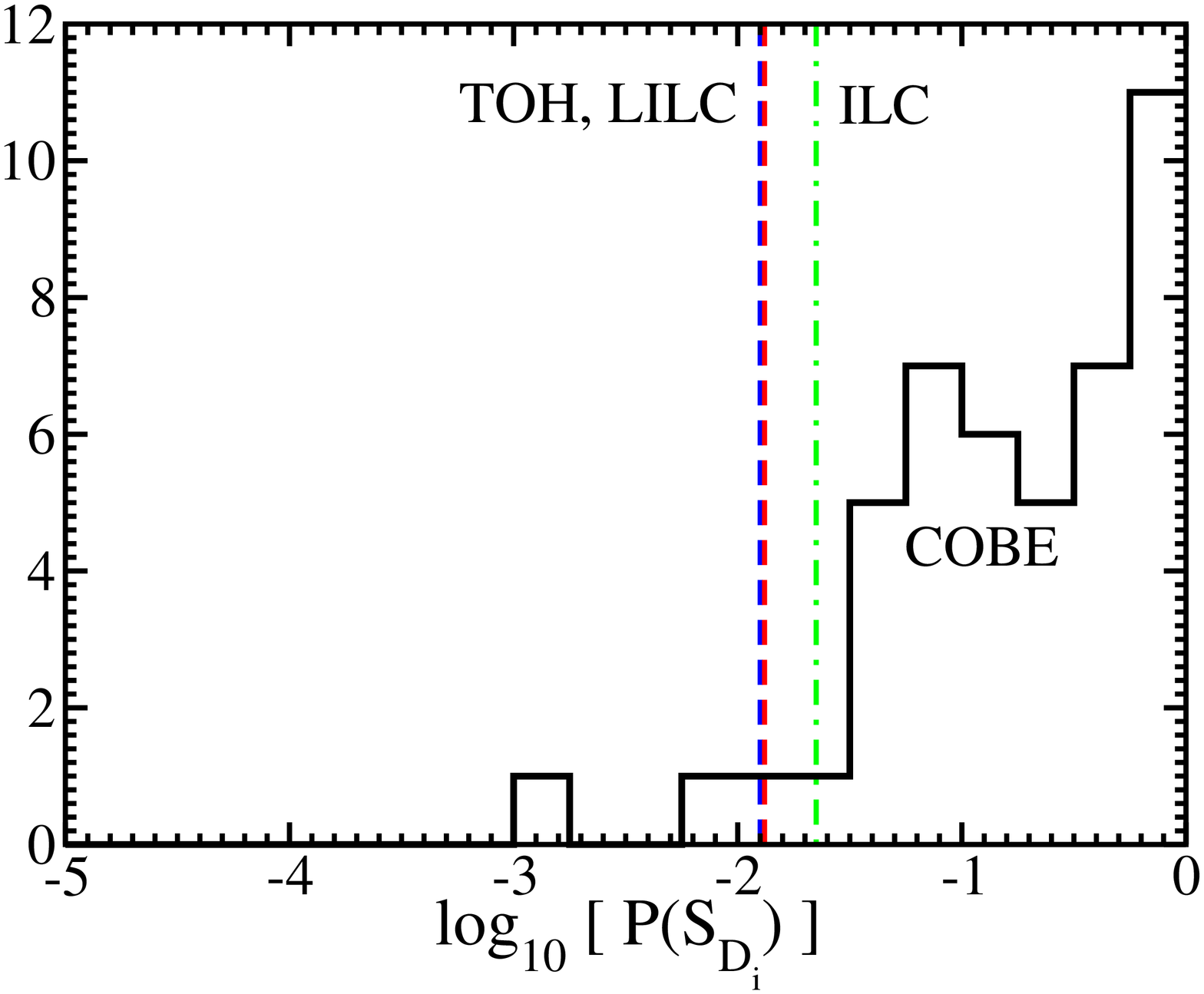}\hspace{-0.2cm}
  \includegraphics[width=1.7in,angle=-90]{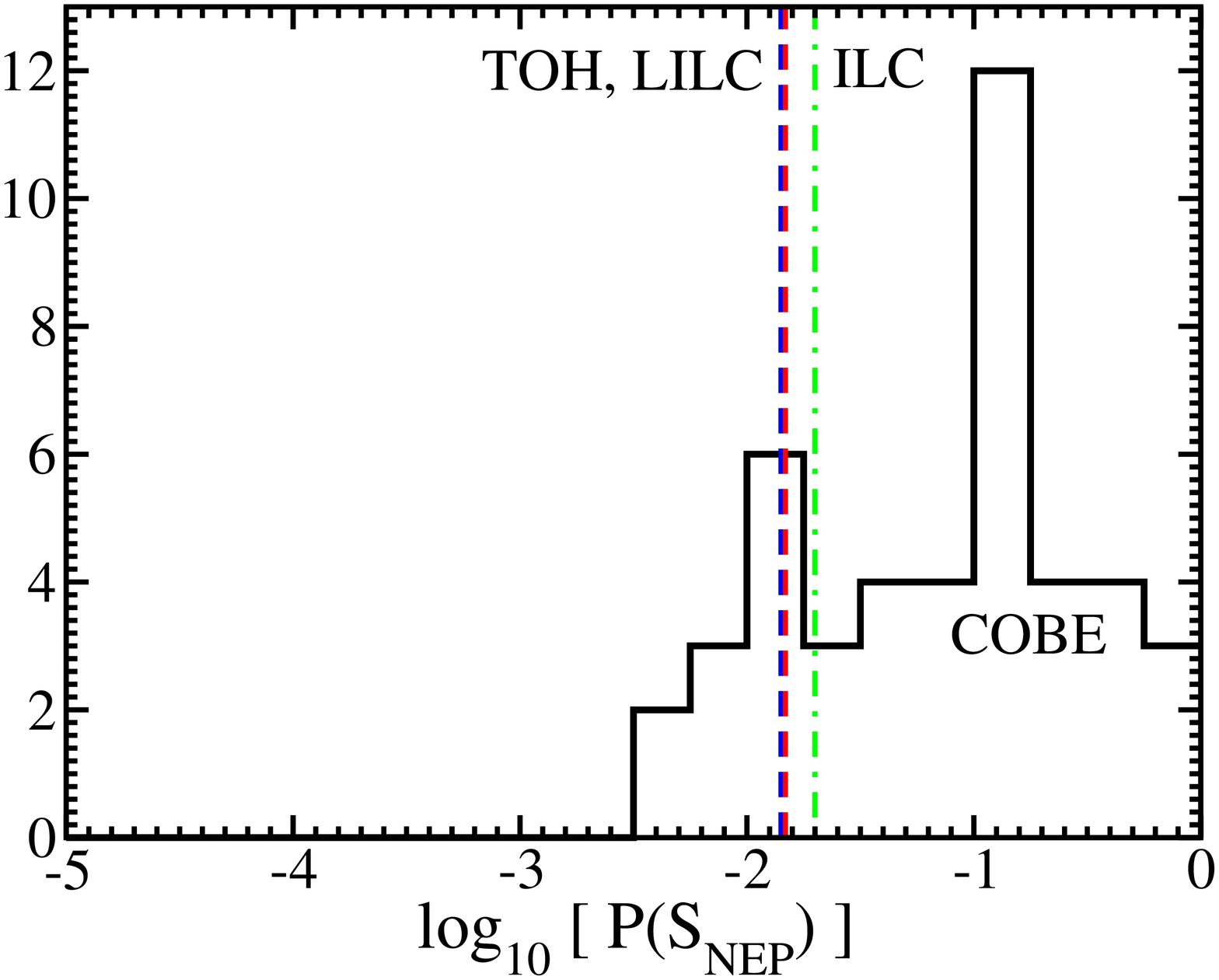}\\
  \includegraphics[width=1.7in,angle=-90]{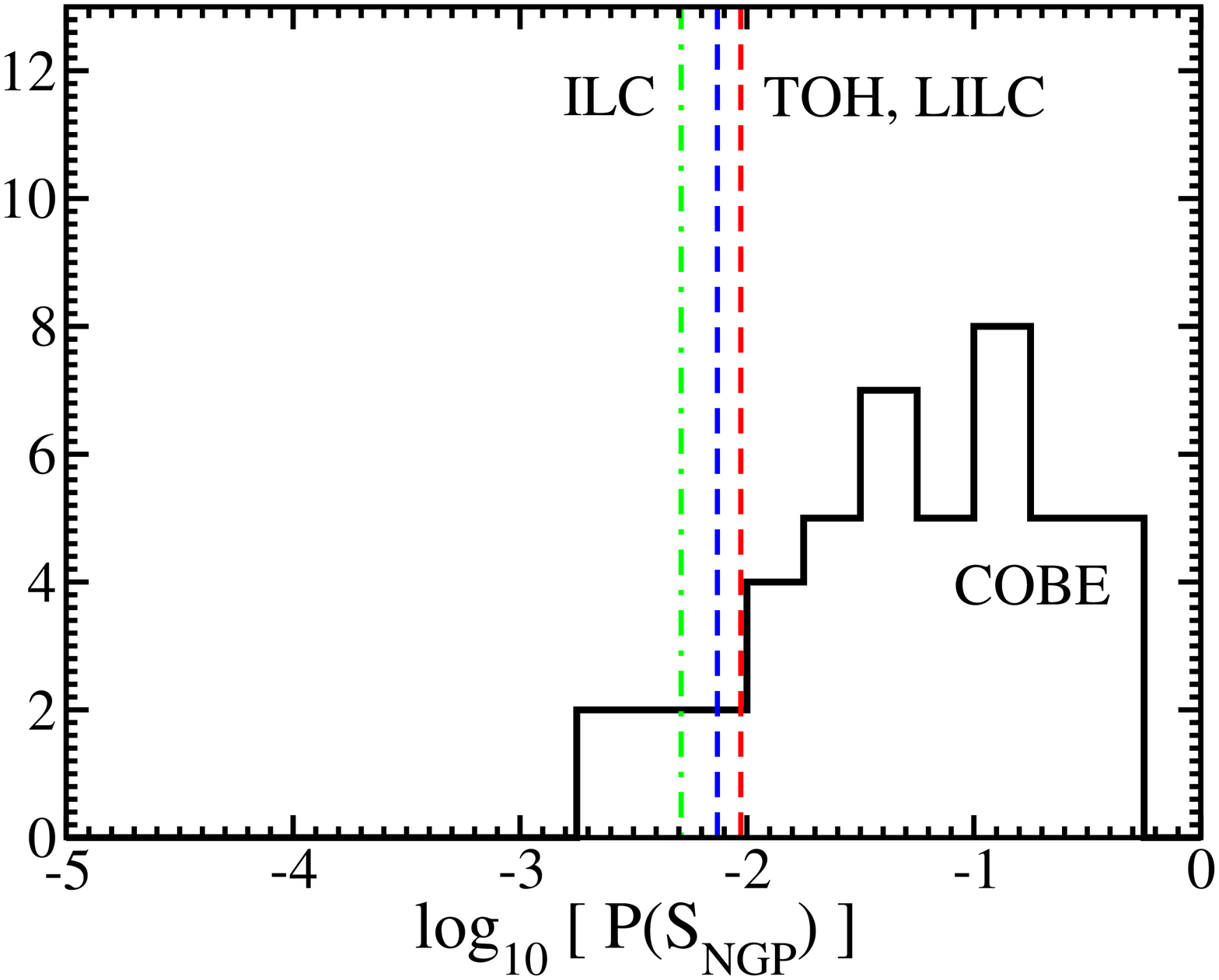}\hspace{-0.2cm}
  \includegraphics[width=1.7in,angle=-90]{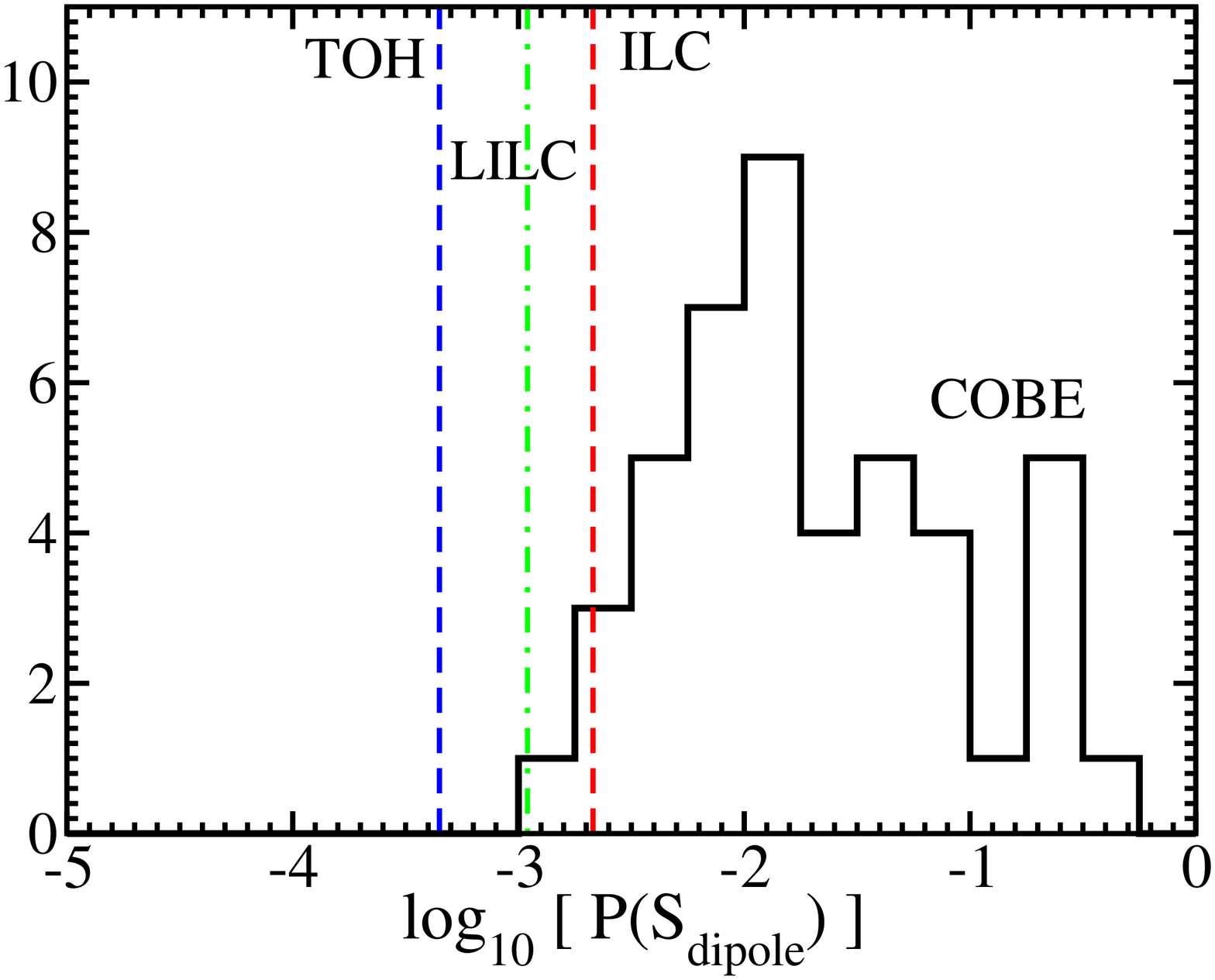}\hspace{-0.2cm}
  \includegraphics[width=1.7in,angle=-90]{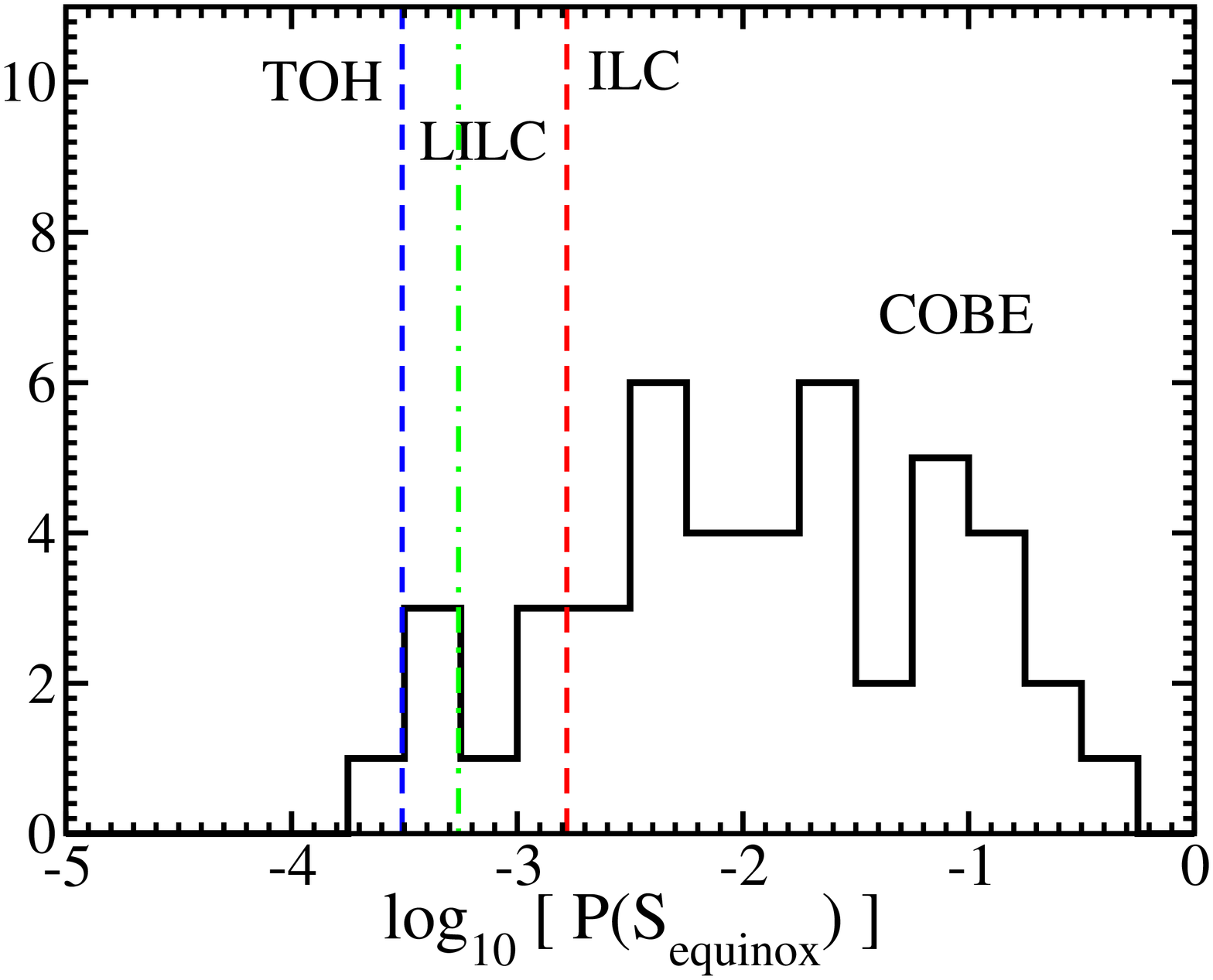}
  \caption{The oriented area statistics from full-sky WMAP maps
    compared to those from Markov chain Monte Carlo realizations of the
    COBE full sky based on COBE-DMR cut-sky data.  We consider
    the $S$ statistic applied to the dot-products of quadrupole-octopole 
    area vectors ($A_i$) and the normals ($D_i$), as well as   
    the sum (i.e.\ the $S$-statistic) of the four dot-products of the 
    area vectors, one quadrupole and three octopole, with the special 
    directions or planes --- the ecliptic plane, NGP, dipole and equinox. 
    The WMAP maps considered for the TOH, ILC and LILC as
    shown by the vertical lines. The histograms are from a total of 45 MCMC
    COBE maps from \citet{Wandelt:GEM} (their maps 1000, 
    1200, \ldots, 9800) which are sufficiently separated in the chain so as to be
    essentially uncorrelated.}
  \label{fig:COBE}
\end{figure*}

The results are shown in Fig.~\ref{fig:COBE}. The $x$-axis in each panel shows
the logarithm of the probability of $S$ (the probability is simply its rank
relative to 100,000 Monte Carlo realizations of the Gaussian random,
statistically isotropic sky). The vertical lines show values for the TOH, ILC
and LILC maps. The histogram shows values for the 45 COBE maps produced using
the MCMC Gibbs sampler. Obviously, there is significant variation in the COBE
statistics which traces to the fact that the MCMC maps are based on incomplete
sky information. Nevertheless, we see that the statistically significant
alignments found with WMAP are in most cases consistent to the results from
COBE. For example, 11 out of 45 COBE maps show the alignment with the north
ecliptic pole that is equally or less likely than that in the WMAP maps, while
4 to 8 COBE maps (depending on which WMAP map is considered) show equal or less
likely alignment with the equinoxes. Note too that the alignment with the
supergalactic plane differs significantly in the three WMAP maps.
Consequently, the comparison of MCMC-derived COBE maps with WMAP cleaned
full-sky maps shows that {\it COBE data are consistent with WMAP in regards to
all alignments found in \citet{Schwarz2004}}.

\section{Angular Power spectrum --- ecliptic plane versus  ecliptic poles}
\label{sec:plane_poles}

There are at least three points in the binned WMAP angular TT power spectrum
that deviate from the predictions of the best fit $\Lambda$CDM model at
comparable or greater statistical significance to the power deviation of the
quadrupole.  These are clearly seen in Fig.~12 of \citet{Bennett2003} in the
bins $\ell=20\mbox{--}24$, $\ell=37\mbox{--}44$ and $\ell=201\mbox{--}220$,
which are visibly low, high and low respectively.  These deviations are
approximately 2, 2.5 and 1.5 times the estimated error in the average $C_\ell$
in each of those bins.  In each case, this is largely cosmic variance
dominated, with only a small fraction of the error being due to statistical
error.  Nevertheless, it should be appreciated that the statistical
significance of these deviations may decline in the second or third year WMAP
maps.

Various explanations have been offered for these deviations in the angular
power spectrum as arising from fundamental physics (e.g.\ \citet{Gordon2004,
Enqvist2004} and references therein).  However, it is instructive to examine
the angular power spectrum computed separately using data from the ecliptic
plane versus data from the ecliptic pole. This is presented in Fig.~7 of
\citet{Hinshaw2003_orig}. [Note that this Figure has been replaced in the
final, published version of the paper \citep{Hinshaw2003} with a cross-band
power spectrum that shows some of the same features.]  The figure caption
instructs the reader to ``note that some of the `bite' features that appear in
the combined spectrum are not robust to data excision.''  Specifically, the
three deviations at $\ell\approx 22$, $40$ and $210$ are found \textit{only} in
the ecliptic polar data and \textit{not} in the ecliptic planar data.  This
suggests that there are some continuous parameters --- the latitude of the
planar-polar division, the orientation of the reference plane relative to the
ecliptic, \textit{etc.} --- as a function of which the separation is
approximately maximized by the WMAP team's choice of an ecliptic planar-polar
division.

A more detailed study is merited after future data releases.  In the meantime,
the dip at the first peak, while the least
significant, has the virtue of being in a region of the angular power
spectrum that has previously been probed by multiple experiments --- TOCO
\citep{Torbet1999,Miller}, Saskatoon \citep{Saskatoon1996}, Boomerang
\citep{Boomerang,Boomerang2005}, Maxima \citep{Maxima}, and Archeops
\citep{Archeops,Tristram:2004ke}.  Of these, only Archeops saw evidence of a
dip in the power at the first peak, though it was of limited statistical
significance.  Other experiments, saw no such dip with greater statistical
confidence.  However, other than WMAP only Archeops covered a significant
portion of the north ecliptic hemisphere; the others covered much less sky or
the southern-sky.  Thus if the dip in the first peak is localized to the
northern sky, especially to the region near the north ecliptic pole, all the
experiments could be consistent.
Notably, Archeops also shows evidence of excess of power at $\ell\simeq 40$.

Figure 7 of \citet{Hinshaw2003_orig} contains further perplexing anomalies in the
low-$\ell$ sky --- the angular power spectra extracted from the ecliptic planar
and polar regions disagree for $\ell<10$.  The differences are much larger than
would be expected from statistics alone, these low-$\ell$ $C_\ell$'s being very
well measured (as can be seen from the error bars in Fig.~8 of the same paper):
\newline\noindent 1) There is a nearly complete absence of ecliptic polar power
in the $\ell=6-7$ bin that is highly reminiscent of the findings in tables
\ref{tab:single_ell} and \ref{tab:angmom-dispersion} that $\ell=6$ is very
``planar.''  This suggests further that said planarity is closely aligned with
the ecliptic, and that this connection extends to $\ell=7$ as well. This is
also reflected by the fact that the best aligned vectors for $\ell = 6$ and
$\ell = 7$ from Sec.~\ref{sec:single_ell} are $46^{\circ}$ and $23^\circ$,
respectively, 
from the ecliptic pole, which already indicates that there is little power in
the (ecliptic) polar cap region defined as $\beta > 30^\circ$ in the WMAP
analysis.
\newline\noindent 2) The aplanarity of
$\ell=5$ (and to a lesser extent $\ell=4$), as seen again in tables
\ref{tab:single_ell} and \ref{tab:angmom-dispersion}, may also be reflected in
the notably even distribution of power between the plane and
poles.\newline\noindent 3) There is a dramatic deficit in ecliptic planar power
compared to ecliptic polar power at $\ell=2$ and $\ell=3$.

If indeed there is contamination of the microwave background from some source
in the ecliptic (north?) polar region that is responsible for the deviations at
$\ell\approx22,40$ and $210$, then it is possible that this contamination is also
the source of the low-$\ell$ microwave background in the polar regions.  In
this case, the maximum cosmic contribution to the CMB at low $\ell$'s would be
the lesser of the ecliptic planar and polar values (barring an exceptional
cancellation).  In particular, the cosmic $C_2$ and $C_3$ would be bounded by
the ecliptic (planar) values.  These are dramatically lower than the customary
values as extracted from the full sky with a Galactic cut, and considerably
less consistent with theoretical expectations than even the current low values.

\section{Conclusion}\label{sec:conclusion}

The multipole vector formalism first introduced to the study of the CMB by
\citet{Copi2004} has proven to be a useful means of studying the structure of
the CMB particularly on large scales. In this work, we have provided an
extensive discussion of the multipole vector formalism highlighting the fact
that the multipole vectors provide an alternative, complete representation of a
scalar function on a sphere (see Sec.~\ref{sec:MV} for details).  In particular,
we have pointed out that the algorithm of \citet{Copi2004} converts the
standard spherical harmonic decomposition into the multipole vector
representation first discussed by Maxwell (\ref{eqn:MMPV-decomp}).  We have
shown how the multipole vector formalism relates to the previously studied
maximum angular momentum dispersion (MAMD) directions, Land-Magueijo vectors,
and temperature minima/maxima directions.  Note that, unlike the multipole
vectors, neither the MAMD nor the minima/maxima directions contain the full
information of a multipole, and are thus not complete representations of the
microwave sky.  The Land-Magueijo vectors and scalars are a complete
representation, but suffer from a rapid proliferation of arbitrary choices for
$\ell>2$.

As noted, the multipole vectors are an excellent way to study alignments and
correlations in the microwave sky.  We have provided a qualitative description
of the striking properties of the quadrupole and octopole in
Sec.~\ref{sec:quad_oct}.  We note that there are strange properties for the
quadrupole and octopole individually as well as jointly.  Not all of these
unexpected properties are independent of each other and an explanation, whether
statistical fluke, residual foreground contamination, or real CMB features,
remains to be determined.

By eye the properties of the quadrupole and octopole multipole vectors seen in
Figs.~\ref{fig:map:tegmark:2}--\ref{fig:map:tegmark:2+3} are striking.  To
quantify these correlation we have used the $S$ and $T$ statistics for the
oriented area and normal vectors (see Sec.~\ref{sec:ST:def}) and applied
them to the quadrupole and octopole (see Sec.~\ref{sec:ST:QO}).  We confirm
the alignment of the quadrupole and octopole planes at greater than
$99$\% C.L\@. We also confirm that the quadrupole-octopole planes are aligned
with the geometry and direction of motion of the solar system.  In particular, they are
perpendicular to the ecliptic plane at approximately $98$\% C.L. and to the
dipole and equinox at $>99.8$\% C.L\@.  They are also perpendicular to the
Galactic poles at $>99$\% C.L.

We have shown that the alignment with the ecliptic plane remains at 99\%
C.L. when the quadrupole-octopole alignment is taken as given (for the TOH-DQ
map; similar or stronger results hold for the other maps, see
Sec.~\ref{sec:ecvsgal}).  The correlations with the dipole and equinox remain
at approximately the 95\% C.L\@.  However, the correlations with other
directions, such as the Galactic poles, do not persist.  This strongly supports
the reality of the ecliptic correlation in the data and suggests that the
aforementioned alignment with the Galactic poles is accidental.

We further stress that the 99\% C.L. correlation of the quadrupole-octopole
planes with the ecliptic plane is a lower bound.  As is evident from
Fig.~\ref{fig:map:tegmark:2+3} and discussed in more detail in
Sec.~\ref{sec:ecvsgal} the ecliptic plane carefully threads its way between the
temperature minima and maxima of the $\ell=2+3$ map separating the weak power
in the northern ecliptic sky from the strong power in the southern ecliptic
sky.  This extra feature that is manifest in the multipole vectors is not
contained in our statistics of oriented area (nor normal) vectors.  Thus the
analyses discussed in this work and in the literature which rely solely on
dot-products of the oriented area (or normal) vectors are not using all the
information available in the multipole vectors.  Dot-products of oriented area
(and normal) vectors are well suited for identifying and defining planes but do
so at the expense of the information of the structure with respect to these
planes. (For the quadrupole the area vector contains only three out of four
pieces of information, the normal only two.  For the octopole the complete
information is contained in the three normals, but the dot-product statistic
misses several degrees of freedom.)  We have estimated that including this
extra structure strengthens the 99\% C.L. bound on the correlation of the
quadrupole-octopole with the ecliptic plane to between $99.93$\% C.L. (for the
ILC map) and $99.996$\% C.L. (for the LILC map).

In this work, we have continued to use cleaned, full-sky maps produced from the
first year WMAP data.  The concern with using these maps is the potential for
residual Galactic foregrounds biasing the results.  Though it is difficult to
see how Galactic contaminations can lead to ecliptic correlations, we have
studied this in two different ways: we have explored the properties of the
foreground multipole vectors (in Sec.~\ref{sec:foregr}) and we have explored
how the multipole vectors and our results change when we perform a symmetric
cut across the Galactic (and other) planes (in Sec.~\ref{sec:cutsky}).
As expected a Galactic foreground is dominated by the $Y_{20}$ and
$\mathrm{Re}(Y_{31})$ modes.  These are very different from the modes that
dominate in the full-sky maps.  As shown in Fig.~\ref{fig:add_foregr} this
corresponds to the multipole vectors and normals for the full-sky maps being in
very different locations than those for the foregrounds.  As foreground
contamination is slowly added to the full-sky maps we see how the full-sky
multipole vectors move to the foreground multipole vectors.  We have found that
large foreground contaminations ($|c|\approx 0.3$ for the quadrupole,
$|c|\approx 1\mbox{--}3$ for the octopole) are required to make the full-sky
multipole vectors look like those from the known foreground maps.

An alternative to using foreground maps is to mask out all information in the
regions of the sky dominated by foregrounds.  This is a more conservative
approach but throws away information about the CMB in some regions of the sky.
For this reason any results from such a cut sky analysis will be weaker than
the corresponding full-sky analysis.  In Sec.~\ref{sec:cutsky}, we considered
symmetric cuts across the Galactic (and other) planes to access the effect on
the multipole vectors and correlations we have found from the full-sky
analyses.  As seen in Fig.~\ref{fig:cutsky} even small cuts lead to large
uncertainties in our results.  This is true independent of the plane about
which we cut.  The correlations we report from the full sky remain consistent
in the cut sky analysis but are weakened as expected.  This result is
consistent with the power equalization reconstruction by \citet{Bielewicz2005}
the cut sky analysis of \citet{SS2004}. 

As a final comparison of the quadrupole and octopole alignments we
calculated the correlations for COBE maps (see Sec.~\ref{sec:COBE}).
Again the results are consistent with those from the WMAP full-sky maps (as
shown in Fig.~\ref{fig:COBE}) but do not have the statistical significance.

The lack of power on the largest angular scales first observed by COBE and more
recently confirmed by WMAP has motivated much of the study of the low multipole
moments, in particular the quadrupole and octopole.  However, these are not the
only multipole bands where there are peculiar features in the power spectrum.
We have extended some of the studies to higher multipoles
(Secs.~\ref{sec:single_ell} and \ref{sec:angmom:statistic}).  Our tests
suggest that there may exist peculiarities in these multipole ranges not
solely in the power, but also in the structure of the multipoles.  These
studies, however, are not complete and thus it is not possible to assign
statistical significance to them.  They do, however, point the way for future
work.

To conclude, using the multipole vector decomposition we have shown that
the quadrupole and octopole of the microwave background sky are correlated
with each other at a level that is excluded from being chance in excess of
99\%.  This comes about from a preponderance of peculiar correlations and
is statistically independent of their observed lack of power.  This
observation is in bold contradiction to the predictions of pre-existing
cosmological model, and argues against an inflationary origin for these
fluctuations.  In addition, there is strong evidence (again of greater than
99\% confidence) that the microwave background at these multipoles is
correlated with the geometry and direction of motion of the solar system.  The observed
signal is most unlikely to be due to residual contamination of the
full-sky microwave background maps by known Galactic foregrounds.

These results strongly suggest that either there is additional, unexplained
foreground contamination of the microwave background, potentially from a source
local to our solar system or its neighborhood, or that there is an unexpected
systematic error in the WMAP maps.  We remain convinced by the WMAP team's
arguments that there is no unexpected systematic error (\citealt{Bennett_foregr};
see also \citealt{Finkbeiner2003}).  In particular, it is very hard to see how a
north-south ecliptic asymmetry, or a quadrupole-octopole plane perpendicular to
the ecliptic could be induced in the WMAP instrument or analysis pipeline.
COBE, with largely independent error sources, saw compatible correlations.
There is also the tantalizing suggestion by Archeops of a deficit in power near
the first peak which is localized on the sky to the region of the ecliptic
north pole.

The astute reader will note that we have persisted in our failure to offer either
a satisfactory possible explanation for an ecliptic-correlated foreground
(especially one apparently concentrated in a plane perpendicular to the
ecliptic) or a prediction that can be convincingly tested.  Both are
failings which we intend to remedy in the near future.  However, we note
that should indeed the low-$\ell$ microwave background prove to be
dominated by a new foreground, this would imply that, barring an unexpected
foreground alignment, the power in the underlying cosmic contribution at
these multipoles is likely to be suppressed below even the currently observed
too-low value.  It is at least amusing to note that the scale on which the
lack of large scale correlations is then manifested is comparable to the
horizon scale at the onset of cosmic acceleration.  At the least this
profound lack of large-angle correlations would further challenge generic
inflationary models, maybe even general relativity on the scale of the
observable universe will need to be reconsidered.    

Whatever the origin of these low-$\ell$ correlations, it is clearly necessary
to reconsider any inferences drawn from the low-$\ell$ WMAP data, including the
temperature-polarization cross-correlation.  For example, our work suggests
that the evidence for early reionization of the universe, resting as it does on
the low-$\ell$ TE angular power spectrum should be viewed with a
healthy dose of skepticism.

On the experimental side, we are eagerly waiting for two new
major and largely independent probes of the large-angle microwave
background radiation that may shed new light on the anomalies discussed in
this paper: polarization measurements by WMAP (though they are expected to
be systematics-dominated on large scales), and measurements of temperature
and polarization by the Planck experiment.

\section*{Acknowledgments}

We thank Tom Crawford, Olivier Dor\'{e}, Hans-Kristian Eriksen, Doug
Finkbeiner, Chris Gordon, Gary Hinshaw, Wayne Hu, Kate Land, Petra Lutter, Joao
Magueijo, Stephan Meyer, Hiranya Peiris, Aleksandar Raki\'c, Syksy
R\"as\"anen, An\v{z}e Slosar, Uro\v{s} Seljak, David Spergel, and Igor Tkachev for useful
conversations; Jeff Weeks for useful comments and suggesting the use of $S$ and
$T$ statistics, and the Aspen Physics Center for hospitality.  DH is supported
by the NSF Astronomy and Astrophysics Postdoctoral Fellowship under Grant No.\
0401066. DJS acknowledges travel support from the DFG. Over the course of this
work GDS has been supported by CERN and by NASA ATP and DOE grants to the CWRU
particle-astrophysics theory group, and by the John Simon Guggenheim Memorial
Foundation.  We have benefited from using the publicly available
\textsc{Healpix} package~\cite{healpix}. GDS also thanks Maplesoft for use of
the \textsc{Maple} software.

\newcommand{\apj}{ApJ}
\newcommand{\apjs}{ApJS}
\newcommand{\mnras}{MNRAS}
\newcommand{\physrev}{Phys. Rev.}
\newcommand{\physrevlett}{Phys. Rev. Lett.}

\bibliographystyle{mn2e}
\bibliography{lowl2}


\bsp

\label{lastpage}

\end{document}